\documentclass[11pt]{article}
\pdfoutput=1
\usepackage{amsmath}
\usepackage{amssymb}
\usepackage{graphicx,bbm,mathrsfs}
\usepackage{nicefrac}
\usepackage{slashed}
\usepackage{bbm}
\usepackage{geometry}
\usepackage{empheq}
\usepackage{stackrel}
\usepackage{ulem}
\usepackage[shortlabels]{enumitem}
\usepackage{placeins}
\usepackage{setspace}
\usepackage{diagbox}

\usepackage{jheppub}

\usepackage{mathtools}
\usepackage{enumitem}
\usepackage{dcolumn}   
\usepackage{bm}        
\usepackage{graphicx,mathrsfs}
\usepackage{nicefrac}
\usepackage{multirow}
\usepackage{color}
\usepackage{makecell}

\usepackage{environ} 
\usepackage{lipsum} 
 \NewEnviron{Smaller11}{
           \scalebox{1.1}{$\BODY$} 
 } 
 \NewEnviron{Smaller08}{
           \scalebox{0.8}{$\BODY$} 
 } 

\usepackage{tikz}
\newcommand*\circled[1]{\tikz[baseline=(char.base)]{
            \node[shape=circle,draw,inner sep=2pt] (char) {#1};}}

\newcommand{\be}{\begin{equation}}
\newcommand{\ee}{\end{equation}}
\newcommand{\bea}{\begin{eqnarray}}
\newcommand{\eea}{\end{eqnarray}}

\newcommand{\n}{n}
\newcommand{\K}{K}

\renewcommand{\b}{b}

\newcommand{\C}{C}
\newcommand{\bC}{\bar C}
\newcommand{\A}{\alpha}
\newcommand{\bA}{\bar\alpha}
\newcommand{\B}{\beta}
\newcommand{\bB}{\bar\beta}
\newcommand{\bAUV}{\bar\alpha_{\rm UV}}
\newcommand{\bBUV}{\bar\beta_{\rm UV}}

\newcommand{\alphaUV}{\alpha_{\rm UV}}
\newcommand{\betaUV}{\beta_{\rm UV}}

\usepackage{wasysym}
\usepackage{hhline,colortbl}

\newcommand\mydiagbox[2]{\hbox{\tabcolsep=\arraycolsep\diagbox{$#1$}{$#2$}}}

\newcommand{\nn}{\nonumber}

\renewcommand{\d}{d } 
\newcommand{\g}{g}

\newcommand{\M}{M}

\usepackage{titlesec}

\titleformat*{\section}{\Large\bfseries}
\titleformat*{\subsection}{\large\bfseries}
\titleformat*{\subsubsection}{\large\bfseries}
\titleformat*{\paragraph}{\large\bfseries}
\titleformat*{\subparagraph}{\large\bfseries}

\makeatletter
\DeclareFontFamily{OMX}{MnSymbolE}{}
\DeclareSymbolFont{MnLargeSymbols}{OMX}{MnSymbolE}{m}{n}
\SetSymbolFont{MnLargeSymbols}{bold}{OMX}{MnSymbolE}{b}{n}
\DeclareFontShape{OMX}{MnSymbolE}{m}{n}{
    <-6>  MnSymbolE5
   <6-7>  MnSymbolE6
   <7-8>  MnSymbolE7
   <8-9>  MnSymbolE8
   <9-10> MnSymbolE9
  <10-12> MnSymbolE10
  <12->   MnSymbolE12
}{}
\DeclareFontShape{OMX}{MnSymbolE}{b}{n}{
    <-6>  MnSymbolE-Bold5
   <6-7>  MnSymbolE-Bold6
   <7-8>  MnSymbolE-Bold7
   <8-9>  MnSymbolE-Bold8
   <9-10> MnSymbolE-Bold9
  <10-12> MnSymbolE-Bold10
  <12->   MnSymbolE-Bold12
}{}

\let\llangle\@undefined
\let\rrangle\@undefined
\DeclareMathDelimiter{\llangle}{\mathopen}%
                     {MnLargeSymbols}{'164}{MnLargeSymbols}{'164}
\DeclareMathDelimiter{\rrangle}{\mathclose}%
                     {MnLargeSymbols}{'171}{MnLargeSymbols}{'171}
\makeatother

\begin{document}

\vspace*{4mm}

\begin{center}

\thispagestyle{empty}
{
\LARGE\sc
Gravity-Induced Photon Interactions
 \\ \vspace{0.5cm}and Infrared Consistency in any Dimensions
}\\[12mm]

\renewcommand{\thefootnote}{\fnsymbol{footnote}}

{\large  
Pedro~Bittar$^{\,a}$ \footnote{pedro.bittar.souza@usp.br}\,,
Sylvain~Fichet$^{\,b}$ \footnote{sylvain.fichet@ufabc.edu.br}\,, 
Lucas~de~Souza$^{\,c}$  \footnote{souza.l@ufabc.edu.br }
}\\[12mm]
\end{center} 
\noindent

${}^a\!$ 
\textit{Department of Mathematical Physics, University of S\~ao Paulo,\\
\indent \, S\~ao Paulo, 05508-090 SP, Brazil}

${}^b\!$ 
\textit{CCNH, Federal University of ABC,}  \textit{Santo Andr\'e, 09210-580 SP, Brazil}

${}^c\!$ 
\textit{CMCC, Federal University of ABC,} \textit{Santo Andr\'e, 09210-580 SP, Brazil}

\addtocounter{footnote}{-1}

\vspace*{15mm}

\begin{center}
{  \bf  Abstract }
\end{center}
\begin{minipage}{15cm}
\setstretch{0.97}
\small

We compute the four-photon ($F^4$) operators  generated by loops of charged particles of spin $0$, $\frac{1}{2}$, $1$ in the presence of gravity  and in any spacetime dimension $d$. 
To this end, we expand the one-loop effective action via the heat kernel coefficients, which capture both the gravity-induced renormalization of the $F^4$ operators and the low-energy Einstein-Maxwell effective field theory (EFT) produced by massive charged particles. 
We set  positivity bounds on the $F^4$ operators using standard arguments from extremal black holes 
(for  $d\geq 4$)
and from infrared (IR) consistency of four-photon scattering (for  $d\geq 3$). 
We find that both approaches yield nearly equivalent results, even though in the amplitudes we discard the graviton $t$-channel pole  and use the vanishing of the Gauss-Bonnet term  at quadratic order for any $d$.

The  positivity bounds 
constrain the charge-to-mass ratio of the heavy particles.  
If the Planckian $F^4$ operators are sufficiently small or negative,  such bounds produce a version of the  $d$-dimensional Weak Gravity Conjecture (WGC) in most, but not all, dimensions. 
In the special case of $d=6$, the gravity-induced beta functions of $F^4$ operators from charged particles of any spin are positive, leading to  WGC-like bounds with a logarithmic enhancement. 
In $d=9,10$,  the WGC fails to guarantee  extremal black hole decay in the infrared EFT, thereby requiring the existence of sufficiently large Planckian $F^4$ operators.

    \vspace{0.5cm}
\end{minipage}

\newpage
\setcounter{tocdepth}{2}
\tableofcontents

\newpage

\section{Introduction}
\label{se:intro}

In the quest to unravel the mysteries of quantum gravity, one route involves a thorough examination of gravitational effective field theories (EFTs) that appear  below the Planck scale.
Such gravitational EFTs are constrained by black hole physics and, like non-gravitational ones, by infrared consistency conditions based on  unitarity and causality.   
In the presence of gravity,   ultra\-violet (UV) and infrared (IR) scales seem to feature intricate connections, already at the classical level as hinted by black hole properties.  
This implies that, even though the completion of quantum gravity lies far in the  UV, we can  hope to gain insights  by scrutinizing gravitational EFTs in the IR.

The notion of IR consistency of EFTs, that implies bounds on certain Wilson coefficients,  has been introduced  in \cite{Pham:1985cr,Ananthanarayan:1994hf,Adams:2006sv}. It  has led to many subsequent developments, see e.g. 
\cite{Arkani-Hamed:2020blm,Alberte:2020jsk,Alberte:2020bdz,Henriksson:2021ymi,Henriksson:2022oeu,Bellazzini:2020cot,Caron-Huot:2021enk,Bellazzini:2021oaj,Caron-Huot:2020cmc,Caron-Huot:2021rmr,Davighi:2021osh,deRham:2021fpu,Caron-Huot:2022ugt,Tolley:2020gtv,deRham:2021bll,deRham:2022hpx,Chiang:2022ltp,Caron-Huot:2022jli,Haring:2022sdp,Hamada:2023cyt,Bellazzini:2023nqj,Eichhorn:2024wba,Knorr:2024yiu}. 
 Black hole physics is also   used to put bounds on gravitational EFTs 
\cite{Kats:2006xp,Cheung:2018cwt,Loges:2019jzs,Goon:2019faz,Jones:2019nev,Loges:2020trf,Arkani-Hamed:2021ajd,Cao:2022iqh,DeLuca:2022tkm}. 
The consistency of EFTs with the UV completion of quantum gravity
 has been explored via  ``swampland'' conjectures, see e.g. the first  weak gravity conjecture \cite{Arkani-Hamed:2006emk}  and recent reviews \cite{vanBeest:2021lhn,Grana:2021zvf, Agmon:2022thq}. 
Conversely, the IR consistency of gravitational EFTs 
constrains UV completions of quantum gravity, and  thus has an interplay  with 
swampland conjectures
\cite{Cheung:2014ega,Bellazzini:2015cra, 
Cheung:2016wjt,Hamada:2018dde, Bellazzini:2019xts,
Chen:2019qvr, 
Arkani-Hamed:2021ajd,Bern:2021ppb,Knorr:2024yiu}.
  The present work is in the spirit of  the latter approach: carving out the space of gravitational EFTs  from the IR, using consistency conditions from both scattering  amplitudes and black holes.

Our focus in this work is on gravitational EFTs  that feature an Abelian gauge symmetry. We refer to the Abelian gauge field as the photon. 
We consider an EFT arising  below the Planck scale $M_{\rm P}\equiv M$, the \textit{ultraviolet} EFT, that  features a  charged particle. 
We consider  charged particles with spin $0,\frac{1}{2},1$ and arbitrary spacetime dimensions $\d$.

The sub-Planckian EFT features, in general, local four-photon operators that we denote here  collectively as $F^4$. The $F^4$ coefficient  $\alpha_{\rm UV}$ encapsulates the subPlanckian effects from the  superPlanckian UV completion.
Depending on the spacetime dimension, loops of charged particles may renormalize the $F^4$ operators, in which case  the value $\alpha_{\rm UV}$ is understood as the value of the coefficient at the Planck scale. As a first step, we will compute at one-loop  this $F^4$  renormalization flow, that occurs regardless of whether the particle is massless or massive.

Additionally, when the charged particle is massive, it can be integrated out when the renormalization scale is much  lower than the particle mass. This produces another, \textit{infrared} EFT  whose only degrees of freedom are the photon and the graviton.   EFTs of this kind are usually referred to as Einstein-Maxwell theory, here we mostly use the term IR EFT.   The 
IR EFT contains a $F^4$ operator with coefficient
\be
\alpha_{\rm IR}=\alpha_{\rm UV} +\Delta \alpha \, \label{eq:alphaIR_schem}
\ee
where the $\Delta\alpha$ corrections  take schematically  the form
\be
\Delta \alpha= a \frac{g^4q^4}{m^{8-\d}}  
+  b \frac{g^2q^2}{m^{6-\d} \M^{\d-2}}
+ \frac{c}{ m^{4-\d}\M^{2\d-4}} \,,
\label{eq:DeltaCorrections}
\ee
where $m$ is the charged particle mass. In certain dimensions, some of the coefficients are enhanced by  $\log(\frac{M}{m})$ terms produced by the renormalization flow.

The various scales and EFTs are summarized as follows: 
\begin{figure}[h!]
    \centering
    \includegraphics[width=0.8\linewidth,trim={0cm 3cm 0cm 3.5cm},clip]{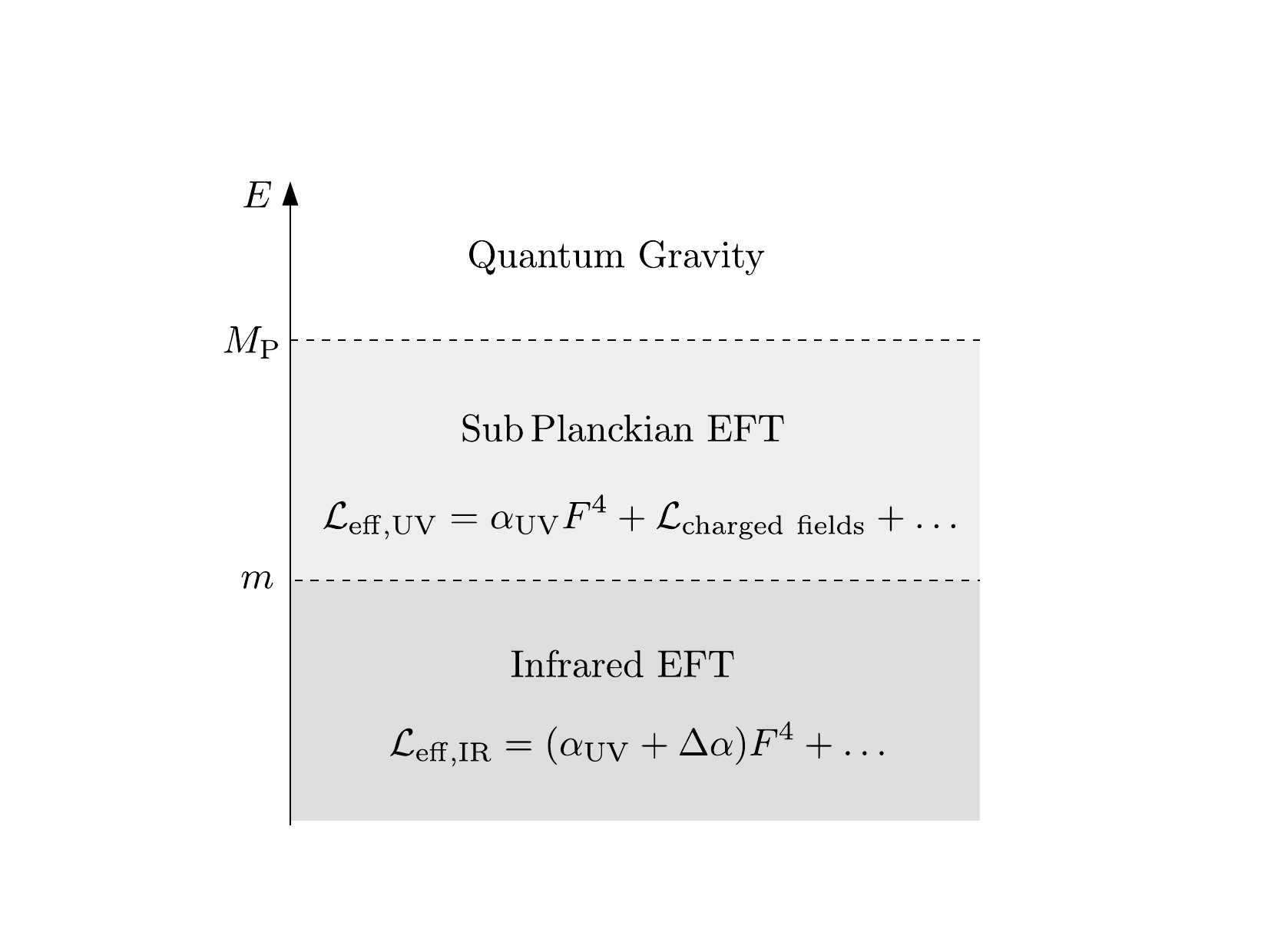}
    \label{fig:EFT}
\end{figure}

 Our approach in this work is  to remain fully agnostic about the UV completion of quantum gravity. 
We work with quantum field theory --- and do not perform actual quantum gravity calculations. The only explicit matching to a specific string theory is given in App.\,\ref{se:strings} as an example. The strength of the EFT approach is that our ignorance of the UV completion of quantum gravity  is encoded into the values of the $\alphaUV$ coefficients, which are treated here  as free parameters. 

 We perform the computation of both the $F^4$ one-loop beta functions and of  the IR EFT directly from the one-loop effective action. 
We use the background field method and the heat kernel formalism  (see \cite{Gilkey_original, Vassilevich:2003xt,vandeVen:1984zk},  other useful references are \cite{Fradkin:1983jc,Metsaev:1987ju, Bastianelli:2008cu,Ritz:1995nt,Hoover:2005uf})
combined with standard EFT techniques.

The sequence of EFTs can be used to compute the contribution of the charged particle to the physical process of four-photon scattering $\gamma\gamma\to \gamma\gamma$ at low energy. 
Such a process is subject to IR consistency bounds from unitarity and causality. In an appropriate basis, IR consistency implies positivity of the $F^4$ coefficients, schematically\,\footnote{
For simplicity, we use the most basic positivity bound from \cite{Adams:2006sv}, that we extend to higher dimensions. 
More refined approaches have been developed, see e.g. \cite{Arkani-Hamed:2020blm,Alberte:2020jsk,Alberte:2020bdz,Henriksson:2021ymi,Henriksson:2022oeu,Bellazzini:2020cot,Caron-Huot:2021enk,Bellazzini:2021oaj,Caron-Huot:2020cmc,Caron-Huot:2021rmr,Davighi:2021osh,deRham:2021fpu,Caron-Huot:2022ugt,Tolley:2020gtv,deRham:2021bll,deRham:2022hpx,Chiang:2022ltp,Caron-Huot:2022jli,Haring:2022sdp,Hamada:2023cyt,Bellazzini:2023nqj}, that are not the focus of this work.  }
\be
\alpha_{\rm IR}\geq 0\,. \label{eq:positivity_schem}
\ee

The IR EFT can also be used to compute the metric of 
non-rotating charged black holes with  large enough radius
\cite{Abe:2023anf,Barbosa:2025smt}.  Requiring that such extremal black holes be able to decay produces  a positivity bound  similar to \eqref{eq:positivity_schem}.\,\footnote{ The extremal black hole decay condition is sometimes referred to as the \textit{black hole} WGC. Here we do not use this naming, the term WGC only refers to the condition on the charged particle, \eqref{eq:WGC_schem}.}

Combining the positivity bound \eqref{eq:positivity_schem}  with Eqs.\,\eqref{eq:alphaIR_schem}\,,\eqref{eq:DeltaCorrections}, one can notice that for appropriate values and signs of $\alpha_{\rm UV}$, $a$, $b$, $c$,  a lower bound appears on  the charge-to-mass ratio,
\be
\frac{g|q| }{m}\M^{\frac{\d-2}{2}}\equiv z \geq z_*\, \label{eq:WGC_schem}\,,
\ee
with $z_*$ a dimensionless number  dependent on $\alpha_{\rm UV}$, $a$, $b$, $c$. 
Eq.\,\eqref{eq:WGC_schem} is precisely the parametric form of the Weak Gravity Conjecture (WGC) in $\d$-dimensions (see \cite{Palti:2019pca}), in which case the bound is $z_*=\sqrt{\frac{\d-3}{\d-2}}$. 
For $z_*$ being a generic  $O(1)$ coefficient, we refer to bounds of the form \eqref{eq:WGC_schem} as \textit{WGC-like}. 
This nontrivial connection between IR and UV consistency was first  pointed out in \cite{Cheung:2014ega} for $d=3,4$. 
Because the pattern of signs and divergences of the $a,b,c$ coefficients is dimension-dependent, the generalization of this phenomenon to arbitrary $d$ is nontrivial and requires a thorough investigation. Here we explore the IR consistency/WGC connection  in arbitrary dimension and revisit the $d=3,4$ cases.

Why might one study arbitrary spacetime dimensions in the first place? While the known real world displays $d=3+1$ dimensions, it is plausible that extra dimensions exist ---  in particular,  string theory requires $d=11$ for consistency. These dimensions may be hidden from us, either because they are compact or because our matter is confined to a three-brane within a higher-dimensional bulk spacetime. 
More specifically, in  this work, we extend  to higher dimensions as a tool to probe the relations among several concepts: the WGC,  the decay of extremal black holes (see Sec. \ref{se:BH}), and the IR consistency of scattering  amplitudes. 
We investigate, for any $d\geq 4$, to what extent the WGC implies the decay of extremal black holes of any size, and whether IR consistency implies extremal black hole decay and the WGC. 
In our approach, testing whether a relation between two concepts holds for any $d$ serves as a check of its robustness. 
We take the viewpoint that, if a given relation qualitatively changes with spacetime dimension, it is unlikely to reflect a deep physical principle, and should instead regarded as coincidental. 
 In sum, the extension to arbitrary $d$
 provides a diagnostic for sharpening and validating our understanding of gravity in $d=4$.

\subsection*{Outline}

In section \ref{se:EM} we review  EFT beyond tree-level from the viewpoint of the quantum effective action. We define the Einstein-Maxwell EFT and show how to reduce it using a property of the Gauss Bonnet term valid in any dimensions. 
In section \ref{se:F4}, we review the bound from extremal black hole decay and
generalize simple positivity bounds from four-photon scattering to any dimension. 
In  section \ref{se:OneLoopEFT}, after reviewing the heat kernel coefficients, we derive and reduce the general one-loop effective action 
obtained from integrating out charged particles of spin $0$, $\frac{1}{2}$ and $1$. 
In section \ref{se:beta_functions}, we present the $F^4$ beta functions and discuss their interplay with IR consistency. 
In section \ref{se:finite_corrections}, we analyze in detail the IR consistency of the infrared EFT, with a systematic discussion from $d=3$ to $11$.  
Section \ref{se:conclusion} contains a detailed summary, and the Appendix contains 
some examples of UV realizations of the $F^4$ operators (\ref{se:strings}), 
  the detailed analysis of the bound from infrared consistency  (\ref{app:IR}) and   extremal black hole decay (\ref{app:BH}), and the complete heat kernel coefficients (\ref{app:HK}).

\section{Einstein-Maxwell EFT in Any Dimensions} \label{se:EM}

We briefly review the notion of loop-level low-energy EFT from the viewpoint of the quantum effective action in subsection \ref{se:EFT}. We then define Einstein-Maxwell EFT in subsection \ref{se:EMEFT} and show how to reduce it  to describe photon scattering in subsections \ref{se:GB}, \ref{se:EMEFT_reduced}.

\subsection{Effective Action and Effective Field Theory}
\label{se:EFT}

Consider a  theory with light fields $\Phi_\ell$ and heavy fields $\Phi_h$. Assume that our interest lies in  the scattering amplitudes of the light fields. Such scattering amplitudes are obtained by taking functional derivatives of the generating functional of connected correlators with respect to sources probing the light fields $J_\ell$. This generating functional is $W[J_\ell]=i\log Z[J_\ell]$ with  the partition function
\be
Z[J_\ell] = \int {\cal D}\Phi_\ell{\cal D}\Phi_h e^{iS[\Phi_\ell,\Phi_h]+i\int d^dx \Phi_\ell J_\ell}\,. 
\ee
We perform the $\Phi_h$ field integral in the partition function. This defines a ``partial'' quantum effective action $\Gamma_h[\Phi_\ell]$, with  
\be
Z[J_\ell] =\int {\cal D}\Phi_\ell  e^{i\Gamma_h[\Phi_\ell]+i\int dx^d \Phi_\ell J_\ell}\,.
\ee

Let us consider the low-energy regime for which the external momenta of the $\Phi_\ell$ amplitudes are much smaller than the mass of the heavy fields, noted $m$. In this limit, the  quantum effective action $\Gamma_h$ can be organized as an expansion in powers of derivatives over $m$.
This is conveniently expressed as an effective Lagrangian ${\cal L }_{\rm eff}$  
\be
\Gamma_h[\Phi_\ell] \equiv \int d^dx \sqrt{-g}\, {\cal L}_{\rm eff}[\Phi_\ell] \ee
where  ${\cal L }_{\rm eff}$  is made of monomials of $\Phi_\ell$ and its derivatives, suppressed by powers of $m$. Schematically,
\be {\cal L}_{\rm eff}[\Phi_\ell]\sim \sum_{a,b} \frac{\Phi_\ell^a(\partial \Phi_\ell)^{2b}}{m^{a+4b-4}}\,. \ee
In practice, ${\cal L }_{\rm eff}$   is typically truncated at some order of the derivative expansion $\partial/m$. This defines an infrared effective field theory (EFT)  that encodes all the effects of the $\Phi_h$ field at energies below $m$, within the accuracy of the truncation of ${\cal L}_{\rm eff}$. 

The derivative expansion applies at each order of the loop expansion of $\Gamma_h$, $\Gamma_h=\Gamma^{(0)}_h+\Gamma^{(1)}_h+\ldots$  Hence the effective Lagrangian can be organized with respect to this loop expansion: ${\cal L}_{\rm eff}={\cal L}^{(0)}_{\rm eff}+{\cal L}^{(1)}_{\rm eff}+\ldots .$ The ${\cal L}^{(0)}_{\rm eff}$ term arises from the tree diagrams involving  $\Phi_h$ encoded in $\Gamma^{(0)}_h$. 
The ${\cal L}^{(1)}_{\rm eff}$ arises from the one-loop diagrams involving $\Phi_h$  encoded in $\Gamma^{(1)}_h$, etc. 

In this paper, we work  at the one-loop level. The finer details of EFT at loop level can be found in \cite{Manohar:1996cq,Manohar:2018aog}.

\subsection{Einstein-Maxwell EFT} 
\label{se:EMEFT}

Consider a gravitational theory with a $U(1)$ gauge symmetry and  massive matter fields.\,\footnote{Throughout this work we use the conventions of Misner-Thorne-Wheeler  \cite{Misner:1973prb}, which include the mostly-plus  metric signature ${\rm sgn}(g_{\mu\nu})=(-,+,\ldots,+)$ and positive scalar curvature for spheres. }
Our interest is in  the scattering amplitudes of the photons  of this theory, i.e. the photon is coupled to a source $J_\gamma$ that generates the amplitudes. 
As explained in section \ref{se:EFT}, we can always integrate out the matter fields exactly, defining a partial quantum effective action $\Gamma_{\rm mat}[F_{\mu\nu}, R_{\mu\nu\rho\sigma}]$.

In the regime for which the external momenta of amplitudes are smaller than the matter field masses $m$, the  quantum effective action can be written as   
\be
\Gamma_{\rm mat}[F_{\mu\nu}, R_{\mu\nu\rho\sigma}] \equiv \int d^dx \sqrt{-g}\, {\cal L}_{\rm eff}[F_{\mu\nu}, R_{\mu\nu\rho\sigma}] \ee
where ${\cal L}_{\rm eff}$ is made of monomials of $F_{\mu\nu}$ and $R_{\mu\nu\rho\sigma}$. 
This defines a low-energy EFT that encodes all the effects of the matter fields at energies below $m$. In this subsection, we refer to this EFT as the Einstein-Maxwell EFT, with ${\cal L}_{\rm eff}\equiv {\cal L}_{\rm EM}$. 
At $\partial^4$ order, the most general  Einstein-Maxwell Lagrangian takes the form\,\footnote{
We assume a spacetime background with no boundary, so that total derivative terms in ${\cal L}_{\rm EM}$ can be ignored,  and operators related by integration by parts are considered redundant.  }  
\begin{eqnarray}
{\cal L}_{\rm EM} =  {\cal L}_{\rm kin} &&
+ \alpha_1 (F^{\mu\nu}F_{\mu\nu } )^2 + \alpha_2 F^{\mu\nu}F_{\nu\rho } F^{\rho \sigma } F_{\sigma \mu} \nn
\\
&& \nn 
 + \alpha_3 \hat R^2 + \alpha_4 \hat R_{\mu\nu}\hat R^{\mu\nu } + \alpha_5 \hat R_{\mu\nu\rho\sigma}\hat R^{\mu\nu\rho\sigma}
\\
&& \nn  
+ \alpha_6 \hat R F^{\mu\nu}F_{\mu\nu}+ \alpha_7 \hat R_{~\,\mu}^\nu F^{\mu \rho}F_{\nu \rho }  + \alpha_8 \hat R_{~~~\mu\nu}^{\rho\sigma} F^{\mu \nu}F_{\rho \sigma} 
 \\
&& \nn
+ \alpha_9 (D_\rho F_{\mu\nu})(D^\rho F^{\mu\nu})+ \alpha_{10} (D_\rho F_{\mu\nu})(D^\mu F^{\rho\nu})  + \alpha_{11} (D^\mu F_{\mu\nu})^2
 \\
&&
+O(\hat R^3, \hat R^2 F^2, \hat RF^4,   F^6) \label{eq:EMEFT_lag1}
\end{eqnarray}
with the kinetic term
\be
{\cal L}_{\rm kin} =  -\frac{1}{4}F_{\mu\nu}F^{\mu\nu} +\frac{1}{2}\hat R\,. 
\ee
We introduced the normalized Riemann tensor $\hat R_{\mu\nu\rho\sigma} \equiv  M^{d-2} R_{\mu\nu\rho\sigma}$.

At tree-level, the effective operators in ${\cal L}_{\rm EM}$ contribute  to the four-photon amplitude as follows: 
\be 
	\includegraphics[width=0.15\linewidth,trim={0cm 0cm 0cm 0cm},clip]{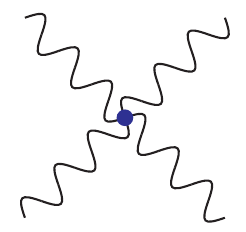} \hspace{1cm}
 	\includegraphics[width=0.22\linewidth,trim={0cm 0cm 0cm 0cm},clip]{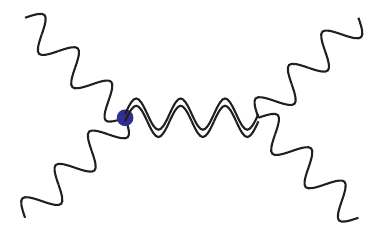}\hspace{1cm}
   	\includegraphics[width=0.22\linewidth,trim={0cm 0cm 0cm 0cm},clip]{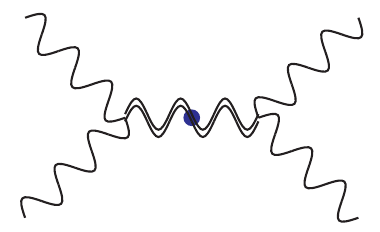}
 \label{diag:F4_EM}
 \ee
The  dots represent the effective vertices from  ${\cal L}_{\rm EM}$ and the double wiggles represent gravitons. Notice that the curvature operators contribute indirectly via modifications of the graviton-photon vertices and to the graviton propagator. 

These diagrams can be simplified using that the physical scattering amplitudes   are invariant under field redefinitions. Following the general lessons of EFT, a subset of such field redefinitions amounts to using the leading order equations of motion in the effective Lagrangian \cite{Manohar:2018aog}. 
In the present case, the equation of motion that we can use are  the Maxwell and Einstein equations $D^\mu F_{\mu\nu}=0$ and $R_{\mu\nu}-\frac{1}{2}R g_{\mu\nu} =\frac{1}{M^{d-2}} T_{\mu\nu} $. The latter implies 
\be
\hat R_{\mu\nu} = T_{\mu\nu}-\frac{1}{d-2}T g_{\mu\nu}\,,\quad\quad \hat R= \frac{2}{2-d}T \label{eq:EEs}
\ee
with 
\be
T_{\mu\nu}= - F_{\mu\rho}F^{\rho}_{~~\nu}- \frac{1}{4}g_{\mu\nu}(F_{\rho\sigma})^2\,,\quad\quad T= \frac{4-d}{4} (F_{\mu\nu})^2
\label{eq:T_identities}
\ee

Using the Maxwell equation, the last operator in ${\cal L}_{\rm EM}$ vanishes. Furthermore, the two other operators involving $D_\mu F_{\rho\sigma}$ can be transformed into combinations of the remaining terms in ${\cal L}_{\rm EM}$ using  the Bianchi identities  $F_{[\mu\nu;\rho]}=0$, $R_{\mu[\nu\rho\sigma]}=0$, and the Ricci identity $[D_\mu,D_\nu]F_{\rho\sigma}=R_{\mu\nu\rho}^{~~~~\lambda} F_{\lambda\sigma}+ R_{\mu\nu\sigma}^{~~~~\lambda} F_{\rho\lambda} $.\,\footnote{For example, one finds  \be
D_{\rho}F_{\mu\nu}D^{\rho}F^{\mu\nu}=  R_{\mu\nu\rho\sigma}F^{\mu\nu}F^{\rho\sigma} -2 R_{\mu}^{~\nu}F^{\mu\rho}F_{\nu\rho} +2 (D^{\mu}F_{\mu\nu})^2
+{\rm total ~derivative } 
\ee}
Finally, the identities from \eqref{eq:EEs}  can be used to  eliminite the $R$ and $R_{\mu\nu}$ in the remaining operators of ${\cal L}_{\rm EM}$. The traceless part of the Riemann tensor, i.e. the Weyl tensor, 
\be
C_{\mu\nu\rho\sigma}= 
R_{\mu\nu\rho\sigma} -\frac{2}{d-2}\left(
g_{\mu[\rho}R_{\sigma] \nu} - g_{\nu[\rho}R_{\sigma] \mu}
\right) + 
\frac{2}{(d-1)(d-2)}R g_{\mu [\rho}g_{\sigma] \nu}
\ee
still remains in ${\cal L}_{\rm EM}$ in the form of operators $C^2$ and  $CF^2$.

\subsection{Reducing the  Curvature Squared Terms}
\label{se:GB}

We can further reduce the basis of operators by noticing that, for the four-photon diagrams of our interest, the graviton self-interactions are irrelevant since only the graviton propagator appears in \eqref{diag:F4_EM}. 
Let us  inspect  the quadratic curvature corrections to the graviton propagator. 

We know that the Riemann tensor goes as $R_{\mu\nu\rho\sigma}\propto \partial_{\mu}\partial_{\sigma}  h_{\nu\rho}+\ldots$ upon the expansion of the metric $g_{\mu\nu}=\eta_{\mu\nu}+h_{\mu\nu}$. 
It is thus sufficient to keep the linear term in each curvature term to obtain the quadratic vertices that correct the graviton propagator. 
We have 
\be
R_{\mu\nu\rho\sigma}=\frac{1}{2}\left( \partial_\mu\partial_\sigma h_{\nu\rho}- \partial_\nu\partial_\sigma h_{\mu\rho} - \partial_\mu\partial_\rho h_{\nu\sigma}+ \partial_\nu\partial_\rho h_{\mu\sigma} \right)+O(h^2)
\ee
\be
R_{\nu\sigma}=\frac{1}{2}\left( \partial_\mu\partial_\sigma h_{\nu}^\mu + \partial_\nu\partial^\mu h_{\mu\sigma}  - \partial_\nu\partial_\sigma h - \square h_{\nu\sigma}\right)+O(h^2)
\ee
\be
R= \partial_\mu\partial_\nu h^{\mu\nu}-\square h
\ee
with $h_{\mu}^\mu=h$, $\partial_\mu\partial^\mu=\square$. 
Going to  Fourier space for simplicity, we find the curvature-squared terms
\be
R^2= h^{\mu\nu}h^{\alpha\beta} {\cal O}^{(1)}_{\mu\nu,\alpha\beta}
\ee
\be
(R_{\mu\nu})^2= h^{\mu\nu}h^{\alpha\beta} {\cal O}^{(2)}_{\mu\nu,\alpha\beta}
\ee
\be
(R_{\mu\nu\rho\sigma})^2= h^{\mu\nu}h^{\alpha\beta} {\cal O}^{(3)}_{\mu\nu,\alpha\beta}
\ee
where
\be
 {\cal O}^{(1)}_{\mu\nu,\alpha\beta}= \left(
 p_\mu p_\nu-p^2\eta_{\mu\nu}
 \right)
 \left(
 p_\alpha p_\beta-p^2\eta_{\alpha\beta}
 \right) \label{eq:O1GB}
\ee
\be
 {\cal O}^{(2)}_{\mu\nu,\alpha\beta}=\frac{1}{4}\left(2p_{\mu}p_{\nu}p_{\alpha}p_{\beta}+p^4 (\eta_{\mu\nu}\eta_{\alpha\beta}+\eta_{\mu\alpha}\eta_{\nu\beta})
 - p^2(p_\mu p_\nu \eta_{\alpha\beta}+p_\alpha p_\beta \eta_{\mu\nu}+p_\mu p_\beta \eta_{\nu \alpha}
 +p_\nu p_\alpha \eta_{\mu \beta})
 \right) \label{eq:O2GB}
\ee
\be
 {\cal O}^{(3)}_{\mu\nu,\alpha\beta}=\frac{1}{4}\left(4 p^4 \eta_{\mu\alpha}\eta_{\nu\beta}+4p_\mu p_\nu p_\alpha p_\beta - 2 p^2(\eta_{\mu\alpha}p_\nu p_\beta + \eta_{\mu\beta}p_\nu p_\alpha + \eta_{\nu\alpha}p_\mu p_\beta
 + \eta_{\nu\beta}p_\mu p_\alpha)\right)
 \label{eq:O3GB}
\ee

Inspecting Eqs.\,\eqref{eq:O1GB},\,\eqref{eq:O2GB},\,\eqref{eq:O3GB}, we find that  the following combination  vanishes at quadratic order in \textit{any} dimension: 
\be
(R_{\mu\nu\rho\sigma})^2-4(R_{\mu\nu})^2+R^2 = 0+O(h^3) \,.
\label{eq:GBapprox}
\ee
This is the familiar Gauss-Bonnet (GB) combination. The fact that it vanishes at  $O(h^2)$ for arbitrary $d$ was first noticed in \cite{ZWIEBACH1985315} in the context of the low-energy limit of string theories.\,\footnote{ 
The GB combination vanishes exactly in $d=3$ due to the exact vanishing of the Weyl tensor. The combination is a total derivative in $d=4$, the Euler number density, and is thus again irrelevant for EFT. } 

\subsection{The Reduced Einstein-Maxwell EFT}

\label{se:EMEFT_reduced}

We conclude that, at least when the relevant physical observable is the four-photon amplitude,  we can reduce the Einstein-Maxwell EFT using the $O(h^3)$-vanishing of the Gauss-Bonnet term and Einstein's equation. 
The final result is 
\be
{\cal L}_{\rm EM,red} = {\cal L}_{\rm kin}
+ \hat\alpha_1 (F^{\mu\nu}F_{\mu\nu } )^2 + \hat\alpha_2 F^{\mu\nu}F_{\nu\rho } F^{\rho \sigma } F_{\sigma \mu} + \gamma \hat C_{~~~\mu\nu}^{\rho\sigma} F^{\mu \nu}F_{\rho \sigma}  +O(F^6, \ldots )\nn
\label{eq:EMEFT_lag_red}
\ee
with 
\begin{eqnarray}
        \hat\alpha_1&=&  \alpha_1+ \frac{(d-4)^2}{4(d-2)^2}\alpha_3
    + \frac{8-3d}{4(d-2)^2} \alpha_4
     - \frac{d^2+4d-16}{4(d-2)^2}\alpha_5 +
    \frac{4-d}{4-2d}\alpha_6
    \nn \\ &&
     -
    \frac{1}{2d-4} \alpha_7 -\frac{3}{(d-1)(d-2)}\alpha_8 +\frac{(d-4) }{(d-1) (d-2)}\left(\alpha_9+\frac{\alpha_{10}}{2}\right)
\\   \hat\alpha_2&=&\alpha_2+\alpha_4+4\alpha_5 -\alpha_7 + \frac{4}{d-2}\alpha_8 + \frac{2 d}{d-2} \left(\alpha_9+\frac{\alpha_{10}}{2}\right) \\
    \gamma &=& \alpha_8      \,  .
\end{eqnarray}

\section{Bounds from Extremal Black Holes and Photon Scattering}

We review the positivity bound produced by the condition that extremal black holes must decay, for any dimension $d\geq 4$. 
We then review the four-photon ($4\gamma$) amplitude generated by $F^4$ operators for any $d\geq 3$. It will be shown in section \ref{se:finite_corrections} and appendix \ref{app:BH} that the  bounds obtained from $4\gamma$ amplitudes upon discarding the $t$-channel graviton pole match approximately the black hole bound.

\label{se:F4}

\subsection{Positivity Bound from Extremal Black Holes}

\label{se:BH}

The non-rotating charged black hole (Reissner-Nordström) solution is  parametrized  by a mass $M_{\displaystyle \circ}$ and total charge $Q_{\displaystyle \circ}$ in Planck mass units. 
In Einstein gravity, the charge-to-mass ratio is bounded from above as
\begin{equation}
  Z_{\displaystyle \circ} \equiv \frac{|Q_{ \circ}|}{M_{ \circ}}\leq  Z_*\,, \qquad Z_* =  \sqrt{\frac{d-3}{d-2} } \,,
\end{equation}
beyond which the Reissner-Nordström solution would feature a naked singularity. A black hole saturating this bound, i.e. $Z_{\displaystyle \circ} = Z_*$, is said to be \textit{extremal}. 

There are compelling arguments that all black holes, including extremal ones, must be able to decay \cite{Adams:2006sv,Kats:2006xp}. This conjecture is sometimes referred to as the black hole WGC, however we avoid using this term here to prevent naming confusion.

Extremal black holes can decay if the particle spectrum satisfies the WGC, i.e. if there is  at least one  particle satisfying \eqref{eq:WGC_schem}  in the spectrum, in which case the black hole discharges via Schwinger effect \cite{Hiscock:1990ex}. However, when no such particle is present in the theory, 
which is the case in our IR EFT where all massive particles are integrated out, the extremal black hole should still be able to decay. 
In such a situation,  the  extremal black hole can only decay into smaller black holes. This is kinematically allowed if the extremality  bound deviates from the GR one. The most general condition is that the charge-to-mass ratio decrease with the mass \cite{Barbosa:2025smt}.

 For  extremal black holes with sufficiently large radius $r_h$, satisfying the condition
 \be r_h\gg\frac{g |q|M}{m^2} \label{eq:r_critical} \,,
 \ee 
  the electromagnetic field is weak at the horizon, such that the Einstein-Maxwell EFT is valid. In this regime,  the Einstein-Maxwell operators  given in \eqref{eq:EMEFT_lag1} induce a deviation to  extremality,  see e.g. \cite{Kats:2006xp,  Jones:2019nev} and also \cite{DeLuca:2022tkm, Barbosa:2025uau} for higher order.
It follows that the charge-to-mass ratio bound in our  IR Einstein-Maxwell EFT takes the form $ Z_{\displaystyle \circ}\leq Z $
where 
\be \dfrac{Z}{Z_*} = 1 + \C_{\rm IR} \frac{4(d-2)(d-3)^2}{(3d-7)}\left(\frac{(d-2)(d-3)}{Q^2_{\displaystyle \circ}} \frac{4\pi^{d-1}}{\Gamma\left(\frac{d-1}{2}\right)}\right)^\frac{1}{d-3} \,. 
\label{eq:ExtBlackHoleRelation} 
\ee
Since the extremal black holes can decay if $Z$ is 
a decreasing function of $Q^2_{\displaystyle \circ} \sim M^2_{\displaystyle \circ}$, \eqref{eq:ExtBlackHoleRelation} implies 
 the positivity bound
\be
C_{\rm IR}>0\,. \label{eq:CIR_pos}
\ee

We can write 
\be
\C_{\rm IR} = \C_{\rm UV} + \Delta\C \label{eq:relationCIRDeltaC}
\ee
where $\C_{\rm UV}$ is the contribution from the UV operators and $\Delta C$ the contribution produced upon integrating the massive charged particles. 
The expression of $\Delta C$ as a function of the EFT operator coefficients is given in App.\,\ref{app:BH}.  In that calculation we use the basis \eqref{eq:EMEFT_lag1}, and not the reduced basis \eqref{eq:EMEFT_lag_red}. This is because  for arbitrary $d>4$ the Gauss-Bonnet combination  does not vanish beyond quadratic order in general, hence the reduction step in section \ref{se:GB} is not allowed in the black hole case.

\subsection{Four-Photon EFT}

\label{se:F4_EFT}

In $d=2$, the photon does not propagate, hence our analysis does not apply.  We focus on $d\geq 3$ for which the photon has $d-2$ physical polarizations. 

For $d=3$, the photon has a single polarization. There is a single independent $F^4$ operator which can be chosen to be $(F_{\mu\nu} F^{\mu\nu})^2$. The other possible $F^4$ structure satisfies $F_{\mu\nu}F^{\nu\rho}F_{\rho\sigma}F^{\sigma\mu}  = \frac{1}{2}(F_{\mu\nu} F^{\mu\nu})^2$.

For $d > 3$, the EFT contains two independent Lorentz structures, 
\be
{\cal L}_{F^4} = \hat\alpha_1 (F_{\mu\nu} F^{\mu\nu})^2 +\hat\alpha_2 F_{\mu\nu}F^{\nu\rho}F_{\rho\sigma}F^{\sigma\mu} = \A  {\cal O} +\B \tilde {\cal O} \label{eq:Leff}
\ee
with
\be{\cal O}= (F_{\mu\nu} F^{\mu\nu})^2 \,,\quad \quad
\tilde {\cal O}=4 F_{\mu\nu}F^{\nu\rho}F_{\rho\sigma}F^{\sigma\mu} -2 (F_{\mu\nu} F^{\mu\nu})^2  \label{eq:OOtildebasis}
\ee
where the ${\cal O}$, $\tilde {\cal O}$ basis is introduced for further convenience. The translation between the two bases is given by \be \hat\alpha_1=\A-2\B\,, \quad \hat\alpha_2 = 4\B\,. \label{eq:translations} \ee Notice that for $d=3$, we have $\tilde {\cal O}=0$ algebraically.  
In $d=4$ we have $\tilde {\cal O}=(F_{\mu\nu} \tilde F^{\mu\nu})^2 $ where the dual tensor is $\tilde F_{\mu\mu}=\frac{1}{2}\epsilon^{\mu\nu\rho\sigma}F_{\rho\sigma}$.

\subsection{Positivity Bounds from Photon Scattering}

\label{se:F4_bound}
 
\subsubsection{General considerations}

In the absence of gravity, positivity bounds on the $F^4$ operators can be derived using unitarity of forward amplitudes or causality.
In the presence of gravity, exploiting the 
forward amplitudes is complicated  due to singular $t$-channel graviton exchange,
\be
\includegraphics[scale=0.4]{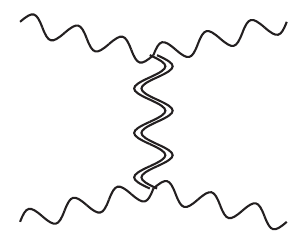}   \label{eq:grav_tree}
\ee
see \cite{Adams:2006sv,Cheung:2014ega,Bellazzini:2015cra,Bellazzini:2019xts,Hamada:2018dde}.  
A  workaround to eliminate the unwanted graviton pole may be to perform an appropriate spatial compactification that removes the $t$-channel infrared singularity, see  \cite{Bellazzini:2019xts}. This approach suggests that the  $t$-channel graviton pole may simply be discarded in the proof of the positivity bounds. 
It was, however, argued  that the obtained results appear to be overly strong \cite{Alberte:2020jsk,Alberte:2020bdz}, at least in the 4d case.  
Another approach to the graviton pole is to work at finite impact parameter   and focus on appropriate sum rules \cite{Caron-Huot:2021enk}.  
In contrast, causality bounds from low-energy photon propagation  apply without extra complication in the presence of gravity. The standard $F^4$ positivity bounds in $d=4$ are independently obtained from causality arguments, see  \cite{CarrilloGonzalez:2023cbf} and also \cite{Eichhorn:2024wba}.

Our approach in the present work is to use the most standard $F^4$ positivity bounds, already presented in \cite{Adams:2006sv},\,\footnote{For more refined positivity bounds, see e.g. \cite{Arkani-Hamed:2020blm,Alberte:2020jsk,Alberte:2020bdz,Henriksson:2021ymi,Henriksson:2022oeu,Bellazzini:2020cot,Bellazzini:2021oaj,Caron-Huot:2020cmc,Caron-Huot:2021rmr,Davighi:2021osh,deRham:2021fpu,Caron-Huot:2022ugt,Tolley:2020gtv,deRham:2021bll,deRham:2022hpx,Chiang:2022ltp,Caron-Huot:2022jli,Haring:2022sdp,Hamada:2023cyt,Bellazzini:2023nqj}. }
to avoid any technical digression. 
 In doing so, we assume a priori that the $t$-channel pole can be neglected in the case of photon scattering, as hinted by $d=4$ causality bounds.
 This  IR consistency bound will be compared to the one from extremal black hole decay, and we will find they are consistent with each other.

\subsubsection{Bounds}

Infrared consistency bounds on the $F^4$ operators are easily extended to any dimension as follows. We follow the approach of \cite{Remmen:2019cyz}. 
We consider the four photon amplitude ${\cal A}_{\gamma\gamma\to\gamma\gamma}$ with ingoing (outgoing) momentum $p_{1,2}$ ($p_{3,4}$) and ingoing(outgoing) polarization vectors $\epsilon_{1,2}$ ($\epsilon_{3,4}$). 
We then take the forward limit 
\be
{\cal A}_{\gamma\gamma\to\gamma\gamma}^{\rm fw}={\cal A}_{\gamma\gamma\to\gamma\gamma}(p_1=p_3,p_2=p_4,\epsilon_1=\epsilon_3,\epsilon_2=\epsilon_4) 
\ee
and require positivity of ${\cal A}_{\gamma\gamma\to\gamma\gamma}^{\rm fw}$ for all $\epsilon_{1,2}$.

\subsubsection*{$d>3$ case.}

The $d>3$ case is analogous to $d=4$. We obtain
\begin{align}
{\cal A}_{\gamma\gamma\to\gamma\gamma}^{\rm fw} & = 16 \hat\alpha_1 s^2 (\epsilon_1\cdot \epsilon_2)^2 
+ 4 \hat\alpha_2 s^2 ((\epsilon_1\cdot \epsilon_2)^2 +(\epsilon_1)^2(\epsilon_2)^2) \\
& = 16 \A s^2 (\epsilon_1\cdot \epsilon_2)^2 
+ 16 \B s^2 ( (\epsilon_1)^2(\epsilon_2)^2- (\epsilon_1\cdot \epsilon_2)^2) \label{eq:A4gam}
\end{align}
From the second line, the requirement  ${\cal A}_{\gamma\gamma\to\gamma\gamma}^{\rm fw}>0$ for all $\epsilon_{1,2}$ implies positivity of the Wilson coefficients in the ${\cal O},\tilde {\cal O}$ basis defined in \eqref{eq:Leff}:\,\footnote{
The $F^4$ positivity bounds presented in e.g. \cite{CarrilloGonzalez:2023cbf,Eichhorn:2024wba}  take the form 
$ 4\hat\alpha_1-3\hat\alpha_2>|4\hat\alpha_1-\hat\alpha_2|$. 
This is equivalent to \eqref{eq:positivity} upon translation to the ${\cal O},\hat {\cal O}$ basis given in  \eqref{eq:translations}. 
}
\be
\A|_{d>3}\geq 0\,\quad\quad \quad  \B|_{d>3}\geq 0 \,. 
\label{eq:positivity}
\ee

\subsubsection*{$d=3$ case.}

For $d=3$, the photon has a single polarization, i.e. it is equivalent to a scalar.\,\footnote{In $d=3$ the $F^{\mu\nu}$ transforms as a vector of $SO(3)$. This can be seen by computing  the dual tensor $F^{\mu\nu}\epsilon_{\mu\nu \rho} \equiv\partial_\rho \phi $, where  the scalar $\phi$ is  the only degree of freedom of $F^{\mu\nu}$. } We can use \eqref{eq:A4gam} with $(\epsilon_1\cdot \epsilon_2)^2=1$. ${\cal A}_{\gamma\gamma\to\gamma\gamma}^{\rm fw}>0$ implies that 
\be
\A|_{d=3}\geq0 \label{eq:positivityd3}
\ee
while the term  multiplying $\B$ vanishes identically, in accordance with the property of the ${\cal O},\tilde{\cal O}$ basis \eqref{eq:Leff}.

\section{The One-Loop EFT of Charged Particles} 

\label{se:OneLoopEFT}

We consider  the  gravitational EFT of fields with spin $0,\frac{1}{2},1$ and with $U(1)$ charge $q$. It is described by the effective Lagrangian ${\cal L}_{\rm eff, UV}$ that contains local higher dimensional operators involving $F_{\mu\nu}$, $R_{\mu\nu\rho\sigma}$ as well as the charged fields.

Our focus here being on the four-photon interactions induced by ${\cal L}_{\rm eff, UV}$, it is enough to write explicitly the  local $F^4$ operators, while neglecting the other higher-dimensional operators. The UV operator involving the charged fields would contribute only at higher order, while the $R_{\mu\nu\rho\sigma}$ can be reduced along the lines of section \ref{se:EM}.  
We have therefore the ultraviolet EFT Lagrangian
\be
{\cal L}_{\rm eff, UV}={\cal L}_{\rm kin}+{\cal L}_{F^4,{\rm UV}} + {\cal L}_{\rm matter}
\ee
with
\be
{\cal L}_{F^4,{\rm UV}} = \alpha_{{\rm UV},1}
(F^{\mu\nu}F_{\mu\nu } )^2 + \alpha_{{\rm UV},2} F^{\mu\nu}F_{\nu\rho } F^{\rho \sigma } F_{\sigma \mu}
+ \gamma_{\rm UV} \hat C_{~~~\mu\nu}^{\rho\sigma} F^{\mu \nu}F_{\rho \sigma}
\,. 
\ee
  The $\alpha_{{\rm UV},i}$, $\gamma_{\rm UV}$ coefficients  are free parameters in the UV EFT. They encapsulate the   effects  of the dynamics of the UV completion in the sub-Planckian four-photon scattering. See appendix \ref{se:strings}
for a few known examples of contributions. In the following we remain agnostic about  $\alpha_{{\rm UV},i}$, $\gamma_{\rm UV}$.

The charged particles with spin $s=0,\frac{1}{2},1$ are described by the following matter Lagrangians.

\paragraph{Spin $0$.}
The Lagrangian is
\be
{\cal L}_{0}= -|D_\mu\Phi|^2 -m^2|\Phi|^2  - \xi|\Phi|^2R\,,
\label{eq:Lag0}
\ee
where $\Phi$ is a complex scalar. We have $D_\mu\Phi=\partial_\mu \Phi +igq A_\mu\Phi$. A conformally coupled scalar has $\xi =\frac{d-2}{4(d-1)} $ in addition to $m=0$.  

\paragraph{Spin $\frac{1}{2}$.}
The Lagrangian is
\be
{\cal L}_{1/2}=-\frac{1}{2}\bar\Psi (\slashed D-m)\Psi  \,,
\label{eq:Lag_half}
\ee
where $\Psi$ is  a Dirac spinor. We have $\slashed D = \gamma^\mu D_\mu$ with $\gamma^\mu$ the $n\times n$ Dirac matrices in $d$ dimensions, with $n=2^{[d/2]}$ the dimension of spinor space \cite{Strathdee:1986jr,Hoover:2005uf}. 

\paragraph{Spin $1$.}

In order to  consistently couple a massive vector to the photon, we consider a 
nonlinearly  realized theory with gauge group $SU(2)$ broken to $U(1)$. The charged gauge boson lives in the $SU(2)/U(1)$ coset. See e.g. \cite{Fichet:2013ola} for  details.\,\footnote{This is analogous to the $W$ boson of the Standard Model upon decoupling the $U(1)_B$ gauge field.}
This approach fixes unambiguously the $U(1)$ magnetic moment of the charged vector.  
The Lagrangian, including a $R_\xi$-type gauge fixing, is
\be
{\cal L}_1+ {\cal L}^{\rm  gf}_1= -\frac{1}{2} |\hat W^{\mu\nu}|^2   + i g q  F^{\mu\nu} W_\mu W_\nu^* - \frac{1}{\xi_1}|D_\mu W^\mu|^2 - m^2 |W^\mu|^2 \,,
\label{eq:Lag1}
\ee
where  $W_\mu$ is the complex vector field. The field strength $\hat W^{\mu\nu}$ is defined as  $\hat W^{\mu\nu} = D^\mu W^\nu- D^\nu W^\mu$ where $D_\mu$ is the $U(1)$ covariant derivative. In the following, we choose the Feynman gauge $\xi_1=1$. 

The coefficient of the $U(1)$ magnetic moment operator $iF^{\mu\nu} W_\mu W_\nu^*$ can  be generalized to other values since it is  invariant under $U(1)$ gauge transformations. In this work we only use the value shown in \eqref{eq:Lag1}, which is the one  enforced by the underlying non-abelian gauge symmetry.

\subsection{Integrating Out Charged Particles at One-Loop }

The leading contribution of the charged particle to the four-photon interaction is through one-loop diagrams. Three kinds of contributions appear, that are respectively proportional to $ (g q)^4$, $ \frac{(g q)^2 }{M^{d-2}} $ and~$ \frac{ 1 }{M^{2d-4}} $:  
\be 
	\includegraphics[width=0.18\linewidth,trim={0cm 0cm 0cm 0cm},clip]{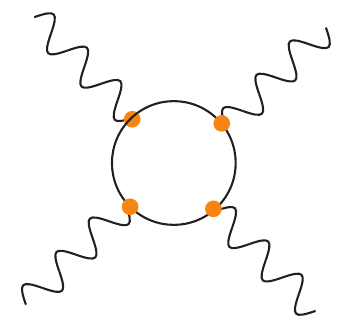} \hspace{1cm}
 	\includegraphics[width=0.25\linewidth,trim={0cm 0cm 0cm 0cm},clip]{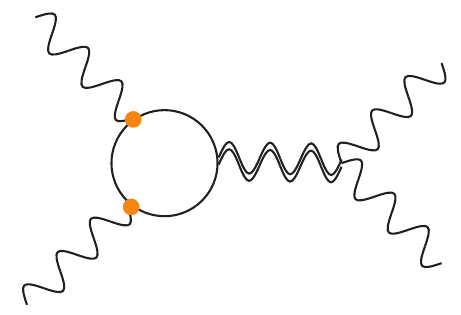}\hspace{1cm}
   	\includegraphics[width=0.35\linewidth,trim={0cm -0.2cm 0cm 0cm},clip]{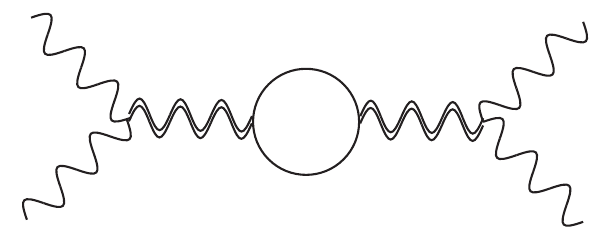}
 \label{eq:loop_diagrams}
 \ee
Here, the  dots represent the $U(1)$ charges, the double wiggles represent gravitons, and the internal bubbles can be  of any of the charged particles.

At energy scales below the charged particle mass, the ${\cal A}_{\gamma\gamma\to \gamma\gamma}$  amplitude can be described by an infrared EFT in which the  massive charged particle is integrated out, as explained in section \ref{se:EFT}. 
 The  generic form of the effective Lagrangian is given in Eq.\,\eqref{eq:EMEFT_lag1}.

For even spacetime dimensions, some of the diagrams in \eqref{eq:loop_diagrams} contain UV divergences. These divergences renormalize the local $F^4$ operators already present in the UV effective Lagrangian. The initial value of the $\alphaUV$, $\betaUV$ coefficients is assumed to be defined from the ${\cal A}_{\gamma\gamma\to \gamma\gamma}$ process  at Planckian energy $E\sim M$, such that the running produces logarithmic corrections of the form $\log\frac{M}{m}$ in the $F^4$ operators of the IR EFT.\,\footnote{The $U(1)$ gauge coupling is not renormalized in the IR EFT, as  can be verified by dimensional analysis. }

Both the renormalization flow and the finite effects from the loops of the charged particle are encoded into the one-loop effective action. An efficient way to extract both of these bits of information is to use the well-known expansion of the effective action into  heat kernel coefficients.  See  \cite{Gilkey_original, vandeVen:1984zk} for seminal papers and \cite{Vassilevich:2003xt} for a review. Other useful references are \cite{Fradkin:1983jc,Metsaev:1987ju,Hoover:2005uf}). Our main technical references are \cite{Vassilevich:2003xt,Hoover:2005uf}.

\subsubsection{Expanding the One-Loop Effective Action}

The one-loop effective action induced by the matter fields takes the form 
\be
\Gamma^{(1)}_{\rm mat} = (-)^F \frac{i}{2}{\rm Tr}\log\left[\left( -\square +m^2 +X \right)_{ij}\right]  
\ee
with $\square =g_{\mu\nu}D^\mu D^\nu $ the Laplacian built from background-covariant derivatives. 
The covariant derivatives give rise to a background-dependent field strength $\Omega_{\mu\nu}=[D_\mu,D_\nu]$, encoding both gauge and curvature connections. It takes the general form
\be
\Omega_{\mu\nu} = -i F^a_{\mu\nu}t_a-\frac{i}{2}R_{\mu\nu}^{~~~~\rho\sigma}J_{\rho\sigma}
\ee
where $t_a$ and $J_{\rho\sigma}$ are the generators of the gauge and spin representation of the quantum fluctuation. 
$X$ is the ``field-dependent mass matrix'' of the quantum fluctuations, it is a local background-dependent quantity. 
The effective field strength $\Omega_{\mu\nu}$ and the effective mass $X$ are, together with the curvature tensor, the building blocks of the heat kernel coefficients. 
Using the heat kernel method reviewed in App.\,\ref{app:HK}, $\Gamma^{(1)}_{\rm mat} $ takes the form
\begin{align}
& \Gamma^{(1)}_{\rm mat} =  (-)^F \frac{1}{2}
\frac{1}{(4\pi )^{\frac{d}{2}}}
\int_{\cal M} d^{d}x \sqrt{\g} \sum^\infty_{r=0} \frac{ \Gamma(r-\frac{d}{2})   }{m^{2r-d}}{\rm tr}\,b_{2r}(x) 
\label{eq:Gam1_b}
\end{align}
with $\rm tr$ the trace over internal (non-spacetime) indexes. Analytical continuation in $d$ has been used, the expression is valid for any dimension. The local quantities $b_{2r}$ are referred to as the heat kernel coefficients.

For odd dimension,  all the  terms in Eq.\,\eqref{eq:Gam1_b} are finite.
For even dimension, there are log-divergences. These log divergences renormalize  ${\cal L}^{(0)}_{\rm eff}$. 
The terms with negative powers of masses in Eq.\,\eqref{eq:Gam1_b}  are finite. They amount to an expansion for large $m$ and give rise to 
the one-loop contribution to the effective Lagrangian ${\cal L}^{(1)}_{\rm eff}$,
\be
 {\cal L}^{(1)}_{\rm eff} = (-)^F
\frac{1}{2}
\frac{1}{(4\pi )^{\frac{d}{2}}}
\sum^\infty_{r=[d/2]+1} \frac{ \Gamma(r-\frac{d}{2})   }{m^{2r-d}} {\rm tr}\, b_{2r}(x)\,. 
\label{eq:Leff_oneloop}
\ee
Only the first heat kernel coefficients are explicitly known,  the coefficients up to $b_8$ contribute to the observables considered in this work.

\subsubsection{Spin $0$}

The one-loop effective action following from the Lagrangian \eqref{eq:Lag0} is 
\be
\Gamma^{(1)}_0= \frac{i}{2} {\rm Tr}\log\left[\left( -\square +m^2 + \xi R  \right)\right]\,.  
\ee
The geometric invariants are 
\be
X= \xi R I\,,\quad \Omega_{\mu\nu}=-i gq F_{\mu\nu}\,.
\ee

\subsubsection{Spin $1/2$}

The one-loop effective action following from the Lagrangian \eqref{eq:Lag_half} is 
\be
\Gamma^{(1)}_{1/2}= - \frac{i}{4}{\rm Tr}\log\left[\left( -\square +m^2 + \frac{1}{4}R+  S^{\mu\nu} gq F_{\mu\nu } \right)\right]  
\ee
with $S^{\mu\nu}=\frac{i}{4}[\gamma^\mu,\gamma^\nu]$. 
The geometric  invariants are
\be X=\frac{1}{4}R+  \frac{i}{2}\gamma^\mu\gamma^\nu gq  F_{\mu\nu } \,,  \quad  \Omega_{\mu\nu}= {- i } gq F_{\mu\nu} +\frac{1}{4}\gamma^\rho \gamma^{\sigma}R_{\rho\sigma\mu\nu}
\,. \ee

\subsubsection{Spin $1$}

For the massive spin $1$ particle, the contributions from the ghosts and the Goldstone boson must be included. In the Feynman gauge, these degrees of freedom are degenerate and do not mix. The ghosts contribute as $-2$  times a scalar adjoint with $\xi=0$.   
Similarly,  the Goldstone contributes as $+1$ the scalar term 
 (see e.g. \cite{Hoover:2005uf,Fichet:2013ola}).
 As a result, the one-loop effective action following from the Lagrangian \eqref{eq:Lag1} is 
\be
\Gamma^{(1)}_{1}=  \frac{i}{2} {\rm Tr}\log\left[\left( (-\square +m^2)\delta^\mu_{~\,\nu} + R^\mu_{~\,\nu}   +2 igq F^\mu_{~\,\nu} \right)\right] -  \frac{i}{2} {\rm Tr}\log\left[\left( -\square +m^2   \right)\right]  
\ee
where the last term is the ghost+Goldstone contribution. 
The geometric  invariants of the vector fluctuation are
\be
    X^\mu_{~\,\nu}= R^\mu_{~\,\nu}   +2 igqF^\mu_{~\,\nu}
    \,,\quad 
    (\Omega_{\mu\nu})^{\rho}_{~~\sigma}= - R^\rho_{~~\sigma\mu\nu} { - i}\delta^\rho_\sigma gq F_{\mu\nu}
\ee

\subsection{The Coefficients}

The complete expressions of the heat kernel coefficients are given in App.\,\ref{app:HK}. Only a subset of terms is relevant to our study. 
Terms which are total derivatives can be ignored since they are irrelevant for scattering amplitudes. As explained in section \ref{se:EM}, in the EFT framework, we can use the leading order equations of motion to reduce the effective Lagrangian. 

The relevant pieces to compute the Einstein-Maxwell EFT are the following.

\subsubsection{$R^2$ Terms}

The curvature squared contributions from the $b_4$ coefficient,
\begin{align}
b_4=&\frac{1}{360}\Big(5 R^2-2R_{\mu\nu}R^{\mu\nu}+2R_{\mu\nu \rho\sigma}R^{\mu\nu \rho\sigma} \Big)I +\ldots\label{eq:b4F4}
\end{align}
with $I$ the identity matrix for internal indexes. For our purposes, these can be further reduced using the $O(h^3)$  vanishing of the Gauss-Bonnet term, see section \ref{se:GB}. 

\subsubsection{$R F^2$ Terms}

The $RF^2$ contributions come from the $b_6$ coefficient. 
These are those with three powers of $X$, two powers of $X$ and two derivatives, and one curvature and two powers of $X$. We have thus
\begin{align}
b_6 =  \frac{1}{360}\bigg( &
8 D_{\rho}\Omega_{\mu\nu}D^{\rho}\Omega^{\mu\nu}
+2 D^{\mu}\Omega_{\mu\nu} D_{\rho}\Omega^{\rho \nu}
+12 \Omega_{\mu\nu}\square \Omega^{\mu\nu}
-12 \Omega_{\mu\nu}\Omega^{\nu\rho}\Omega^{~~\mu}_{\rho}  \nn  \\ \nn 
&{ +}6 R_{\mu\nu \rho\sigma}\Omega^{\mu\nu}\Omega^{\rho\sigma}
-4 R_{\mu}^{~\nu}\Omega^{\mu\rho}\Omega_{\nu\rho} + 5 R\Omega_{\mu\nu}\Omega^{\mu\nu} \\ & 
 +60 X\square X+ 30 D_\mu X D^\mu X - 60 X^3 - 30 X \Omega_{\mu\nu}\Omega^{\mu\nu}  + 30 XX R 
\bigg)   +\ldots
\label{eq:b6F4}
\end{align}
To reduce $b_6$, we use the photon EOM, the Bianchi identities and the Ricci identity, as detailed in section \ref{se:EMEFT}.

\subsubsection{$F^4$ Terms}

The $F^4$ coefficients come from the $b_8$ heat kernel coefficient, which can be found in Ref.~\cite{Metsaev:1987ju}. 
Converting to Minkowski space 
we have 
\bea
b_8&=&\frac{1}{24}\biggl(
\gamma^{(s)}_1(F_{\mu\nu}F^{\mu\nu})^2 +
\gamma^{(s)}_2 F_{\mu\nu}F^{\nu\rho}F_{\rho\lambda}F^{\lambda\nu}
\biggr) +\ldots
\eea

\bea
(\gamma^{(0)}_{1}, \gamma^{(0)}_{2})&=&\left( \frac{1}{12},\frac{7}{105} \right)\\
(\gamma^{(1/2)}_{1}, \gamma^{(1/2)}_{2})&=&\left(\frac{1}{3}n\,,-\frac{14}{15}n \right)\\
(\gamma^{(V)}_{1}, \gamma^{(V)}_{2})&=&\left( \frac{d-48}{12}\,, \frac{d+240}{15}\right)
\eea
and  $(\gamma^{(1)}_{1}, \gamma^{(1)}_{2})= (\gamma^{(V)}_{1}, \gamma^{(V)}_{2})-(\gamma^{(0)}_{1}, \gamma^{(0)}_{2})$ for the massive spin $1$ particle.  

\subsection{The Reduced Einstein-Maxwell  Effective Action}\label{se:ReducedEFT}

Putting together the results from previous subsections, we obtain the low-energy Einstein-Maxwell effective action generated by integrating out the charged particles of spin $s=0,\frac{1}{2},1$. The leading contributions are encoded in the  one-loop effective action, $\Gamma_s^{(1)}$.

We apply the reduction computed in \eqref{eq:EMEFT_lag_red}.
The reduced Einstein-Maxwell  effective action is
 \be \Gamma_{\rm IR}= \Gamma^{(0)} +\Gamma_s^{(1)}+\ldots \ee with 
\be
\Gamma^{(0)}= \int d^d x \left( {\cal L}_{\rm kin}+{\cal L}_{F^4,{\rm UV}} \right)
\ee
\be
\Gamma_s^{(1)}= \int d^d x \left(  
\Delta\hat\alpha^{(s)}_1 (F_{\mu\nu} F^{\mu\nu})^2 +
\Delta \hat\alpha^{(s)}_2 F_{\mu\nu}F^{\nu\rho}F_{\rho\sigma}F^{\sigma\mu}
+\Delta \gamma^{(s)} \hat C_{~~~\mu\nu}^{\rho\sigma} F^{\mu \nu}F_{\rho \sigma}  \right)
\ee
\begin{align}
 \Delta \hat \alpha^{(s)}_{1,2} =     \frac{1}{(4\pi)^{d/2}} \Bigg[    & \frac{g^4 q^4}{m^{8-d}} \Gamma\left(4-\frac{d}{2}\right) a^{(s)}_{1,2} 
 \\ & \nonumber
 + \frac{g^2 q^2}{m^{6-d}M^{d-2}}\Gamma\left(3-\frac{d}{2}\right) b^{(s)}_{1,2} + 
\frac{1}{m^{4-d}\M^{2d-4}}\Gamma\left(2-\frac{d}{2}\right) c^{(s)}_{1,2} ~~ \Bigg]    
\\
\Delta \gamma^{(s)} =  \frac{1}{(4\pi)^{d/2}}   &\frac{g^2 q^2}{m^{6-d}\M^{d-2}}  \Gamma\left(3-\frac{d}{2}\right) d^{(s)} 
\end{align}

The coefficients for each spin  are 
\begin{subequations}
\label{eq:coefs_a}
\begin{alignat}{5}
    &a^{(0)}_1 &&= \frac{1}{288}\,, \hspace{2cm} &&a^{(0)}_2 &&= \frac{1}{360}\,,  \\ 
    &a^{(1/2)}_1 &&= -\frac{n}{144}\,, \hspace{2cm} &&a^{(1/2)}_2 &&= \frac{7n}{360}\,,\\
    &a^{(1)}_1 &&= \frac{d-49}{288}\,, \hspace{2cm} &&a^{(1)}_2 &&= \frac{d+239}{360}\,.
\end{alignat}
\end{subequations}
\begin{subequations}
\label{eq:coefs_b}
\begin{alignat}{5}
    &b^{(0)}_1 &&= \frac{1}{720} \left[\left(30 \xi-5+\frac{4}{(d-1)(d-2)}\right)(d-4)+\left( 4 +\frac{8}{d-2} \right)\right] \frac{1}{ (d-2)}\,,\\
    &b^{(1/2)}_1 &&= - \frac{n}{720}\left[\left(-5+\frac{4}{(d-1)(d-2)}\right) \frac{d-4}{2 (d-2)}- \frac{13(d-2)-4}{(d-2)^2}\right]\,,\\
    &b^{(1)}_1 &&= \frac{1}{720}  \left[\left(\frac{4(d + 59)}{(d-1)(d-2)}- 5\left(d-31\right)\right)\frac{d-4}{d-2}+\frac{4(d-1)(d+120)}{(d-2)^2}\right]\,,\\ 
    &b^{(0)}_2 &&= - \frac{1}{360} \left( 4 +\frac{8}{d-2} \right)\,,\\
    &b^{(1/2)}_2 &&= -\frac{n}{360}\left(-\frac{4}{d-2} +13\right) \,,\\
    &b^{(1)}_2 &&= -\frac{1}{360} \left(4(d+119)+\frac{8(d + 59)}{d-2}\right)\,.
\end{alignat}
\end{subequations}

\begin{subequations}
\label{eq:coefs_c}
\begin{alignat}{5}
    &c^{(0)}_1 &&= \frac{1}{720}\left[\frac{6+\xi\left(\xi-\tfrac{1}{3}\right)}{4}  \frac{(d-4)^2}{(d-2)^2}  - \frac{3(3d-8)}{(d-2)^2} \right], \hspace{1.5cm} && c^{(0)}_2 &&= \frac{1}{60}\,,\\ 
    &c^{(1/2)}_1 &&= -\frac{n}{960}\left(  \frac{3(3d-8)}{(d-2)^2} + \frac{(d-4)^2}{(d-2)^2}  \right), \hspace{1.5cm} && c^{(1/2)}_2 &&= \frac{n }{80}\,,\\
    &c^{(1)}_1 &&= \frac{1}{240}\left[\frac{(d-11)(d-4)^2}{2(d-2)^2}- \frac{(d+9)(3d-8)}{(d-2)^2}\right], \hspace{1.5cm} &&c^{(1)}_2 &&= \frac{d+9}{60}\,.
\end{alignat}
\end{subequations}
\begin{subequations}
\label{eq:coefs_d}
\begin{alignat}{5}
    &d^{(0)} &&= - \frac{1}{180}\,,\\
    &d^{(1/2)} &&=  \frac{n}{360}\,,\\
    &d^{(1)}  &&= -\frac{d+59}{180}\,.
\end{alignat}
\end{subequations}

Before reduction of the Einstein-Maxwell Lagrangian,  our results for the $RF^2$ operators from spin $\frac{1}{2}$ in $d=3$ and from spin $0,\frac{1}{2}$ in $d=4$ match respectively those found in \cite{Ritz:1995nt,Drummond80} and \cite{Drummond80,Bastianelli:2008cu}. 
 For $d=3$,  the $a^{(0)}_{1,2}$ coefficients with $\xi=0$ match with \cite{Chen:2019qvr} and for $d=4$, the $a^{(s)}_{1,2}$ coefficients match those from \cite{EH_Lag, Dunne:2004nc, Fichet:2014uka}, upon appropriate conversion to the $(\cal O, \tilde {\cal O})$ basis using \eqref{eq:Leff}.

\subsubsection{Beta Functions and Logarithmic Corrections}
\label{eq:beta_computation}

The Gamma functions in $\Delta \hat \alpha_{1,2} $, $\Delta \gamma$ diverge for certain even dimensions. 
These are physi\-cal divergences, appearing here via the framework of dimensional regularization. These  divergences imply that  there is a renormalization group equation associated to the corresponding local operators contained in $\Gamma^{(0)}$.
 In that situation, the values  $\alpha_{{\rm UV},i}$ or  $\gamma_{\rm UV}$ are  understood as the values at the initial condition  of the renormalization flow that we choose to be the Planck scale. That is, $\mu=\M$ with $\mu$ the renormalization scale, and
 $\alpha_{{\rm UV},i}\equiv \alpha_i(\M)$,  $\gamma_{{\rm UV}}\equiv \gamma(\M)$. 
These are the values appropriate to study the physical process ${\cal A}_{\gamma\gamma\to\gamma\gamma}$ with energy scale near $\M$. 
To study ${\cal A}_{\gamma\gamma\to\gamma\gamma}$ at lower energies, the renormalization scale must be changed accordingly to minimize higher-order contributions to the one-loop prediction. 

Let us  compute the beta functions explicitly.
Divergences occur when $r-\frac{d}{2}\sim -n$ i.e. $d\sim 2n+2r$ with $n\in \mathbb{N}$. We define $\epsilon = 2n+2r-d $. Introducing the renormalization scale $\mu$ in the Lagrangian, we have
\be
\frac{\mu^\epsilon}{m^\epsilon}\Gamma\left(r-\frac{d}{2}\right)  \xrightarrow{\epsilon\to 0}  \frac{(-1)^n}{n!}2\left(\frac{1}{\epsilon}+\log\left(\frac{\mu}{m}\right)\right)\,.
\ee
Such terms from the one-loop effective action $\Gamma^{(1)}$ combine with the coefficients of the local operators in  $\Gamma^{(0)}$.  One absorbs the $1/\epsilon$ constant into the definitions of the coefficients, leaving only the $\log(\mu)$ dependence. The physical parameter is identified (at one-loop order) as
\be
\alpha_i^{\rm phys}=\alpha_i(\mu)+B_i  \log\frac{\mu}{m}\, 
\ee
where the generic $B_i$ coefficient is computed from \eqref{eq:coefs_a}-\eqref{eq:coefs_d}, and analogously for $\gamma$.   Requiring $\frac{d}{d\mu} \alpha^{\rm phys}_i =0  $ determines the one-loop beta function for the Lagrangian parameter \be \boldsymbol{\beta}_{\alpha_i}\equiv \frac{d}{d \log \mu}\alpha_i=-B_i+O({\rm higher~order~loops})\,. \ee

The beta functions for the couplings of the $F^4$ and $CF^2$
 operators are presented in section \ref{se:beta_functions}. 

Finally, when the renormalization flow is caused by a massive particle, it stops at the scale $\mu=m$. Below this scale, we work with the IR EFT (i.e. the Einstein-Maxwell EFT) in which the only remainder of the charged particles is the set of finite contributions to the local operators (see \cite{Manohar:1996cq}). The coefficients in the IR EFT take the form
 \be
\alpha_{{\rm IR},i} = \alpha_{{\rm UV},i}+ \Delta \alpha^{\rm finite}_i+ B_i \log\frac{M}{m} \, 
\ee
\label{eq:EFT_evenDimensions}
and analogously for $\gamma$.

\section{The $F^4$ Beta Functions and Infrared Consistency}

\label{se:beta_functions}

In this section we assume that the charged particle is exactly massless.  We  compute the 1-loop beta functions and discuss the  $F^4$ renormalization flow.

\subsection{The Beta Functions}

We  compute the 1-loop beta functions of the $F^4$ and $CF^2$ operators along the lines presented in 
section \ref{eq:beta_computation}. 
  For a massless spin-1 particle,  the corresponding heat kernel coefficients are given by $a_i^{(1)}=a_i^{(V)}-2 a_i^{(0)}$. 
At zero mass, the only beta functions of the $F^4$ operators appear for $d=4$, $6$ and $8$. They are given in Tabs.\,\ref{tab:beta_zeromass_Alpha},~\ref{tab:beta_zeromass_Beta},~\ref{tab:beta_zeromass_Gamma}.

In $d=4$,  graviton and photon loops produce an additional contribution  to the $R^2$ operators. Note these are the loops that cause the $d=4$ conformal anomaly in the IR EFT (see e.g. \cite{birrell1984quantum, Vassilevich:2003xt}). Following section \ref{se:EMEFT}, in the Einstein-Maxwell EFT, these loops of gravitons and photons contribute to the renormalization flow of the $F^4$ operators via diagrams such as
\be 	\includegraphics[width=0.35\linewidth,trim={0cm 0cm 0cm 0cm},clip]{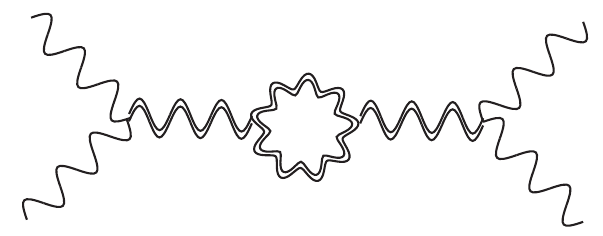} \hspace{1cm}
 \nn 
 \ee
The corresponding graviton contribution to the $b_4$ coefficient is 
\begin{align}
b^{\rm grav}_4 & = \frac{1}{180}\left(
414 R_{\mu\nu}^2+\frac{553}{4}R^2
\right) \nn
\\ &  = \frac{23}{10 }\left(F^{\mu\nu}F_{\nu\rho } F^{\rho \sigma } F_{\sigma \mu}-\frac{1}{4}(F^{\mu\nu}F_{\mu\nu } )^2\right) 
\end{align}
where we have used the Gauss-Bonnet identity \eqref{eq:GBapprox} and {the leading-order Einstein equation \eqref{eq:EEs}}.
This corresponds to $4 {\bm\beta}^{\rm grav}_{\hat \alpha_1}= -{\bm\beta}^{\rm grav}_{\hat \alpha_2} = \frac{23}{160\pi^2} $. 
Translating to the ${\cal O}, \tilde {\cal O}$ basis we obtain the contributions \be {{\bm\beta}^{\rm grav}_{\alpha}}={{\bm\beta}^{\rm grav}_{\beta}}=-\frac{23}{640\pi^2}\,.\ee
 The photon loop contributes by an additional $\frac{1}{2}a^{(1)}_i$, where the $\frac{1}{2}$ accounts for the photon being a real vector. 
 Due to the fact that graviton and photon are massless, such loops do not contribute to the running of $F^4$ in other dimensions. 

\begin{table}[t]
$$
\begin{array}{|c||c||c||c|}
\hline
 \hbox{\mydiagbox{\quad \,\, d}{\text{Spin }}} & 0  & \frac{1}{2} & 1  \\ \hline \hline

4 & -\frac{19}{480\pi^2}\frac{1}{\M^4} & -\frac{13}{320\pi^2}\frac{1}{\M^4} & 
-\frac{29}{640\pi^2}\frac{1}{\M^4}\\

6 & \frac{ 23-50 \xi }{76800 \pi ^3}\frac{g^2 q^2}{\M^4} & \frac{13}{4800\pi^3}\frac{g^2 q^2}{\M^4} & 
\frac{661}{38400\pi^3}\frac{g^2 q^2}{\M^4} \\ 

8 & -\frac{7 }{184320 \pi ^4}g^4 q^4  & -\frac{1}{2880 \pi ^4}g^4 q^4 & 
-\frac{47}{30720 \pi ^4}g^4 q^4\\\hline
\end{array}
$$
        \caption{Beta function of the $\A$ coefficient. }
    \label{tab:beta_zeromass_Alpha}
\end{table}

\begin{table}[t]
$$
\begin{array}{|c||c||c||c|}
\hline
 \hbox{\mydiagbox{\quad \,\, d}{\text{Spin }}} & 0  & \frac{1}{2} & 1  \\ \hline \hline

4 & -\frac{19}{480\pi^2}\frac{1}{\M^4} & -\frac{13}{320\pi^2}\frac{1}{\M^4} & 
-\frac{29}{640\pi^2}\frac{1}{\M^4}\\

6 & \frac{1}{7680 \pi ^3}\frac{g^2 q^2}{\M^4} & \frac{1}{480\pi^3}\frac{g^2 q^2}{\M^4} & 
\frac{13}{960\pi^3}\frac{g^2 q^2}{\M^4} \\ 

8 & -\frac{1}{184320 \pi ^4}g^4 q^4  & -\frac{7}{11520 \pi ^4}g^4 q^4 & 
-\frac{41}{30720\pi^4}g^4 q^4\\\hline
\end{array}
$$
        \caption{Beta function of the $\B$ coefficient.}
    \label{tab:beta_zeromass_Beta}
\end{table}

\subsection{Discussion}

Whenever the charged particle is massless or if $m\ll M$, 
the coefficients of the $F^4$ operators at low-energy  scales are controlled by their beta functions. The renormalization flow washes away any finite correction, that becomes negligible compared to  large logarithms. 
When $m=0$, the running coefficients at $\mu\ll M$ is 
\be \alpha_i(\mu)\approx B_i \log \frac{M}{\mu}
\label{eq:alphaRG}
\,.\ee

From Tabs.\,\ref{tab:beta_zeromass_Alpha},\ref{tab:beta_zeromass_Beta}, 
we can see that ${\bm\beta}|_{d=4}$ and ${\bm\beta}|_{d=8}$ are negative for loops of all spins.  Therefore, in $d=4,8$, the coefficients of the ${\cal O}$, $\tilde {\cal O}$ operators tend to grow positively when the theory flows towards the infrared. It would be tempting to conclude that the  renormalization flow tends to make the theory infrared-consistent. However, the  positivity bounds we are using strictly require $m\neq 0$,  and thus do not apply in the present case.  

In fact, the opposite behavior appears in $d=6$.
The spin $\frac{1}{2}$ and 1 beta functions are \textit{positive} for both $\alpha$ and $\beta$ coefficients.
The scalar beta function for the $\beta$ coefficient is positive, as  is the beta function for $\alpha$ if $\xi<\frac{23}{50}$. This includes  the  conformal coupling value in  $d=6$, $\xi=\frac{1}{5}$.
These positive beta functions imply that the coefficients of the ${\cal O}$, $\tilde {\cal O}$ operators are driven towards \textit{negative} values at sufficiently low energy scales. If positivity bounds applied at $m=0$, they would be necessarily violated in the deep IR, implying that the gravitational EFTs of massless particles in $d=6$ are infrared-inconsistent. Such a strong conclusion is avoided if none of the positivity bounds apply for $m=0$. 

 Nevertheless, we will see in next section that, in the presence of a small nonzero mass,  the positivity of ${\bm\beta}|_{d=6}$  tends to create a tension with the positivity bounds. It would be interesting to find if other ingredients like a gravitino or non-minimal couplings can make the beta function negative,  along the lines of \cite{Arkani-Hamed:2021ajd}.\,\footnote{
We mention that the  $a$-theorem in $d=6$ also presents an unexpected behavior compared to $d=2,4$, see 
\cite{Grinstein:2014xba,Grinstein:2015ina}.
}

Finally, we find that the sign of the beta function of the $CF^2$ operator, given in Tab.\,\ref{tab:beta_zeromass_Gamma}, depends on the spin.  This has no consequence for the positivity bounds we are using. 
The  sign of the $\gamma$ coefficient is irrelevant in the positivity bounds given in  e.g.\,\cite{Bellazzini:2019xts}, which involve $|\gamma|$, and in the present work we use a simpler positivity bound that is independent on $\gamma$.

\begin{table}[t]
$$
\begin{array}{|c||c||c||c|}
\hline
{\rm Spin}  & 0  & \frac{1}{2} & 1  \\ \hline \hline

\multicolumn{1}{c|}{} & \frac{1}{5760 \pi ^3}\frac{g^2 q^2}{\M^4} &-  \frac{1}{1440\pi^3}\frac{g^2 q^2}{\M^4} & \frac{13}{1152\pi^3}\frac{g^2 q^2}{\M^4} \\

\cline{2-4}
\end{array}
$$
        \caption{Beta function of the $\gamma$ coefficient in $d=6$.}
    \label{tab:beta_zeromass_Gamma}
\end{table}

The beta function for the $C$ coefficient of the extremality relation, defined in \eqref{eq:ExtBlackHoleRelation}, follows the same sign pattern as $\beta_\alpha$ and $\beta_\beta$. 

\FloatBarrier

\section{Finite Corrections  and Infrared Consistency}

\label{se:finite_corrections}

In this section, we assume that the charged particle is massive, $m>0$. Photon scattering at energy scales below $m$ is described by the infrared EFT, that encodes the finite corrections induced when integrating out the charged particle.

For any even dimension, at least some of the coefficients of the IR EFT receive loga\-rithmic corrections that are large when $m\ll M$. 
Some of these logarithmic corrections correspond to the beta functions presented in section \ref{se:beta_functions}. 
The renormalization flow of the massless case is recovered  when taking $m\to 0 $ at finite energy or finite $\mu$,  the only difference occurs in the spin-1 case since there is no Goldstone boson in the massless case.\,\footnote{The other corrections simply do not exist in the $m\to 0$ limit, since the charged particle is not integrated out.}

The contributions to the $F^4$ Wilson coefficients in the IR EFT take the form 
\be
\Delta \alpha= a \frac{g^4q^4}{m^{8-\d}}  
+  b \frac{g^2q^2}{m^{6-\d} \M^{\d-2}}
+ \frac{c}{ m^{4-\d}\M^{2\d-4}} \,.
\label{eq:DeltaCorrections1}
\ee
We define a reduced notation that is used throughout this section. 

\subsection{Reduced Notation}

The corrections induced  by the charged particles are second order polynomials in $q^2$, see \eqref{eq:DeltaCorrections1}. In terms of the 
charge-to-mass ratio $z$ introduced in \eqref{eq:WGC_schem}, we have \be 
\Delta \alpha = \dfrac{1}{m^{4-d}\M^{2d-4}}\left(a z^4 + bz^2 + c \right),\quad\quad z = \frac{g|q|}{m} \M^{\frac{d-2}{2}}\,.
\label{eq:z_DeltaCorrections}
\ee 
Note that $[g]=2-\frac{d}{2}$ hence $z$ is dimensionless for any $d$. 

We further define the loop factor
\be
    \K_d = \left\{\begin{array}{cc}
         2^{d}\pi^{\frac{d}{2}} \, & \text{if } d \text{ even} \\
         2^{d}\pi^{\frac{d-1}{2}} \, &  \text{if } d \text{ odd}
        \end{array}\right.
\ee
and work with a scaled dimensionless version of \eqref{eq:z_DeltaCorrections}, given by
\be
    \Delta\bA =\bar a z^4 + \bar b z^2 + \bar c \equiv \K_d \,m^{4-d}\M^{2d-4}  \Delta\A \,.
    \label{eq:Reduced_z_DeltaCorrections}
\ee
Similar
definitions hold for $\bA_{\rm IR}$, $\bA_{\rm UV}$, $\boldsymbol{\beta}_{\bA}$, and  for the $\beta$ coefficients (i.e. $\Delta \bB$, $\bB_{\rm IR}$, $\bB_{\rm UV}$ and $\boldsymbol{\beta}_{\bB}$). 
The  infrared consistency condition \eqref{eq:positivity} is equivalent to 
\be
\bA_{\rm IR} \geq 0\,,\quad \bB_{\rm IR} \geq 0\,. 
\label{eq:ReducedIRboundschem}
\ee

\subsection{General Analysis of Positivity  }

\label{se:general_analysis}

The $\bar\alpha_{\rm IR}(z)$  polynomial is defined on $\mathbb{R}_+$.
Due to this restricted domain, studying the positivity of $\bar\alpha_{\rm IR}(z)$ requires distinguishing various cases that we classify here.

Let $\bA^*$ be the rightmost extremum of the quartic polynomial $\Delta\bA(z)$, namely
\be
\bA^* = \left\{\begin{array}{cc}
         \bar c \,, & \text{if } \bar a \bar b \geq 0\\
         \frac{4\bar a \bar c -\bar b^2}{4 \bar a} \,, & \text{if } \bar a \bar b < 0 \,,
        \end{array}\right.
\ee
\label{eq:RightCritical}
and let $0 \leq z_1 \leq z_2$ be the two roots of  $\bA_{\rm IR}(z)$ in $\mathbb{R}_+$. 

Depending on the coefficients, the positivity constraint \eqref{eq:ReducedIRboundschem} imposes restrictions on the charge-to-mass ratio $z$, which are classified into the following cases.   
\begin{enumerate}[{\it(1)}]
    \item  Case $\bar a > 0$  

       \begin{itemize}
        \item  If $\bA_{\rm UV} \geq -\bA^*$, then we have that $\bA_{\rm IR} \geq 0$ holds for all $ z\geq 0$.
   \end{itemize}

    \begin{enumerate}[{\it(a)}]
        \item Case $\bar b \geq 0$  

            \begin{itemize}
                \item If $\bA_{\rm UV} < -\bA^* = -\bar c$, then there is a lower bound on  $z$ in the form $z \geq z_2>0$.

            \end{itemize}
            
        \item Case $\bar b < 0$

            \begin{itemize}
                \item If $\bA_{\rm UV} < -\bar c$, then there is a lower bound on $z$ in the form $z \geq z_2 >0$.
        
                \item  If $-\bar c \leq \bA_{\rm UV} < -\bA^*$, then there is another allowed region for $z$, so that $z \in \left[0,z_1\right] \cup \left[z_2, \infty\right)$.

           \end{itemize}
            
    \end{enumerate}
\end{enumerate}

\begin{enumerate}[{\it(1)}]
 \setcounter{enumi}{ 1}
    \item Case $\bar a<0$

    \begin{itemize}
        \item If $\bA_{\rm UV} < -\bA^*$, then the infrared consistency condition is violated (i.e. $\bA_{\rm IR} < 0$ for all $z\geq0$). In this case, we say that this value for $\bA_{\rm UV}$ is \textit{excluded}.
    \end{itemize}
    
\begin{enumerate}[{\it(a)}]
    \item Case $\bar b \geq 0$

    \begin{itemize}
        \item If $\bA_{\rm UV} \geq -\bar c$, then there is an upper bound on $z$ in the form $0 \leq z \leq z_2$.

            \item If $-\bA^* \leq \bA_{\rm UV} < - \bar c $, then there is a lower and an upper bound on $z$ in the form $z_1 \leq z \leq z_2$.

    \end{itemize}

    \item Case $\bar b < 0$

    \begin{itemize}
        \item If $ \bA_{\rm UV} \geq -\bA_* = -\bar c$, then there is an upper bound on $z$ in the form $0 \leq z \leq z_2$.

    \end{itemize}
\end{enumerate}

\end{enumerate}

Case \textit{(1)} with sufficiently small $\bAUV$
implies the existence of a WGC-like bound on $z$. 
Conversely, in case \textit{(2)} with sufficiently large $\bAUV$,
 $z$ gets always bounded in a finite region.

For $d > 3$ we need to consider these conditions for both $\cal O$, $\tilde{\cal O}$ operators, i.e. for both $\bar\alpha_{\rm IR}$, $\bar\beta_{\rm IR}$ coefficients,
 so that the most restrictive condition dominates.
{ The above analysis applies whether or not  logarithmic contributions are present. This is because the logarithms  depend only on the scale ratio $\frac{m}{M}$, which can be treated as an independent quantity with respect to the charge-to-mass ratio $z$. Also, for $d\geq 8$, the logarithms factor out of the entire polynomial and are thus irrelevant for the positivity analysis.  }

\subsection{Positivity Bounds and the WGC}
\label{se:results_summary}

This subsection presents the synthesis of our results for the relations between positivity bounds and the WGC.
We consider the positivity bounds from infrared consistency of $\gamma\gamma \to \gamma\gamma $ ($\alpha_{\rm IR}\geq 0$, $\beta_{\rm IR}\geq 0$),  discussed in section \ref{se:F4_bound}, and those  from  extremal black hole decay ($C_{\rm IR}>0$), discussed in section \ref{se:BH}. 
We remind that, here,  WGC means specifically  that there exists a nonzero $O(1)$ lower bound for the charge-to-mass ratio, i.e.  $z\geq z_*>0 $ (see \eqref{eq:WGC_schem}).  

One of our focus is the  dependence of the various relations with respect to spacetime dimension.
As argued in section \ref{se:intro}, we take the viewpoint that dimensional dependence is a test of robustness. A relation that holds in any $d$ might be profound, while one that only holds for certain $d$ may
 be viewed as more coincidental.   
 The systematic analyses for each dimension are collected in Apps.\,\ref{app:IR} and  \ref{app:BH}.

The positivity bounds from IR consistency and black hole decay are fundamentally different. The latter does not apply for $d=3$, and is  still a conjecture, while the former is rigorous.\,\footnote{Another difference is that the IR consistency bounds are non-strict while the black hole bound is strict. This has no practical consequences.} Yet, both $C_{\rm IR}$ and $\alpha_{\rm IR}$, $\beta_{\rm IR}$  are quadratic in $z^2$, hence we can apply the analysis of Sec.\,\ref{se:general_analysis} to all of them. 
Their relations to the WGC-like bounds turn out to be very similar, hence we summarize them together.

\subsubsection{Results for $d\leq 11$}

The results for the $\Delta \alpha$, $\Delta \beta$ and $\Delta C$,  and the condition on $\bAUV$, $\bBUV$ for the WGC-like bounds on $z$ to exist,
are  collected and discussed systematically
in Apps.\,\ref{app:IR} and  \ref{app:BH} for spacetime dimension from $d=3$ to $d=11$.  
 In the scalar case, the positivity bounds are presented as exclusion regions in the $z\xi$-plane, see Figs.\,\ref{fig:A_D5}-\ref{fig:A_D8}.

We summarize the results using the following diagram:
\\~
\\
\begin{tikzpicture}
    \node[anchor=south west,inner sep=0] at (0,0) {\includegraphics[width=1.05\linewidth,trim={4.5cm 4cm 3cm 4.2cm},clip]{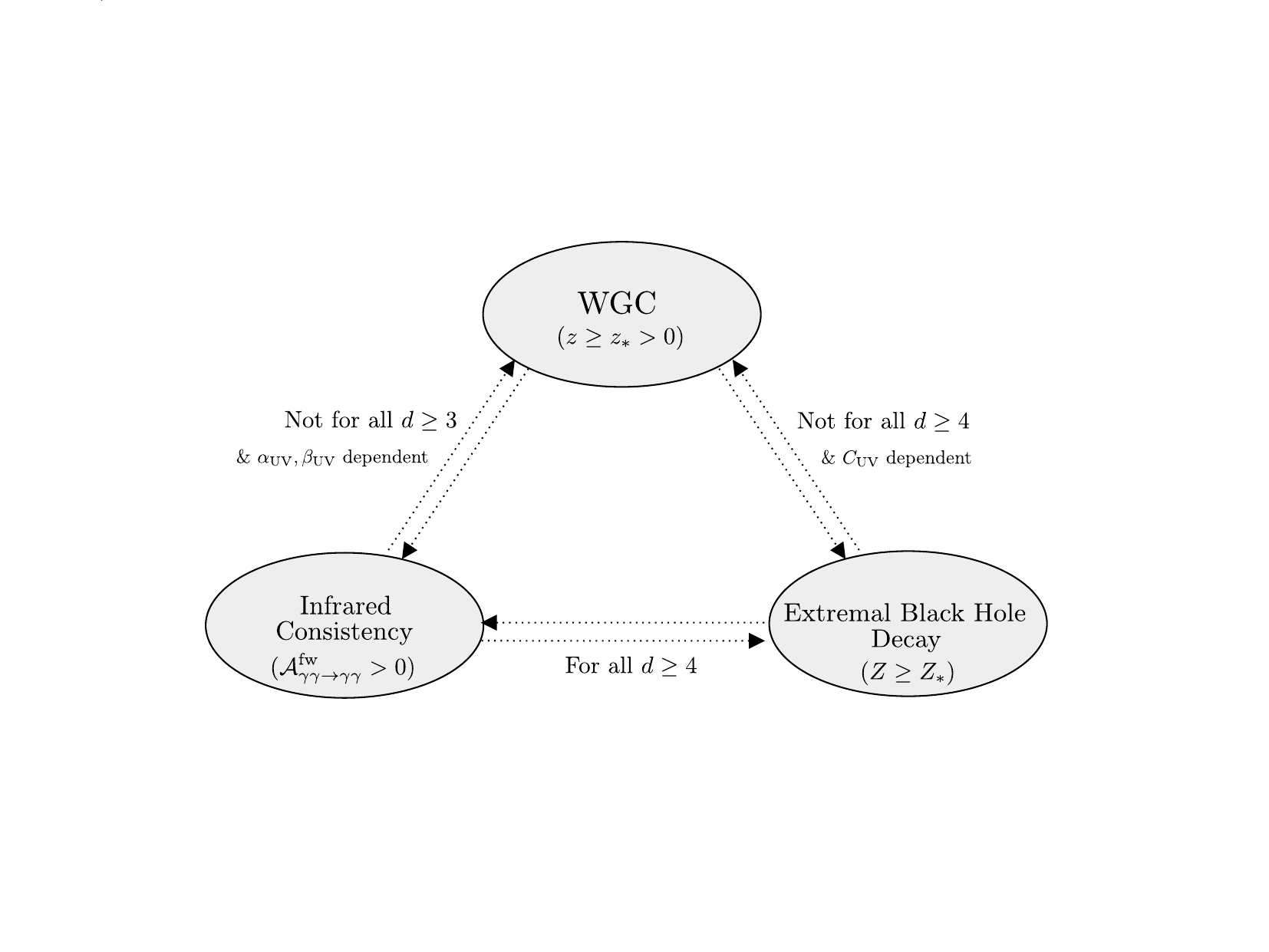}};
    \node[] at (4.8,6.7) {\tiny {\circled{1} } };
    \node[] at (4.,3.9) {\tiny {\circled{2} } };
    \node[] at (9.6,6.8) {\tiny {\circled{1} } };
    \node[] at (10.5,3.9) {\tiny {\circled{2} } };
    \node[] at (5.5,2.8) {\tiny {\circled{4} } };
    \node[] at (9.0,1.9) {\tiny {\circled{3} } };
\end{tikzpicture}
\\
Each of the implication arrows in this diagram come with conditions that are detailed below.  The annotations summarize the essential message.

\begin{enumerate}[label=\protect\circled{\arabic*}]
\item Infrared consistency  and extremal black hole decay both imply a WGC-like bound for $d=3,4,5,7,8,11$,  for any spin, if  $\bAUV$, $\bBUV$, $\bC_{\rm UV}$ are sufficiently small or negative.  
 In these dimensions, the $\Delta\alpha$, $\Delta\beta$, $\Delta C$ are positive for large enough $z$, i.e. the cases \textit{1a)} or \textit{1b)} of the general analysis (see \ref{se:general_analysis}) apply. 
In sharp contrast, for $d=9,10$, IR consistency does \textit{not} implies a WGC-like bound.
 In these dimensions, the $\Delta\alpha$, $\Delta\beta$, $\Delta C$ are negative at large enough $z$, i.e. 
 cases \textit{2a)} or \textit{2b)} apply, such that $z$ is bounded in a finite region.

\item The WGC  implies IR consistency and extremal black hole decay for $d=3,4,5,7,8,11$, unless the $\bAUV$, $\bBUV$, $\bC_{\rm UV}$ are too negative. For $d=9,10$, the positivity bounds are respected only if the UV coefficients are  positive and large enough for a given $z$. A remarkable consequence is that, in $d=9,10$,  the WGC does \textit{not}  imply that extremal black holes can decay in the IR EFT --- this instead  must be ensured by the  presence of sufficiently positive Planckian $F^4$ operators.

\item  Infrared consistency implies approximately the extremal black hole decay condition for any $d\geq 4$,  
 independently of the UV operators.

\item Extremal black hole decay  implies approximately the $\alpha_{\rm IR}>0$ condition from IR consistency for any $d\geq 4$,  
 independently of the UV operators. 

\end{enumerate}

The implications {\scriptsize \circled{3}}, {\scriptsize\circled{4}} 
are approximate in the sense that 
the expressions for $\Delta\alpha$ and $\Delta C$ have $O(1)$ differences in the $z^2$ and $z^0$ coefficients, however without qualitative implications.
This can be seen directly at the level of the expressions for  $\Delta \alpha$, $\Delta C$ in Apps.\,\ref{app:IR} and  \ref{app:BH}. As an example, for spin $\frac{1}{2}$  in $d=5$, the bound on $z$ for negligible UV operators is $z \geq 2.77$ from the IR consistency condition (see   Tab.\,\ref{tab:bound_condition_D5}), while it is $z \geq 2.78$  from the black hole condition  (see   Tab.\,\ref{tab:Delta_bar_IR_UV_5D_BlackHole}).\,\footnote{
The implications {\tiny\circled{3}}, {\tiny\circled{4}} can also be studied at the level of the EFT coefficients. In particular, it would be interesting to investigate 
the connection to the compactified bounds of \cite{Bellazzini:2019xts}, that are slightly stronger than Eq.\,\eqref{eq:positivity}.  The generalization to $d>4$  requires a thorough analysis  that we leave for future work.  }

In the special case of $d=6$, we find that implication {\scriptsize\circled{1}} produces  WGC-like  bounds  that are enhanced by a $\sqrt{\log \frac{M}{m}}$ factor. 
This happens for spin $\frac{1}{2}$, $1$,  and $0$   with $\xi>\frac{50}{23}$ in the $\Delta\bA$ case, and is a manifestation of the positive beta functions found in section \ref{se:beta_functions} (see App.\,\ref{app:IR} and Tab.\,\ref{tab:bound_condition_D6_case2}). 
Conversely, from implication {\scriptsize\circled{2}} we conclude that there exists a WGC-like bound that ensures that the $d=6$ IR EFT is infrared-consistent and that extremal black holes can decay.

\subsubsection{Results for $d>11$}

{{For $d>11$, a pattern appears. The higher dimensional cases are qualitatively  analogous to the cases $d=8$, 9, 10, 11, mod $4$. 
 Furthermore, the cases $d=8$ and $11$ (respectively, $d=9$ and $10$) are also similar to each other, up to the overall factor $\log\frac{\M}{m}$ which does not imply significant changes.
Each of the analogous cases has the same sign pattern for $\bar a,$ $\bar b,$ and $\bar c$.

\begin{itemize}

    \item $d=8+4n$ (respectively, $d=11+4n), \,\, n=1,2,3\dots$ is analogous to $d=8$ (respectively, $d=11$). Assuming  $\bAUV, \bBUV, \bar C_{\rm UV}$ are zero, $z$ is unbounded  for any spin.\,\footnote{ For spin $0$,   the bounds on $z$ become independent of $\xi$, i.e. when  $n\geq 1$, the exclusion region analogous to Fig.\,\ref{fig:A_D8} becomes empty.     }

     \item $d=9+4n$ (respectively, $d=10+4n), \,\, n=1,2,3,\dots$ is analogous to $d=9$ (respectively, $d=10$).   Similar to $d=9$ and 10, taking vanishing $\bAUV, \bBUV, \bar C_{\rm UV}$ implies a violation of the positivity constraint for any spin. Nonzero positive $\bAUV$, $\bBUV$ (resp. $\bar C_{\rm UV}$)  are then mandatory to ensure IR consistency (resp.  extremal black hole decay).

\end{itemize}
   }}

\FloatBarrier

\section{Summary}
\label{se:conclusion}

We  computed the four-photon operators generated  by charged particles  in any dimension  in the presence of gravity.
 We then used consistency conditions  from four-photon scattering and extremal black holes 
 to derive a set of bounds on the gravitational EFTs of charged particles.

The general setup is a subPlanckian (i.e. UV) gravitational EFT of a $U(1)$ charged particle  of spin $0$, $\frac{1}{2}$, or $1$. The EFT Lagrangian  features local $F^4$  operators that encode the effects of the  UV completion of quantum gravity on four-photon scattering. We briefly review some realizations of these UV $F^4$ operators from strings and branes.  

We computed the effect of loops of charged particles within the UV EFT, with focus on their contributions to four-photon scattering.  
From a diagrammatic viewpoint, besides non-gravitational  box diagrams  that generate $F^4$ operators, triangles and bubbles of charged particles  attached to gravitons generate $RF^2$ and $R^2$ operators.
We compute the effect of all these loops directly via expansion of the one-loop effective action encoded in the heat kernel coefficients.

The one-loop divergences of the effective action renormalize the $F^4$, $RF^2$ and $R^2$ operators in certain dimensions. Furthermore, in the  case of massive charged particles, the one-loop effective action provides the IR EFT in which charged particles are integrated out.
 Following the standard rules of EFT, the basis of $F^4$, $R F^2$, $R^2$  operators can be reduced, to some extent, using the equations of motion i.e. field redefinitions.

Our focus is ultimately on the physical process of four-photon scattering. Gravitons contribute at one-loop, but without self interactions --- to the exception of  diagrams that renormalize $R^2$ operators in $d=4$.  
We further use that the Gauss-Bonnet combination of $R^2$ operators vanishes at quadratic order of graviton fluctuation in \textit{any} dimension.
 The combination of these two facts implies that the basis of EFT effective operators can be reduced to $F^4$ and $CF^2$ operators  in any dimension in the calculation of $4\gamma$ amplitudes.

We provide the general result for the reduced one-loop  effective action for charged particles of spin $0$, $\frac{1}{2}$, or $1$ in any dimension. 
Gravity induces a renormalization flow of the $F^4$ operator  in $d=4,6,8$ and of $CF^2$ in $d=6$ dimensions. 
We verified the consistency of our results with some independent results on $d=3,4$  Einstein-Maxwell theory
from \cite{Drummond80,Ritz:1995nt, Bastianelli:2008cu, Dunne:2004nc, Fichet:2014uka}.

Turning to positivity bounds, we compute four-photon scattering in any  $d\geq 3$, and apply  a standard infrared consistency argument that provides positivity bounds on the $F^4$ operators. 
We also compute the bound produced by the condition that extremal black holes can decay in any dimension $d\geq 4$ \cite{Kats:2006xp}. 
 We find that both approaches yield nearly equivalent results, even though in the amplitudes we discard the graviton $t$-channel pole  and use the vanishing of the Gauss-Bonnet term  at quadratic order for any $d$.  
  The bound obtained without the  graviton $t$-channel is also supported by independent results from causality in $d=4$ \cite{CarrilloGonzalez:2023cbf}.

The infrared consistency of four-photon scattering and the decay of extremal black holes put bounds on the UV EFT of charged particles, our results are as follows.  

In  $d=4$ and $d=8$, the $F^4$ beta functions are negative, driving $F^4$ to positive values in the infrared.
In contrast,  the $d=6$ beta function from spin $0$, $\frac{1}{2}$ and $1$   drives  the  $F^4$ towards negativity. 
While for a massless particle the $F^4$ operators flows to  arbitrarily large and negative in the infrared, there is no immediate inconsistency because our positivity bounds do not apply for  strictly zero $m$. Still, it would be interesting to find if some additional ingredient can reverse the  beta function sign, for example, due to the gravitino or non-minimal couplings along the lines of \cite{Arkani-Hamed:2021ajd}.

For massive charged particles, we investigate the positivity bounds on the IR EFT in any dimension,  with specific focus on $d$ from $3$ to $11$. 
Our results always depend on the value of the UV $F^4$ operators encapsulating unknown superPlanckian effects.  We remain agnostic  to its  value a priori, but for concreteness we  discuss  cases where it is either negligible or large and positive.

The $F^4$ positivity bounds can constrain the charge-to-mass ratio $z$ provided the UV $F^4$ coefficients are not too large and positive. The quantities constrained are quadratic polynomials in $z^2$ defined on $\mathbb{ R}_+$. A variety of bounds appear depending on the shape of these polynomials. The bounds on $z$ can be from above or below, can be one or two-sided, and disjoint domains  are also  possible.

For $d=3,4,5,7,8,11$, we find that the positivity bounds imply  $O(1)$ lower bounds on $z$ similar to the $d$-dimensional Weak Gravity Conjecture for any spin, and for sufficiently small or negative $\bAUV$, $\bBUV$. 
We systematically present the condition on $\bAUV$, $\bBUV$ for the WGC-like bounds on $z$ to exist.
The bounds on the scalar depend on $\xi$ except in $d=4$, they  are presented as exclusion regions in the $z\xi$-plane.

In the specific case of  vanishing  $\bAUV$, $\bBUV$ or $C_{\rm UV}$, neat WGC-like bounds appear, for example in $d=5$ for all spins. This feature is not  general, however. For instance 
{{in $d=3$,  the}} spin-$0$ and spin-$\frac{1}{2}$ cases remain unbounded. This conclusion differs from the one from \cite{Cheung:2014ega}, but simply because the finite contribution from $R^2$ was ignored or equivalently absorbed into the UV coefficient in this reference.

 For $d=6$, the positivity bounds produce  WGC-like  bounds  that are enhanced by a $\sqrt{\log \frac{M}{m}}$ factor. 
This happens for spin $\frac{1}{2}$, $1$,  and $0$   with $\xi>\frac{50}{23}$ in the $\Delta\bA$ case, and is a manifestation of the positive beta functions found in $d=6$. 

Finally, for $d=9,10$, it turns out that the logic is different. The UV coefficient must be large enough for positivity bounds to be satisfied for a given value of $z$. In these cases,  $z$ is bounded in a finite range. As a result, even if the UV coefficients are very large, there  is necessarily an upper bound on the charge multiplied by a power of mass. For higher dimensions, similar cases arise following a mod $4$ pattern. 
A remarkable implication of
these $d=9,10$ results is that the WGC does \textit{not}  imply that extremal black hole decay in the IR EFT.

{A general takeaway from our study is that the connection between positivity bounds (from IR consistency, extremal black hole decay) and the WGC may not be so profound, as it appears to be  strongly dimension-dependent. }

{On the other hand, the approximate correspondence that we observe in any dimension between the IR consistency bounds and extremal black hole decay deserves further investigation. 
In $d=4$ it can be noticed that an  IR consistency bound obtained via the compactification method of \cite{Bellazzini:2019xts}, which is slightly more stringent than the one used here,  turns out to match precisely the extremal black hole bound once  both electric and magnetic cases are taken into account. 
It would be very interesting to verify whether this correspondence persists in $d>4$, that would likely hold only up to $O(\frac{1}{M^4})$ corrections due to the non-vanishing Gauss-Bonnet contribution  to the extremality relation. This investigation requires, however, a thorough analysis of the compactification method for higher $d$, that we leave for future work.  }

\FloatBarrier

\begin{acknowledgments}

We thank Dmitri Vassilievich  for useful discussions, and  Benjamin Knorr,   Javi Serra, Junsei Tokuda,  Congkao Wen and collaborators for valuable correspondence.  This work was supported in part by the S\~ao Paulo Research Foundation (FAPESP), grant 2021/10128-0. The work of PB was supported by Coordination for the Improvement of Higher Education Personnel – Brazil (CAPES) – Finance Code 88887.816450/2023-00, and LS was supported by grant 2023/11293-0 of FAPESP.

\end{acknowledgments}

\appendix

\section{Examples of $F^4$ Operators from Strings and Branes}
\label{se:strings}

The ultraviolet $F^4$ operators are generated in string theory. For example,   
the four-photon interaction arising at low-energy from perturbative open string amplitude  in $d=10$, at lowest order in the string coupling $g_s$, can be found in \cite{SCHWARZ1982223,Polchinski:1998rr}.
For large string tension $\frac{1}{2\pi\alpha_s'}$ we deduce from  the string amplitude the following effective Lagrangian (see also \cite{deRoo:2003xv}):
\be
{\cal L}_{F^4,\,{\rm string}} = \frac{1}{8}g_s (2\pi\alpha'_s)^2 \left(F^{\mu\nu}F_{\nu\rho } F^{\rho \sigma } F_{\sigma \mu}-\frac{1}{4}(F^{\mu\nu}F_{\mu\nu } )^2\right) +O(\alpha^{'4}_s)
\label{eq:F4_strings}
\ee
The prediction  holds under the compactification of spatial dimensions, however in that case contributions from the Kaluza-Klein modes should also be taken into account, which  likely dominate the low-energy $F^4$ operators.    

The ultraviolet $F^4$ operators also appear in models where the photon arises from a D-brane, and charged particles correspond to open strings attached to the brane. This configuration was shown in \cite{Fradkin:1985qd} to be described by a Born-Infeld action. We deduce the effective Lagrangian (see also \cite{Davila:2013wba})
\,\footnote{In $d=4$ we recover the original BI Lagrangian 
${\cal L}_{{\rm BI}} = \b^2 \left(1-\sqrt{1+\frac{1}{2\b}F^{\mu\nu}F_{\mu\nu}-\frac{1}{4\b^2}(\tilde F^{\mu\nu}F_{\mu\nu})^2}\right)$ \cite{Born:1934gh}.
}
\begin{align}
{\cal L}_{{\rm BI}} 
&=-\b^2 \sqrt{-\det(\eta_{\mu\nu}+\frac{1}{\b}F_{\mu\nu})}
+\b^2 \sqrt{-\det(\eta_{\mu\nu})}
\\ &= -\frac{1}{4}F^{\mu\nu}F_{\mu\nu} +\frac{1}{8\b^2}
\left(F^{\mu\nu}F_{\nu\rho } F^{\rho \sigma } F_{\sigma \mu}-\frac{1}{4}(F^{\mu\nu}F_{\mu\nu } )^2\right)
+ O(b^{-3})
\end{align}
In the result of \cite{Fradkin:1985qd}, upon suitable field normalization one obtains $b^{-2}=g_s (2\pi \alpha'_s)$ that exactly reproduces the overall coefficient in  ${\cal L}_{F^4,\,{\rm string}}$. 

Once expressed in the ${\cal O}$, $\tilde {\cal O}$ basis, the specific combination of $F^4$ operators generated by the string models implies  
\be \A_{\rm UV} = \B_{\rm UV}>0\,
\label{eq:A_P_positivity}
\ee
in any dimension. 
Such a positivity bound is not surprising. Since string theory must be a consistent completion of quantum gravity, the Planckian $F^4$ operators have to satisfy the positivity bound regardless of the presence of light fields in the theory, which implies  Eq.\,\eqref{eq:A_P_positivity}.

The $F^4$ operator can also appear in compact extra dimension models with the photon localized on a brane.  
In that case, a universal tree-level contribution comes from the Kaluza-Klein graviton exchange, i.e. the massive version of the tree diagram \eqref{eq:grav_tree}. Each massive graviton $h^{(n)}_{\mu\nu}$ couples to the brane-localized photon stress tensor, $\int_{\rm brane} {\sqrt{g}_{\rm ind}} h^{(n)}_{\mu\nu} T^{\mu\nu}$. 
The $F^4$ operators are generated by integrating out the massive gravitons, whose propagator is $G^{(n)}_{\mu\nu,\rho\sigma}= \frac{-i}{p^2+m_n^2}\left(
\frac{1}{2}(P^{(n)}_{\mu\rho}P^{(n)}_{\nu\sigma}+P^{(n)}_{\mu\sigma}P^{(n)}_{\nu\rho})-\frac{1}{d-1}P^{(n)}_{\mu\nu}P^{(n)}_{\rho\sigma}
\right)$ with $P^{(n)}_{\mu\nu}=\eta_{\mu\nu}-\frac{p_\mu p_\nu}{m_n^2}$. In the EFT, these massive gravitons generate the operator 
\begin{eqnarray}
{\cal L}_{\rm grav} \propto && \nn 
 T^{\mu\nu}T_{\mu\nu}-\frac{1}{d-1}T^2
\\ && \propto
F^{\mu\nu}F_{\nu\rho } F^{\rho \sigma } F_{\sigma \mu}
-\frac{d+8}{16(d-1)}(F^{\mu\nu}F_{\mu\nu } )^2 \,.
\end{eqnarray}
The subsequent $\alphaUV$, $\betaUV$  coefficients turns out to be positive for any $d\geq 3$. 
In the $d=4$ case the combination becomes  $F^{\mu\nu}F_{\nu\rho } F^{\rho \sigma } F_{\sigma \mu}-\frac{1}{4}F^{\mu\nu}F_{\mu\nu }$ as in the string case. This matches the results from  \cite{Fichet:2013ola, Fichet:2013gsa}.

 Even though the above results suggests positivity of the $\alphaUV$, $\betaUV$ coefficients in the ultraviolet EFT, we should keep in mind that negative contributions do exist, for instance from charged KK modes, as can be seen from our massive results from section \ref{se:finite_corrections}. 
In this work, we do not systematically evaluate the contributions that appear upon compactification. Our approach is  to remain fully agnostic about the values of $\alphaUV$, $\betaUV$. 


\section{Detailed Bounds from Infrared Consistency}
\label{app:IR}

We discuss the consequences of IR consistency for each spacetime dimension.  The results  follow a pattern  at large dimension. Lower dimensions require separate discussion,  we cover the cases from $d=3$ to $d=11$. 
In the following,  we report systematically  the implications of the positivity bounds \eqref{eq:ReducedIRboundschem} while  remaining agnostic about the values of $\alphaUV$, $\betaUV$.

\subsection*{Case $d=3$}

\begin{table}[h]
    $$\begin{array}{|c||c|}\hline 
    \text{Spin} & \Delta\bA \\ \hline\hline
    0 &  \frac{7 z^4}{1920}+\frac{(1-10 \xi ) z^2}{480}  + \frac{60 \xi ^2-20 \xi +3}{480}   \\ 
    \frac{1}{2} & \frac{z^4}{240}-\frac{z^2}{480}+\frac{1}{240} \\ 
    1 &\frac{127 z^4}{960}-\frac{11z^2}{60}+\frac{1}{30}  \\
    \hline 
    \end{array}$$
    
        \caption{Reduced coefficient $\Delta\bA$ in $d=3$.}
    \label{tab:WC_D3}
\end{table}

The values of $\Delta\bA$ are presented in Tab.\,\ref{tab:WC_D3}. 
We remind that there is only one independent operator in $d=3$,  chosen to be $F_{\mu\nu}F^{\mu\nu}$ with reduced coefficient $\bar\alpha_{\rm IR}= \bar\alpha_{\rm UV}+\Delta\bA$.

From Tab.\,\ref{tab:WC_D3}  we have $\bar a>0$, $\bar c>0$ for all spins, $\bar b<0$ for spin $\frac{1}{2}$ and $1$, and ${\rm sign }(\bar b)={\rm sign }(1-10\xi)$  for spin 0.   Therefore  cases \textit{(1a)} and \textit{(1b)} from the classification in section~\ref{se:general_analysis} apply.
It follows that the positivity bound $\bar\alpha_{\rm IR}\geq 0$ can constrain the charge-to-mass ratio $z$ depending on the value of the UV coefficient $ \bar\alpha_{\rm UV}$. 
The exact condition for the existence of a bound on $z$ is shown in Tab.\,\ref{tab:bound_condition_D3}.

We see that for spin $\frac{1}{2}$, and for spin $0$ with any $\xi$, $\bAUV$ has to be sufficiently negative in order for $z$ to be bounded. 
For spin $1$, $z$ can be bounded for small positive $\bAUV$.

As an example, we may consider the specific case where $\bAUV$ is negligible, setting $\bAUV=0$. For spin $\frac{1}{2}$, we have $\bA_{\rm UV} = 0 \geq - \frac{1}{256} = \bA^*$,  thus $\bA_{\rm IR} \geq 0, \forall z \in \mathbb{R}_{+}$, hence $z$ is unbounded. On the other hand, for spin $1$, we have $\bA_{\rm UV} = 0 < \frac{23}{762} = \bA^*$, so that $z$ is bounded on a region $\left[0,z_1\right] \cup \left[z_2,\infty\right)$, corresponding to the second case of \textit{(1b)}. In the spin-$0$ case, 
 for any $\xi$ the   $\bAUV$ coefficient has to be negative to produce a bound, therefore $z$ is unbounded.

Our expressions from Tab.\,\ref{tab:bound_condition_D3} 
reproduce the ones from \cite{Cheung:2016wjt} upon neglecting the $\bar c$ coeffi\-cient  and changing the convention for $z$.\,\footnote{In \cite{Cheung:2016wjt}, for $d=3$, the convention $M=\frac{1}{2}$ is used and $z$ is defined as  $\frac{|q|}{m}$. This  differs from our definition of $z$ by a factor of $\sqrt{2}$.} 
Our conclusions for $\alphaUV=0$ differ from those in \cite{Cheung:2016wjt} due to the $\bar c$  contribution --- originating  from the $R^2$ operator, which is not taken into account in \cite{Cheung:2016wjt}. 
As seen above, the $\bar c$ contribution crucially favors positivity. The case studied in    \cite{Cheung:2016wjt} is  instead exactly recovered from our results by tuning  $\bAUV+ \bar c $ to zero.

\begin{table}[h]
$$
\begin{array}{|c||c||c|}
\hline
 \text{Spin} &\text{Condition for}~z~\text{bounded}  & \text{Bound if}~ \bA_{\rm UV} = 0  \\ \hline \hline

0 & \bAUV < \left\{\begin{array}{cc}
     -\frac{60 \xi ^2 - 20\xi +3}{480 } & \,\text{if } \xi\leq \frac{1}{10} \\ 
     -\frac{ 16 \xi^2- 6\xi +1}{168} & \, \text{if } \xi> \frac{1}{10}
    \end{array}\right. &  \text{Unbounded} \\ 

\frac{1}{2} & \bAUV < - \frac{1}{256 } & \text{Unbounded}\\

1 & \bAUV < \frac{23}{762}  \text{ } \text{ } \text{ }  & z\leq 0.464~\text{or}~z\geq 1.08 \\ \hline
\end{array}
$$
        \caption{Condition  for the existence of IR consistency  bounds on the charge-to-mass ratio $z$  and IR consistency bounds on $z$ if $\bAUV=0$  in $d=3$. }
    \label{tab:bound_condition_D3}
\end{table}

\FloatBarrier

\subsection*{Case $d=4$}

\begin{table}[h]
$$
\begin{array}{|c||c|c|}
\hline
 \text{Spin } & \Delta\bA & \Delta\bB \\ \hline \hline

0 &  \frac{7 z^4}{1440}-\frac{z^2}{180} + \frac{1}{120}\log \frac{M}{m} & \frac{z^4}{1440}-\frac{z^2}{180} + \frac{1}{120}\log \frac{M}{m} \\ 

\frac{1}{2} & \frac{z^4}{90}-\frac{11 z^2}{360} + \frac{1}{40}\log \frac{M}{m} &  \frac{7 z^4}{360}-\frac{11 z^2}{360} + \frac{1}{40}\log \frac{M}{m} \\ 

1 &  \frac{29 z^4}{160}-\frac{31 z^2}{60} + \frac{13}{120}\log \frac{M}{m} & \frac{27 z^4}{160}-\frac{31 z^2}{60} + \frac{13}{120}\log \frac{M}{m} \\ \hline
\end{array}
$$
        \caption{
        Reduced coefficient $\Delta\bA$, $\Delta\bB$ in $d=4$.
        }
    \label{tab:Delta_bar_IR_UV_4D}
\end{table}

The values of $\Delta\bA$, $\Delta\bB$ are presented in Tab.\,\ref{tab:Delta_bar_IR_UV_4D}. 
The scalar case is independent of $\xi$ in $d=4$ due to the vanishing of the trace of the stress tensor, see \eqref{eq:T_identities}.
The $\bar c$ coefficient originating from $R^2$   features a logarithm that corresponds to the effect of the 4d beta function shown in Tabs.\,\ref{tab:beta_zeromass_Alpha},\,\ref{tab:beta_zeromass_Beta}. 

From Tab.\,\ref{tab:Delta_bar_IR_UV_4D} we have that both $\bar a$ and $\bar c$ are positive  and $\bar b$  negative for any spin, for both $\Delta \bA$ and $\Delta \bB$.
We are thus in case \textit{(1b)}.
A bound on $z$ appears if  $\bAUV$ or $\bBUV$ are sufficiently negative, the exact condition is given in Tab.\,\ref{tab:bound_condition_D4_case1}. As an example, we may consider the specific case where $\bAUV$, $\bBUV$ is negligible, setting $\bAUV=0$, $\bBUV=0$. Assuming that $\frac{M}{m}\gtrsim 50$,   $z$ is unbounded for any spin.

Our results reproduces those from \cite{Cheung:2016wjt} for $d=4$ 
when ignoring the logarithmic term (i.e. $\bar c $) or absorbing it into 
$\bAUV$, $\bBUV$, and changing the convention for $z$.\,\footnote{In  \cite{Cheung:2016wjt}, for $d=4$,  the convention $M^2=\frac{1}{2}$ is used. Hence, the  definition of $z$ differs from ours by a factor of $2$.  }
Accordingly, the special case considered in  \cite{Cheung:2016wjt} is exactly reproduced here by tuning $\bA_{\rm UV} + \bar c  = 0$.

\begin{table}[h]
            $$
            \begin{array}{|c||c|} \hline
             \text{Spin} & \text{Condition for}~z~\text{bounded}\\ \hline \hline 
            
            0 & \bA_{\rm UV}  < \frac{1}{630} - \frac{1}{120}\log \frac{\M}{m}~~\text{or}~~ \bB_{\rm UV}  < \frac{1}{90} - \frac{1}{120}\log \frac{\M}{m}\\ 
            
            \frac{1}{2} & \bA_{\rm UV}  < \frac{121}{5760} - \frac{1}{40}\log \frac{\M}{m}~~\text{or}~~ \bB_{\rm UV}  < \frac{121}{10080} - \frac{1}{40}\log \frac{\M}{m}\\

            1 & \bA_{\rm UV}  < \frac{961}{2610} - \frac{13}{120}\log \frac{\M}{m}~~\text{or}~~ \bB_{\rm UV} <\frac{961}{2610} - \frac{13}{120}\log \frac{\M}{m} \\ \hline
            \end{array}
            $$
                \caption{Condition  for the existence of IR consistency  bounds on the charge-to-mass ratio $z$ in $d=4$.}
                \label{tab:bound_condition_D4_case1}
            \end{table}

\FloatBarrier

\subsection*{Case $d=5$}

\begin{table}[h]

    $$\begin{array}{|c||c|c|} \hline
    \text{Spin} & \Delta\bA  & \Delta\bB  \\ \hline\hline 
    0 &  \frac{7 z^4}{2880}-\frac{(3-5 \xi) z^2}{360}  - \frac{60 \xi ^2-20\xi +23}{2160} & \frac{ z^4}{2880}-\frac{z^2}{216}-\frac{1}{120}  \\  \frac{1}{2} & \frac{z^4}{180}-\frac{7z^2}{180}-\frac{4}{135} & \frac{7 z^4}{720}-\frac{7 z^2}{216}-\frac{1}{40} \\ 
    1 &\frac{67 z^4}{720}- \frac{197 z^2}{360} -\frac{151}{1080} & \frac{61 z^4}{720}- \frac{25 z^2}{54} -\frac{7}{60} \\ \hline 
    \end{array}$$
        \caption{Reduced coefficients $\Delta\bA$ and $\Delta\bB$ in $d=5$. }
    \label{tab:WC_D5}
\end{table}

The values of $\Delta\bA$, $\Delta\bB$ are presented in Tab.\,\ref{tab:WC_D5}. 
We have $\bar a>0$, $\bar c<0$  for all spins, for both $\Delta\bar \alpha$, $\Delta\bar \beta$ and any $\xi$. We have $\bar b<0$  for spin $\frac{1}{2}$ and $1$, and for spin 0 we have $\bar b<0$  for $\Delta\bB$ and ${\rm sign }(\bar b)={\rm sign }(3-5\xi)$ for $\Delta\bA$. We are thus in cases \textit{(1a)} and \textit{(1b)}.
A bound on $z$ appears if  $\bAUV$ or $\bBUV$ are sufficiently small, the exact condition is given in Tab.\,\ref{tab:bound_condition_D5}.

\begin{figure}[t]
    \centering

    \includegraphics[scale=0.6]{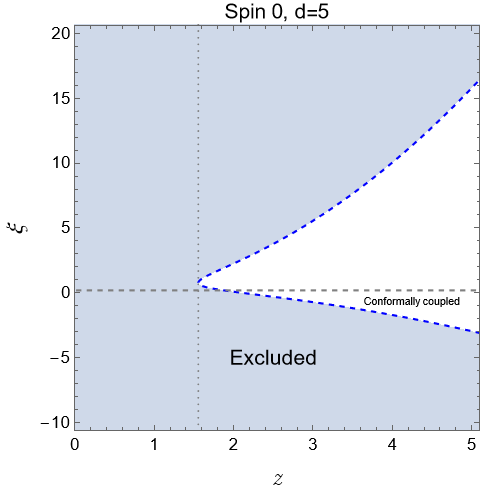}
    
    \caption{Infrared consistency bounds  on  the charged spin $0$ particle in $d=5$.}
    \label{fig:A_D5}
\end{figure}

As an example, we may consider the specific case where $\bAUV$, $\bBUV$ is negligible, setting $\bAUV=0$, $\bBUV=0$.  We obtain WGC-like bounds for all spins, as shown in Tab.\,\ref{tab:bound_condition_D5}. In the spin-$0$ case,  the bound is $\xi$-dependent. Positivity excludes a region in the $z\xi$-plane,  as shown in Fig.\,\ref{fig:A_D5}. For reference, we include in Fig.\,\ref{fig:A_D5} and in analogous figures in higher dimensions the 
value of $\xi$ for which the scalar is conformally coupled if $m\to 0$. 
The allowed region in Fig.\,\ref{fig:A_D5} has a critical point to the left, which imposes a lower bound on the charge-to-mass ratio for all $\xi$. The subsequent WGC bound for all $\xi$ is 
$z  \geq \sqrt{\frac{13+\sqrt{937}}{18}} \approx 1.56$.

\begin{table}[h]
$$
\begin{array}{|c||c||c|c|} \hline
 \shortstack{~ \\ \text{Spin} \\ \vspace{0.05cm}} & \shortstack{\vspace{0.1cm} \\ \text{Condition for} \\ \vspace{0.005cm} \\~$z$~\text{bounded}} & \shortstack{ \vspace{0.1cm} \\ Bound if \\ \vspace{0.005cm} \\ $\bA_{\rm UV} = 0$} &\shortstack{ \vspace{0.1cm} \\ Bound if \\ \vspace{0.005cm} \\ $\bB_{\rm UV} = 0$} \\ \hline \hline 

\multirow{2}{*}{$0$} & \bA_{\rm UV} < \left\{\begin{array}{cc}
     \frac{60 \xi ^2-20\xi +23}{2160} & \, \text{if } \xi\geq \frac{3}{5} \\ 
     \frac{720 \xi ^2-500 \xi +269}{15120} & \, \text{if } \xi< \frac{3}{5}
    \end{array}\right. & \multirow{2}{*}{ \text{Figure } \ref{fig:A_D5}} & \multirow{2}{*}{$z \geq 3.87$} \\ 

&  ~~\text{or}~~ \bB_{\rm UV} < \frac{77}{3240} & & \\

\frac{1}{2} & \bA_{\rm UV} <  \frac{211}{2160} \text{ or } \bB_{\rm UV} <  \frac{337}{6480} &z \geq 2.77 & z \geq 1.99 \\

1 & \bA_{\rm UV} < \frac{136661}{144720} \text{ or } \bB_{\rm UV} < \frac{74029}{98820} &  z \geq 2.48 & z \geq 2.39  \\ \hline
\end{array}
$$
        \caption{Condition  for the existence of IR consistency  bounds on the charge-to-mass ratio $z$ and IR consistency bounds on $z$ if $\bAUV=\bBUV=0$ in $d=5$.
        }
    \label{tab:bound_condition_D5}
\end{table}

\FloatBarrier

\subsection*{Case $d=6$}

\begin{table}[h]
$$
\begin{array}{|c||c|c|}
\hline
 \text{Spin } & \Delta\bA & \Delta\bB\\ \hline \hline

0 & \frac{7 z^4}{1440} - \left(\frac{(23 - 50 \xi) z^2}{1200} + \frac{15 \xi ^2-5\xi +3 }{ 240 }\right)\log \frac{M}{m} &  \frac{z^4}{1440}  - \left(\frac{z^2}{120}+\frac{1}{120}\right)\log \frac{M}{m} \\ 

\frac{1}{2} &   \frac{z^4}{45}  - \left( \frac{13 z^2}{75}+\frac{31}{480}\right)\log \frac{M}{m} &   \frac{7 z^4}{180}  - \left(\frac{2z^2}{15}+\frac{1}{20}\right)\log \frac{M}{m} \\ 

1 &  \frac{55 z^4}{288} - \left(\frac{269 z^2}{240} + \frac{1}{6}\right)\log \frac{M}{m} & \frac{49 z^4}{288}  - \left(\frac{7 z^2}{8}+\frac{1}{8}\right)\log \frac{M}{m}  \\ \hline
\end{array}
$$
        \caption{Reduced coefficients $\Delta\bA$ and $\Delta\bB$ in $d=6$.}
    \label{tab:Delta_bar_IR_UV_6D}
\end{table}

The values of $\Delta\bA$, $\Delta\bB$ are presented in Tab.\,\ref{tab:Delta_bar_IR_UV_6D}. 
The $\bar b$ and $\bar c$ coefficients originating respectively from $F^2R$ and $R^2$   feature a logarithm that corresponds to the effect of the 6d beta function from Tabs.\,\ref{tab:beta_zeromass_Alpha},\,\ref{tab:beta_zeromass_Beta}. 
This logarithm will in turn influence the WGC-like bounds, as discussed below.

From Tab.\,\ref{tab:Delta_bar_IR_UV_6D} we have $\bar a>0$, $\bar c<0$  for all spins, for both $\Delta\bar \alpha$, $\Delta\bar \beta$ and any $\xi$. We have $\bar b<0$ for spin $\frac{1}{2}$ and 1, and for spin 0 we have $\bar b<0$  for $\Delta\bB$ and ${\rm sign }(\bar b)={\rm sign }(50\xi-23)$ for $\Delta\bA$. We are thus in cases \textit{(1a)} and \textit{(1b)}. A bound on $z$ appears if  $\bAUV$ or $\bBUV$ are sufficiently small, the exact condition is given in Tab.\,\ref{tab:bound_condition_D6}.

As an example we may consider the specific case where $\bAUV$, $\bBUV$ is negligible, setting $\bAUV=0$, $\bBUV=0$. We obtain WGC-like bounds for all spins, that depend on $\log\frac{\M}{m}$. We focus on the regime where $\log\frac{\M}{m} \gg 1$, the results are shown in Tab.\,\ref{tab:bound_condition_D6_case2}.  The bounds presented  
are 
weaker than those obtained for small $\log\frac{\M}{m}$ and thus hold for any value of $\frac{\M}{m}$.

We emphasize that the logarithmic enhancement occurring in the bounds is tied to the positive sign of the 6d beta functions. While the positive beta functions lead to  stronger  bounds  on $z$ at large $\log\frac{M}{m}$,   negative beta functions would lead to weaker bounds on $z$, that are independent of the logarithm. 

In the spin-$0$ case, the $\xi$-dependent bound is shown in Fig.\,\ref{fig:A_D6}, where $\log\frac{\M}{m}$ was set to 100 to plot the exclusion region. The allowed region in Fig.\,\ref{fig:A_D6} has a critical point to the left which imposes a lower bound for the charge-to-mass ratio for all $\xi$.
This critical lower bound increases with $\log\frac{\M}{m}$, converging to the limit $\sqrt{\frac{880+5\sqrt{93410}}{1007}}\approx 1.55$. For smaller $\log\frac{\M}{m}$ the bound 
is only slightly weaker, with e.g.  $z \geq 1.27$ if $\log\frac{\M}{m}=1$, which holds for any $\log\frac{\M}{m} \geq 1$ and any $\xi$.
{Finally, if $\xi \leq \frac{23}{50}$, the WGC-like bound strengthens to $z \geq \sqrt{\frac{6 (23-50\xi)}{35} \log\frac{\M}{m}}$, which holds for any value of $\frac{\M}{m}$, analogous to other bounds in Tab.\,\ref{tab:bound_condition_D6_case2}.

\begin{table}[h]
$$
\begin{array}{|c||c|} \hline
 \text{Spin} & \text{Condition for}~z~\text{bounded}\\ \hline \hline 

\multirow{3}{*}{$0$} &\left\{\begin{array}{cc}
     \bA_{\rm UV} < \frac{15 \xi ^2-5\xi +3}{240} \log\frac{\M}{m} &  \,\text{if } \xi > \frac{23}{50} \\ 
     \bA_{\rm UV} < \left(\frac{15 \xi ^2-5\xi +3}{240}+\frac{\left(23-50\xi \right)^2}{28000}\log\frac{\M}{m}\right) \log\frac{\M}{m} &  \,\text{if } \xi \leq \frac{23}{50}
    \end{array}\right. \\ 
    
   & ~~\text{or}~~ \bB_{\rm UV} < \left(\frac{1}{120}+\frac{1}{40}\log\frac{\M}{m}\right)\log\frac{\M}{m} 

\\ 

\shortstack{\vspace{0.3cm}\\$\frac{1}{2}$} & \shortstack{\vspace{0.3cm}\\$\bA_{\rm UV} < \left(\frac{31}{480}+\frac{169}{500}\log\frac{\M}{m}\right)\log\frac{\M}{m}$   ~~\text{or}~~ $\bB_{\rm UV} < \left(\frac{1}{20}+\frac{4}{35}\log\frac{\M}{m}\right)\log\frac{\M}{m}$} \\

1 & \bA_{\rm UV} < \left(\frac{1}{6}+\frac{72361}{44000}\log\frac{\M}{m}\right)\log\frac{\M}{m}  ~~\text{or}~~ \bB_{\rm UV} < \left(\frac{1}{8}+\frac{9}{8}\log\frac{\M}{m}\right)\log\frac{\M}{m}  \\

\hline
\end{array}
$$
        \caption{Conditions for the existence of  bounds on  $z$ in $d=6$.}
    \label{tab:bound_condition_D6}
\end{table}

\begin{table}[h]
$$
\begin{array}{|c||c|c|} \hline
 \text{Spin} &  \text{Bound if}~\bA_{\rm UV} = 0 &  \text{Bound if}~\bB_{\rm UV} = 0 \\ \hline \hline

$0$ & \text{Figure }\ref{fig:A_D6}\, & z \geq 3.46\sqrt{\log \frac{\M}{m}}\\

\frac{1}{2} & z \geq 2.79\sqrt{\log \frac{\M}{m}} & z \geq 1.85\sqrt{\log \frac{\M}{m}}  \\

1 & z \geq 2.42\sqrt{\log \frac{\M}{m}}  &  z \geq 2.27\sqrt{\log \frac{\M}{m}} \\

\hline
\end{array}
$$
        \caption{IR consistency bounds on $z$ if $\bAUV=\bBUV=0$ and $\log\frac{\M}{m} \gg 1$ in $d=6$. }
    \label{tab:bound_condition_D6_case2}
\end{table}

\begin{figure}[h]
    \centering

    \includegraphics[scale=0.6]{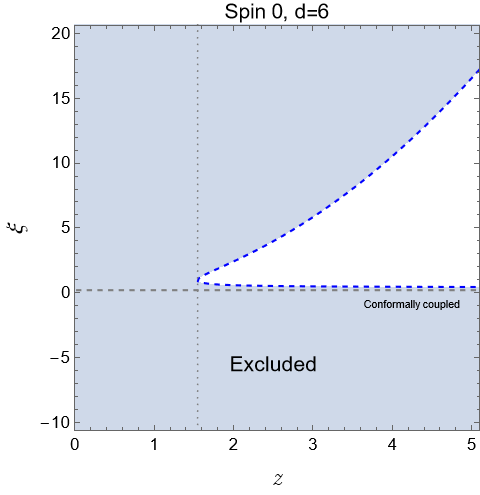}
    
    \caption{Infrared consistency bounds  on  the charged spin $0$ particle in the regime $\log \frac{\M}{m} \gg 1$ (here with $\log\frac{\M}{m} = 100$) 
    in $d=6$.}
    \label{fig:A_D6}
\end{figure}

 \FloatBarrier

\subsection*{Case $d=7$}

\begin{table}[h]
    $$\begin{array}{|c||c|c|} \hline 
    \text{Spin} & \Delta\bA  & \Delta\bB  \\ \hline \hline 
    0 &  \frac{7 z^4}{1440}+\frac{(37 - 90\xi) z^2}{1800} + \frac{540 \xi ^2-180 \xi +83}{9000} & \frac{z^4}{1440}+\frac{7 z^2}{900}+\frac{1}{180}  \\  \frac{1}{2} & \frac{z^4}{45}+\frac{83 z^2}{450}+\frac{17}{375} & \frac{7 z^4}{180}+\frac{61 z^2}{450}+\frac{1}{30} \\  1 &\frac{47 z^4}{240}+\frac{57 z^2}{50}+\frac{287}{2250} & \frac{41 z^4}{240}+\frac{127 z^2}{150}+\frac{4}{45} \\ \hline 
    \end{array}$$
        \caption{
Reduced coefficients $\Delta\bA$ and $\Delta\bB$ in $d=7$.        
        }
    \label{tab:WC_D7}
\end{table}

The values of $\Delta\bA$, $\Delta\bB$ are presented in Tab.\,\ref{tab:WC_D7}. 
We have $\bar a>0$, $\bar c>0$  for all spins, for both $\Delta\bar \alpha$, $\Delta\bar \beta$ and any $\xi$.
We have $\bar b>0$ for spin $\frac{1}{2}$ and 1, and for spin 0 we have $\bar b>0$  for $\Delta\bB$ and ${\rm sign }(\bar b)={\rm sign }(37-90\xi)$ for $\Delta\bA$. We are thus in cases \textit{(1a)} and \textit{(1b)}.
A bound on  $z$ appears if $\bAUV$ or $\bBUV$ are sufficiently small, the exact condition is given in Tab.\,\ref{tab:bound_condition_D7}. 

As an example, we  consider the specific case $\bAUV=0$, $\bBUV=0$.
For spin  $\frac{1}{2}$ we are in case \textit{(1a)} with $\bar\alpha^*>0$, hence $z$ is unbounded. 
For spin $1$ we are in case \textit{(1b)} with $-\bar\alpha^*>0$. As a result, $z$ is bounded to two disjoint regions, the one at larger $z$ being WGC-like. 
For spin $0$, the $\Delta\bB$ does not constrain $z$, while a $\xi$-dependent bound exists from  $\Delta\bA$. This is shown in Fig.\,\ref{fig:A_D7}. 

This region in Fig.\,\ref{fig:A_D7} has critical points on the left and bottom. If $z \leq \sqrt{\frac{11+\sqrt{257}}{10}}\approx 1.64$ then $\bA_{\rm IR} \geq 0, \forall \xi$ and if $\xi \leq \frac{225+\sqrt{26985}}{360} \approx 1.08$ then $\bA_{\rm IR} \geq 0, \forall z$.
Conversely when $\xi> 1.08$ the domain of $z$ is restricted to  two disjoint regions, corresponding to the $-\bar c \leq \bA_{\rm UV} < -\bA^*$ case in \textit{(1b)}. The domain at larger $z$ is WGC-like.

\begin{table}[h]
$$
\begin{array}{|c||c||c|c|} \hline
 \shortstack{~ \\ \text{Spin} \\ \vspace{0.05cm}} & \shortstack{\vspace{0.1cm} \\ \text{Condition for} \\ \vspace{0.005cm} \\~$z$~\text{bounded}} & \shortstack{ \vspace{0.1cm} \\ Bound if \\ \vspace{0.005cm} \\ $\bA_{\rm UV} = 0$} &\shortstack{ \vspace{0.1cm} \\ Bound if \\ \vspace{0.005cm} \\ $\bB_{\rm UV} = 0$} \\ \hline \hline

\multirow{2}{*}{$0$} & \bA_{\rm UV} < \left\{\begin{array}{cc}
     -\frac{540 \xi ^2-180 \xi +83}{9000}  &  \,\text{if } \xi \leq \frac{37}{90} \\ 
     \frac{1080 \xi ^2-1350 \xi +197}{15750} &  \,\text{if } \xi> \frac{37}{90}
    \end{array}\right.  & \multirow{2}{*}{\text{Figure } \ref{fig:A_D7}} & \multirow{2}{*}{\text{Unbounded}}  \\

 & \text{or}~~ \bB_{\rm UV} < -\frac{1}{180}\qquad\qquad\qquad\qquad  & &

\\

\frac{1}{2} & \bA_{\rm UV} <  -\frac{17}{375} \text{ or } \bB_{\rm UV} <  -\frac{1}{30} & \text{Unbounded} &  \text{Unbounded}  \\

1 & \bA_{\rm UV} < -\frac{287}{2250}\text{ or } \bB_{\rm UV} < -\frac{4}{45} & \text{Unbounded}  &  \text{Unbounded}   \\ \hline 
\end{array}
$$
        \caption{Condition  for the existence of IR consistency  bounds on the charge-to-mass ratio $z$  and IR consistency bounds on $z$ if $\bAUV=\bBUV=0$ in $d=7$. }
    \label{tab:bound_condition_D7}
\end{table}

\begin{figure}[h]
    \centering

    \includegraphics[scale=0.6]{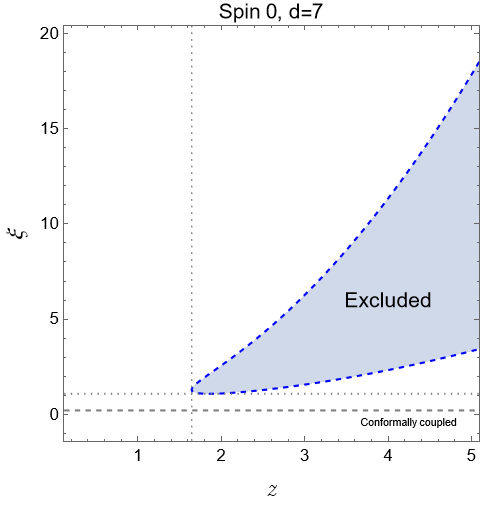}
    
    \caption{Infrared consistency bounds  on  the charged spin $0$ particle in $d=7$.}
    \label{fig:A_D7}
\end{figure}

\FloatBarrier

\subsection*{Case $d=8$}

\begin{table}[h]
$$
\begin{array}{|c||c|c|}
\hline
 \text{Spin } & \Delta\bA & \Delta\bB \\ \hline \hline

0 & \left(\frac{7 z^4}{720}  + \frac{(27 -70 \xi)z^2}{1260} + \frac{15 \xi ^2-5 \xi +2}{ 270 }\right)\log \frac{M}{m} & \left(\frac{z^4}{720}+\frac{z^2}{135}+\frac{1}{240}\right)\log \frac{M}{m} \\ 

\frac{1}{2} & \left(\frac{4 z^4}{45}+\frac{121 z^2}{315}+\frac{19}{270}\right)\log \frac{M}{m} & \left(\frac{7 z^4}{45}+\frac{37 z^2}{135}+\frac{1}{20}\right)\log \frac{M}{m} \\ 

1 & \left(\frac{289 z^4}{720}+\frac{1459 z^2}{1260}+\frac{29}{270}\right)\log \frac{M}{m} & \left(\frac{247 z^4}{720}+\frac{112 z^2}{135}+\frac{17}{240}\right)\log \frac{M}{m}  \\ \hline
\end{array}
$$
        \caption{Reduced coefficients $\Delta\bA$ and $\Delta\bB$ in $d=8$.}
    \label{tab:Delta_bar_IR_UV_8D}
\end{table}


The values of $\Delta\bA$, $\Delta\bB$ are presented in Tab.\,\ref{tab:Delta_bar_IR_UV_8D}. 
All coefficients feature  a logarithm  corresponding to the effect of the 8d beta function shown in Tabs.\,\ref{tab:beta_zeromass_Alpha},\,\ref{tab:beta_zeromass_Beta}.

From Tab.\,\ref{tab:Delta_bar_IR_UV_8D} 
we have $\bar a>0$ and $\bar c>0$ in all cases. We have $\bar b<0$  for spin 1 for both  $\Delta\bA$ or $\Delta\bB$ and $\bar b>0$ for both spin $\frac{1}{2}$ and $0$ for $\Delta \bB$, while for $\Delta\bA$ we have  $\bar b>0$ for spin $\frac{1}{2}$ and ${\rm sign }(\bar b)={\rm sign }(70\xi-27)$ for spin 0. We are thus in cases \textit{(1a)} and \textit{(1b)}. A bound on  $z$ appears if $\bAUV$ or $\bBUV$ are sufficiently small, 
 the exact condition is given in Tab.\,\ref{tab:bound_condition_D8}.

As an example, we consider the specific case $\bAUV=0$, $\bBUV=0$.
For spin  $\frac{1}{2}$ and 1 we are in case \textit{(1a)} for both $\Delta\bA$ and $\Delta\bB$ hence $z$ is unbounded.
For spin $0$, the $\Delta\bB$ does not constrain $z$, while a $\xi$-dependent bound exists from  $\Delta\bA$. It is shown in Fig.\,\ref{fig:A_D8}. The region in Fig.\,\ref{fig:A_D8} has critical points to the left and bottom. If $z \leq\sqrt{\frac{92+5\sqrt{562}}{63}} \approx 1.83$ we have $\bA_{\rm IR}>0$, $\forall \xi$ and if $\xi \leq \frac{113+\sqrt{9835}}{126} \approx 1.68$ we have $\bA_{\rm IR}>0$, $\forall z$.
Conversely,  when $\xi> 1.68 $ the domain of $z$ is restricted to  two disjoint regions, corresponding to the $-\bar c \leq \bA_{\rm UV} < -\bA^*$ case in \textit{(1b)}. The domain at larger $z$ is WGC-like. 

\begin{table}[h!]
$$
\begin{array}{|c||c||c|c|} \hline
 \shortstack{~ \\ \text{Spin} \\ \vspace{0.05cm}} & \shortstack{\vspace{0.1cm} \\ \text{Condition for} \\ \vspace{0.005cm} \\~$z$~\text{bounded}} & \shortstack{ \vspace{0.1cm} \\ Bound if \\ \vspace{0.005cm} \\ $\bA_{\rm UV} = 0$} &\shortstack{ \vspace{0.1cm} \\ Bound if \\ \vspace{0.005cm} \\ $\bB_{\rm UV} = 0$} \\ \hline \hline 

\multirow{2}{*}{$0$} & \bA_{\rm UV} < \log \frac{M}{m}\left\{\begin{array}{cc}
    - \frac{15 \xi ^2-5 \xi +2}{ 270 } &  \,\text{if } \xi \leq \frac{27}{70}\\
     \frac{882 \xi ^2-1582 \xi +163}{37044} & \,\text{if } \xi > \frac{27}{70}  
     
    \end{array}\right.  & \multirow{2}{*}{\text{Figure \ref{fig:A_D8}}}  & \multirow{2}{*}{\text{Unbounded}}  \\

 & \text{or}~~ \bB_{\rm UV} < -\frac{1}{240}\log \frac{M}{m}\qquad\qquad\qquad\qquad\qquad  & &

\\

\frac{1}{2} & \bA_{\rm UV} <  -\frac{19}{270}\log \frac{M}{m} \text{ or } \bB_{\rm UV} <  -\frac{1}{20}\log \frac{M}{m}&  \text{Unbounded} & \text{Unbounded} \\

\shortstack{\vspace{0.1cm}\\$1$\\\vspace{0.08cm}} & \shortstack{\vspace{0.2cm}\\$\bA_{\rm UV} < -\frac{29}{270}\log \frac{M}{m}$ \text{ or }  $\bB_{\rm UV} < -\frac{17}{240}\log \frac{M}{m}$\\~}  &\text{Unbounded}& \text{Unbounded}  
\\ \hline 
\end{array}
$$
        \caption{Condition  for the existence of IR consistency  bounds on the charge-to-mass ratio $z$  and IR consistency bounds on $z$ if $\bAUV=\bBUV=0$ in $d=8$.}
    \label{tab:bound_condition_D8}
\end{table}

\begin{figure}[t]
    \centering

    \includegraphics[scale=0.6]{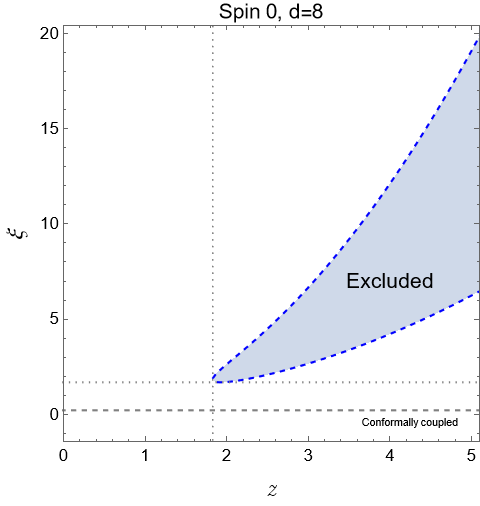}
    
    \caption{Infrared consistency bounds  on  the charged spin $0$ particle in $d=8$.}
    \label{fig:A_D8}
\end{figure}

\FloatBarrier

\subsection*{Case $d=9$}

\begin{table}[h]
    $$\begin{array}{|c||c|c|} \hline 
    \text{Spin} & \Delta\bA  & \Delta\bB  \\ \hline\hline 
    0 & -\frac{7 z^4}{720} - \frac{(37-100 \xi) z^2}{2520} - \frac{1500 \xi ^2-500 \xi +183}{44100}& -\frac{z^4}{720}-\frac{z^2}{210}-\frac{1}{450}  \\ \frac{1}{2} & -\frac{4 z^4}{45}-\frac{83 z^2}{315}-\frac{424}{11025} & -\frac{7 z^4}{45}-\frac{58 z^2}{315}-\frac{2}{75} \\ 
     1 &-\frac{37 z^4}{90}-\frac{247 z^2}{315}-\frac{1397}{22050} & -\frac{31 z^4}{90}-\frac{172 z^2}{315}-\frac{1}{25} \\ \hline
    \end{array}$$
        \caption{Reduced coefficients $\Delta\bA$ and $\Delta\bB$ in $d=9$.}
    \label{tab:Delta_bar_IR_UV_9D}
\end{table}

The values of $\Delta\bA$, $\Delta\bB$ are presented in Tab.\,\ref{tab:Delta_bar_IR_UV_9D}. 
We have $\bar a<0$, $\bar c<0$  for all spins, for both $\Delta\bar \alpha$, $\Delta\bar \beta$ and any $\xi$. We have $\bar b<0$ for spin $\frac{1}{2}$ and 1, and for spin 0 we have $\bar b>0$  for $\Delta\bB$ and ${\rm sign }(\bar b)={\rm sign }(100\xi-37)$ for $\Delta\bA$. We are thus in cases \textit{(2a)} and \textit{(2b)}. The positivity bounds are satisfied if $\bAUV$ and $\bBUV$ are sufficiently large and for a finite range of $z$. That is, $z$ always features an upper bound in $d=9$. The exact condition for having IR consistency for \textit{some} $z$ is given in Tab.\,\ref{tab:bound_condition_D9}.

As a first example, we  consider the specific case $\bAUV=0$, $\bBUV=0$.
For  spin  $\frac{1}{2}$, 1 and for the $\Delta \bB$ coefficient of spin $0$, we are in case \textit{(2b)}  with $-\bA^*=-\bar c>0 $, hence these cases are fully excluded.

A second example is to consider  $\bAUV,\bBUV\gg 1 $. In that case, for any spin, there is an upper bound on $z$. The bounds take the form $z< \bar C( \bAUV^{1/4},\bBUV^{1/4})$ with $\bar C = (-\bar a) ^{-1/4}$, the exact  values are given in Tab.\,\ref{tab:bound_condition_D9}. 
Translating to the non-reduced notation, we have the bound
\be
g|q| m^{\frac{1}{4}}< C \alpha_{\rm UV}^{1/4}\label{eq:upperbound_D9}
\ee
with $C=\bar C K^{1/4}_9$, and similarly for $\beta_{\rm UV}$. 
This upper bound on $g|q|$ is independent on the strength of gravity, depending only on the UV coefficient.   

The bound \eqref{eq:upperbound_D9} may be compared to strong coupling estimates of EFT coefficients. Ignoring all loop factors for simplicity, we have $\alpha_{\rm UV}\sim \Lambda^{-9}$, $m<\Lambda$, with $\Lambda$ the EFT cutoff.   The bound \eqref{eq:upperbound_D9}
is \textit{less} constraining than the strong coupling estimate $g|q| \sim \Lambda^{-5/2} $ except if $m\sim \Lambda$.  Conversely, in a weakly coupled UV completion, with e.g. $\alpha_{\rm UV}\sim \frac{\lambda}{\Lambda^9}$, $\lambda \ll 1$, the bound \eqref{eq:upperbound_D9} can easily be constraining --- while maintaining the assumption $\bAUV\gg 1$, i.e. $\alpha_{\rm UV}\gg \frac{m^5}{M^{14}}$.

\begin{table}[t]
$$
\begin{array}{|c||c||c|c|}  \hline
 \shortstack{~ \\ \text{Spin} \\ \vspace{0.05cm}} & \shortstack{\vspace{0.1cm} \\ \text{Necessary condition} \\ \vspace{0.005cm} \\ \text{for IR consistency}} & \shortstack{ \vspace{0.1cm} \\ Bound if \\ \vspace{0.005cm} \\ $\bA_{\rm UV}=\bB_{\rm UV} = 0$} &\shortstack{ \vspace{0.1cm} \\ Bound if \\ \vspace{0.005cm} \\ $\bA_{\rm UV},\bB_{\rm UV} \gg 1$} \\ \hline \hline

 \shortstack{~ \\ $0$ \\ \vspace{0.8cm}}  & \shortstack{$\bA_{\rm UV} \geq \left\{\begin{array}{cc}
     -\frac{8000 \xi ^2-23000 \xi +1721}{1234800}  &  \,\text{if } \xi \geq \frac{37}{100} \\ 
     \frac{1500 \xi ^2-500 \xi +183}{44100} &  \,\text{if } \xi < \frac{37}{100}
    \end{array}\right.$
    \\
    \text{or}~~$\bB_{\rm UV} \geq \frac{1}{450}$ \qquad\qquad\qquad\qquad\qquad\qquad} & \shortstack{~\\\textit{Excluded}\\\vspace{0.8cm}} & \shortstack{$z\leq 3.18\,\bAUV^{1/4}$ \\ \vspace{0.05cm}\\ $z\leq 5.18\,\bBUV^{1/4}$\\\vspace{0.25cm}}
  
    \\

\shortstack{\vspace{0.15cm}\\$\frac{1}{2}$\\\vspace{0.1cm}} & \shortstack{\vspace{0.15cm}\\$\bA_{\rm UV} \geq  \frac{424}{11025}$ \text{ or } $\bB_{\rm UV} \geq  \frac{2}{75}$\\\vspace{0.1cm}} & 
\shortstack{\vspace{0.05cm}\\\textit{\text{Excluded}}\\\vspace{0.15cm}} &  \shortstack{$z\leq 1.83\,\bAUV^{1/4}$   \\ \vspace{0.05cm}\\  $z\leq 1.59\,\bBUV^{1/4}$}  \\

\shortstack{\vspace{0.1cm}\\$1$\\\vspace{0.1cm}} & \shortstack{\vspace{0.1cm}\\$\bA_{\rm UV} \geq \frac{1397}{22050}$\text{ or } $\bB_{\rm UV} \geq \frac{1}{25}$\\\vspace{0.1cm}}  & \shortstack{\vspace{0.05cm}\\\textit{\text{Excluded}}\\\vspace{0.15cm}}   & 
\shortstack{\vspace{0.3cm}\\$z\leq 1.25\,\bAUV^{1/4}$   \\ \vspace{0.05cm}\\  $z\leq 1.31\,\bBUV^{1/4}$}
\\

\hline 
\end{array}
$$
         \caption{Condition  to have  IR consistency for some $z$ and IR consistency bounds on $z$ if $\bA_{\rm UV}=\bB_{\rm UV} = 0$ or $\bA_{\rm UV},\bB_{\rm UV} \gg 1$  in $d=9$.}
    \label{tab:bound_condition_D9}
\end{table}

\FloatBarrier

\subsection*{Case $d=10$}

\begin{table}[h]
$$
\begin{array}{|c||c|c|}
\hline
 \text{Spin } & \Delta\bA & \Delta\bB \\ \hline \hline

0 &-\left(\frac{7 z^4}{720} + \frac{ (97 - 270 \xi) z^2}{8640}+ \frac{270 \xi ^2-90 \xi +31}{ 11520 }\right)\log \frac{M}{m} & -\left(\frac{z^4}{720}+\frac{z^2}{288}+\frac{1}{720}\right)\log \frac{M}{m} \\ 

\frac{1}{2} & -\left(\frac{8 z^4}{45}+\frac{109 z^2}{270}+\frac{47}{960}\right)\log \frac{M}{m} &  -\left(\frac{14 z^4}{45}+\frac{5 z^2}{18}+\frac{1}{30}\right)\log \frac{M}{m}  \\ 

1 & -\left(\frac{101}{240}+\frac{1721 z^2}{2880}+\frac{499}{11520}\right)\log \frac{M}{m} & -\left(\frac{83 z^4}{240}+\frac{13 z^2}{32}+\frac{19}{720}\right)\log \frac{M}{m}  \\ \hline
\end{array}
$$
        \caption{Reduced coefficients $\Delta\bA$ and $\Delta\bB$ in $d=10$.}
    \label{tab:Delta_bar_IR_UV_10D}
\end{table}

The values of $\Delta\bA$, $\Delta\bB$ are presented in Tab.\,\ref{tab:Delta_bar_IR_UV_10D}. All coefficients feature  a logarithm. The sign pattern is exactly the same as for $d=9$. As a result, similar conclusions follow: IR consistency is satisfied if $\bAUV$ and $\bBUV$ are large enough, see Tab.\,\ref{tab:bound_condition_D10}. Moreover, $z$ is always bounded from above, as exemplified in Tab.\,\ref{tab:bound_condition_D10}. 
Translating to the non-reduced notation, we obtain the upper bound
\be
g|q| m^{\frac{1}{2}}< C \alpha_{\rm UV}^{1/4} \label{eq:upperbound_D10}
\ee
with $C=\bar C K^{1/4}_{10}$ and similarly for $\beta_{\rm UV}$.

\begin{table}[t]
$$
\begin{array}{|c||c||c|c|}  \hline
 \shortstack{~ \\ \text{Spin} \\ \vspace{0.05cm}} & \shortstack{\vspace{0.1cm} \\ \text{Necessary condition} \\ \vspace{0.005cm} \\ \text{for IR consistency}} & \shortstack{ \vspace{0.1cm} \\ Bound if \\ \vspace{0.005cm} \\ $\bA_{\rm UV}=\bB_{\rm UV} = 0$} &\shortstack{ \vspace{0.1cm} \\ Bound if \\ \vspace{0.005cm} \\ $\bA_{\rm UV},\bB_{\rm UV} \gg 1$} \\ \hline \hline

\shortstack{~ \\ $0$ \\ \vspace{0.8cm}} & \shortstack{\small{\text{\(\bAUV \geq \log \frac{M}{m}\)}} \(\left\{\begin{array}{cc}
     -\frac{4860 \xi ^2-29700 \xi +1597}{2903040}\, & {\footnotesize{\hspace{-0.2cm}\text{ if \(\xi > \frac{97}{270}\)}}} \\ 
     \footnotesize{\text{\(\frac{270 \xi ^2-90 \xi +31}{ 11520 }\)}} &  {\footnotesize{\text{\hspace{-0.25cm}\, if \(\xi \leq \frac{97}{270}\)}}}
    \end{array}\right.\)  
    \\
    \text{or}~~$ \bB_{\rm UV} \geq \frac{1}{720}\log \frac{M}{m}$\qquad\qquad\qquad\qquad \\\vspace{0.05cm}} & \shortstack{~\\\textit{Excluded}\\\vspace{1.2cm}} & \shortstack{\vspace{0.1cm}\\$z\leq \left(\frac{720}{7\log\frac{M}{m}}\bAUV \right)^{\frac{1}{4}}$\\$z\leq \left(\frac{720}{\log\frac{M}{m}}\bBUV \right)^{\frac{1}{4}}$ \\\vspace{0.2cm}}

\\

\shortstack{\vspace{0.1cm}\\$\frac{1}{2}$\\\vspace{0.3cm}} & \shortstack{\vspace{0.1cm}\\$\bA_{\rm UV} \geq  \frac{47}{960}\log \frac{M}{m}$ \text{ or } $\bB_{\rm UV} \geq  \frac{1}{30}\log \frac{M}{m}$\\\vspace{0.3cm}}&\shortstack{\vspace{0.1cm}\textit{Excluded}\\\vspace{0.3cm}}&  \shortstack{$z\leq \left(\frac{45}{8\log\frac{M}{m}}\bAUV \right)^{\frac{1}{4}}$ \\ $z\leq \left(\frac{45}{14\log\frac{M}{m}}\bBUV \right)^{\frac{1}{4}}$} \\

\shortstack{\vspace{0.1cm}\\$1$\\\vspace{0.5cm}} & \shortstack{\vspace{0.1cm}\\$\bA_{\rm UV} \geq \frac{499}{11520}\log \frac{M}{m}$~~\text{or}~~ $\bB_{\rm UV} \geq \frac{19}{720}\log \frac{M}{m}$\\\vspace{0.3cm}}   & \shortstack{\vspace{0.1cm}\textit{Excluded}\\\vspace{0.35cm}}   &  \shortstack{\vspace{0.4cm}\\$z\leq \left(\frac{720}{293\log\frac{M}{m}}\bAUV \right)^{\frac{1}{4}}$\\$z\leq \left(\frac{240}{83\log\frac{M}{m}}\bBUV \right)^{\frac{1}{4}}$\\\vspace{-0.35cm}}
 \\

\hline 
\end{array}
$$
        \caption{Condition  to have  IR consistency for some $z$, and IR consistency bounds on $z$ if $\bA_{\rm UV}=\bB_{\rm UV} = 0$ or $\bA_{\rm UV},\bB_{\rm UV} \gg 1$  in $d=10$.}
    \label{tab:bound_condition_D10}
\end{table}

\clearpage

\subsection*{Case $d=11$}

\begin{table}[h]
    $$\begin{array}{|c||c|c|} \hline
    \text{Spin} & \Delta\bA  & \Delta\bB  \\ \hline\hline 
   0 &  \frac{7 z^4}{1080}+\frac{(123-350 \xi) z^2}{20250} + \frac{2940 \xi ^2-980 \xi +323}{255150} & \frac{z^4}{1080}+\frac{11 z^2}{6075}+\frac{1}{1575}  \\ 
   \frac{1}{2} & \frac{16 z^4}{135}+\frac{2216 z^2}{10125}+\frac{2896}{127575} & \frac{28 z^4}{135}+\frac{904 z^2}{6075}+\frac{8}{525} \\ 
   1 &\frac{31 z^4}{108}+\frac{656 z^2}{2025}+\frac{548}{25515} & \frac{25 z^4}{108}+\frac{262 z^2}{1215}+\frac{4}{315} \\ \hline
    \end{array}$$
        \caption{Reduced coefficients $\Delta\bA$ and $\Delta\bB$ in $d=11$. }
    \label{tab:WC_D11}
\end{table}

The values of $\Delta\bA$, $\Delta\bB$ are presented in Tab.\,\ref{tab:WC_D11}. 
We have $\bar a>0$, $\bar c>0$  for all spins, for both $\Delta\bar \alpha$, $\Delta\bar \beta$ and any $\xi$.
We have $\bar b>0$  for spin $\frac{1}{2}$ and 1, and for spin 0 we have $\bar b>0$  for $\Delta\bB$ and ${\rm sign }(\bar b)={\rm sign }(123-350\xi)$ for $\Delta\bA$. We are thus in cases \textit{(1a)} and \textit{(1b)}.
A bound on  $z$ appears if $\bAUV$ or $\bBUV$ are sufficiently small, the exact condition is given in Tab.\,\ref{tab:bound_condition_D11}. 
The sign pattern is similar to $d=7$. 

Consider the specific case $\bAUV=\bBUV=0$. For spin  $\frac{1}{2}$ and 1 we are in case \textit{(1a)} with $\bar\alpha^*>0$, hence $z$ is unbounded.
{For spin $0$, the $\Delta\bB$ does not constrain $z$, while a $\xi$-dependent bound would be expected from  $\Delta\bA$. However, as in $d=3$, for every $\xi$ the coefficient $\bAUV$ must be negative\footnote{Different from $d=3$, 
a curious cancellation of the  $\xi^2$ term  occurs in the condition for $z$ bounded   with $\xi > \frac{123}{350}$. Despite this cancellation, for any $\xi$ in this domain, the $\Delta\bA$ coefficient has to be negative  to produce a bound. 
 } to generate a bound on $z$, hence $z$ is unbounded in this case.}

\begin{table}[h]
$$
\begin{array}{|c||c||c|c|}
\hline
 \shortstack{~ \\ \text{Spin} \\ \vspace{0.05cm}} & \shortstack{\vspace{0.1cm} \\ \text{Condition for} \\ \vspace{0.005cm} \\~$z$~\text{bounded}} & \shortstack{ \vspace{0.1cm} \\ Bound if \\ \vspace{0.005cm} \\ $\bA_{\rm UV} = 0$} &\shortstack{ \vspace{0.1cm} \\ Bound if \\ \vspace{0.005cm} \\ $\bB_{\rm UV} = 0$} \\ \hline \hline
 
\multirow{2}{*}{$0$} & \text{\small{\(\bA_{\rm UV} < \)}}\left\{\begin{array}{cc}
     \text{\small{\(-\frac{2940 \xi ^2-980 \xi +323}{255150}\)}} & \, \text{\hspace{-0.25cm}\footnotesize{\, if \(\xi \leq \frac{123}{350}\)}} \\ 
     \text{\small{\(-\frac{9700 \xi -358}{2278125}\)}} & \, \text{\hspace{-0.2cm}\footnotesize{ if \(\xi> \frac{123}{350}\)}}
    \end{array}\right. &  \multirow{2}{*}{\text{Unbounded}}
     &  \multirow{2}{*}{ \text{Unbounded} } \\ 

 &  \text{or}~~~ \bB_{\rm UV} < -\frac{1}{1575}\qquad\qquad\qquad\qquad &  &   \\

\frac{1}{2} & \bA_{\rm UV} <  -\frac{2896}{127575} ~~~~\text{or}~~~~ \bB_{\rm UV} <  -\frac{8}{525} & \text{Unbounded} &  \text{Unbounded} \\

1 & \bA_{\rm UV} < -\frac{548}{25515} ~~~~\text{or}~~~~ \bB_{\rm UV} <  -\frac{4}{315} & \text{Unbounded} & \text{Unbounded}  \\ \hline
\end{array} $$

         \caption{Condition  for the existence of IR consistency  bounds on the charge-to-mass ratio $z$  and IR consistency bounds on $z$ if $\bAUV=\bBUV=0$ in $d=11$.}
    \label{tab:bound_condition_D11}
\end{table}

\FloatBarrier

\section{Detailed Bounds from Extremal Black Holes}
\label{app:BH}

We present the results from extremal black hole decay for each spacetime dimension. 
The discussion follows  the same structure as  App.\,\ref{app:IR} and does not need to be repeated.  

\subsection{Coefficients}

Translating equation (B.15) of \cite{Kats:2006xp} to our operator basis and  conventions, we obtain the $C_{\rm IR}$ coefficient 
\begin{align} 
C_{\rm IR} &=  \alpha_1 + \frac{\alpha_2}{2} +\frac{(d-4)^2\alpha_3+ \left(2 d^2-11 d+16\right)\alpha_4}{4(d-2)^2} \nn \\  & \quad + \frac{\left(2 d^3-16 d^2+45 d-44\right)\alpha_5}{2(d-3)(d-2)^2} + \frac{(d-4)\alpha_6 + (d-3) (\alpha_7+\alpha_8)}{2(d-2)}\,.
\end{align} \label{eq:CIR}
Using the formalism of section \ref{se:BH}, we have $C_{\rm IR}=C_{\rm UV} +\Delta C$
where the correction $\Delta\C$ produced by charged particles takes the form
\begin{align}
 \Delta \C^{(s)} =     \frac{1}{(4\pi)^{d/2}} \Bigg[    & \frac{g^4 q^4}{m^{8-d}} \Gamma\left(4-\frac{d}{2}\right) a^{(s)}_{\C} \nonumber 
 \\ & 
 + \frac{g^2 q^2}{m^{6-d}M^{d-2}}\Gamma\left(3-\frac{d}{2}\right) b^{(s)}_{\C} + 
\frac{1}{m^{4-d}\M^{2d-4}}\Gamma\left(2-\frac{d}{2}\right) c^{(s)}_{\C} ~~ \Bigg]  \,.  
\end{align}
The coefficients for each spin  are

\begin{subequations}
\label{eq:coefs_a_BlackHole}
\begin{alignat}{5}
    &a^{(0)}_C &&= \frac{7}{1440}\,, \hspace{2cm} &&b^{(0)}_C &&=\frac{ (30 \xi -11)d-120 \xi +38}{720 (d-2)}\,,\\ 
    &a^{(1/2)}_C &&= \frac{n}{360}\,, \hspace{2cm} &&b^{(1/2)}_C &&= -\frac{n}{1440} \frac{19 d - 52}{(d-2)}  \,,\\
    &a^{(1)}_C &&= \frac{7 d+233}{1440}\,, \hspace{2cm} &&b^{(1)}_C &&= - \frac{11 d^2+401 d-1162}{720(d-2)}\,.
\end{alignat}
\end{subequations}

\begin{subequations}
\label{eq:coefs_c_BlackHole}

\begin{alignat}{4}
c^{(0)}_C  =     \frac{1}{1440} \Bigg[    & \frac{2 (3-d) \left(2 d^2-11 d+16\right)}{(d-3)(d-2)^2} 
 \\ & \nonumber +\frac{4 \left(2 d^3-16 d^2+45 d-44\right)}{(d-3)(d-2)^2} + \frac{5(1-6 \xi )^2 (d-4)^2}{(d-2)^2}~~ \Bigg]\,,\\
c^{(1/2)}_C = \frac{n}{11520} & \frac{39 d^3-305 d^2+822 d-760}{(d-3)(d-2)^2}\,, \\ 
 c^{(1)}_C = \frac{1}{1440} \hspace{0.2cm}& \hspace{-0.2cm} \frac{9 d^4+86 d^3-1073 d^2+3118 d-2800}{(d-3)(d-2)^2}\,.
\end{alignat}
\end{subequations}

\subsection{Bounds}

We apply the positivity analysis  performed on 
the $\alpha$ and $\beta$ coefficients in section \ref{se:finite_corrections},
but now on the $\C$ coefficient of the black hole charge-to-mass ratio. We present the tables with the results from $d=4$ to $d=11$, in the same format as those presented in section \ref{se:finite_corrections} for the analysis of amplitudes. 

Finally we present the exclusion region in the $z\xi$-plane for $d=5$ to exemplify the similarity with the results from amplitudes consistency, see Fig.\,\ref{fig:A_D5}. We also show the figures for $d=9$ and $d=10$, which were not presented before because the $\bAUV=0$ condition was considered together with the $\bBUV=0$, which is stronger.

\begin{table}[h]
$$
\begin{array}{|c||c|c|}
\hline
 \text{Spin } & \Delta\bC\,\, \\ \hline \hline

0 &  \frac{7 z^4}{1440}-\frac{z^2}{240} + \frac{1}{120}\log \frac{M}{m} \\ 

\frac{1}{2} & \frac{z^4}{90}-\frac{z^2}{30} + \frac{1}{40}\log \frac{M}{m} \\ 

1 &  \frac{29 z^4}{160}-\frac{103 z^2}{240} + \frac{13}{120}\log \frac{M}{m} \\ \hline
\end{array}
$$
        \caption{
        Reduced coefficient $\Delta\bC$ in $d=4$.}
    \label{tab:Delta_bar_IR_UV_4D_BlackHole}
\end{table}

\begin{table}[h]
$$
\begin{array}{|c||c|c|c|}
\hline
 \text{Spin } & \Delta\bC & \text{Bound if } \bC_{\rm UV} = 0 \,\,\\ \hline \hline

0 &  \frac{7 z^4}{2880}-\frac{(17 - 30 \xi )z^2}{2160} - \frac{12 \xi^2 -4\xi +3}{432} & \text{Figure } \ref{fig:BH}  \\ 

\frac{1}{2} & \frac{z^4}{180}-\frac{43 z^2}{1080} - \frac{5}{216} & z \geq 2.78 \\ 

1 &  \frac{67 z^4}{720}-\frac{559 z^2}{1080}-\frac{13}{72} & z \geq 2.43  \\ \hline
\end{array}
$$
        \caption{
        Reduced coefficient $\Delta\bC$ and extremal black hole decay bounds on $z$ if $\bC_{\rm UV}=0$ in $d=5$.}
    \label{tab:Delta_bar_IR_UV_5D_BlackHole}
\end{table}

\begin{table}[h]
$$
\begin{array}{|c||c|c|c|}
\hline
 \text{Spin } & \Delta\bC & \text{Bound if }\bC_{\rm UV} = 0 \,\,\\ \hline \hline

0 & \frac{7 z^4}{1440} - \left(\frac{(7 - 15 \xi) z^2}{360} + \frac{135 \xi ^2-45 \xi +16 }{ 2160 }\right)\log \frac{M}{m} & \text{Similar to figure } \ref{fig:A_D6}  \\ 

\frac{1}{2} &   \frac{
 z^4}{45}  - \left( \frac{31 z^2}{180}+\frac{101}{2160}\right)\log \frac{M}{m} & z\geq 2.78 \operatorname{log}\frac{\M}{m} \\ 

1 &  \frac{55 z^4}{288} - \left(\frac{41 z^2}{36} + \frac{47}{216}\right)\log \frac{M}{m} &  z\geq 2.44 \operatorname{log}\frac{\M}{m}  \\ \hline
\end{array}
$$
        \caption{Reduced coefficients $\Delta\bC$  and extremal black hole decay bounds on $z$ if $\bC_{\rm UV}=0$ in $d=6$.}
    \label{tab:Delta_bar_IR_UV_6D_BlackHole}
\end{table}

\begin{table}[h]
    $$\begin{array}{|c||c|c|} \hline 
    \text{Spin} & \Delta\bC & \text{Bound if }\bC_{\rm UV} = 0 \,\,\\ \hline \hline 
    0 &  \frac{7 z^4}{1440}+\frac{(13 - 30\xi) z^2}{600} + \frac{45 \xi ^2-15 \xi +4}{750} &  \text{Similar to figure } \ref{fig:A_D7} \\  \frac{1}{2} & \frac{z^4}{45}+ \frac{9 z^2}{50}+\frac{571}{18000} & \text{Unbounded} \\  1 &\frac{47 z^4}{240}+\frac{91 z^2}{75}+\frac{1463}{9000} & \text{Unbounded}\\ \hline 
    \end{array}$$
        \caption{
Reduced coefficients $\Delta\bC$ and extremal black hole decay bounds on $z$ if $\bC_{\rm UV}=0$ in $d=7$.        
        }
    \label{tab:WC_D7_BlackHole}
\end{table}


\begin{table}[h]
    $$\begin{array}{|c||c|c|c|} \hline 
    \text{Spin} & \Delta\bC & \text{Bound if }\bC_{\rm UV} = 0 \,\, \\ \hline \hline 
    0 & \left(\frac{7z^4}{720} + \frac{(5 -12 \xi)z^2}{216} + \frac{300 \xi ^2-100 \xi +23}{ 5400 }\right)\log \frac{M}{m}  & \text{Similar to figure } \ref{fig:A_D8}  \\  \frac{1}{2} & \left( \frac{4z^4}{45} +\frac{10 z^2}{27}+\frac{29}{600}\right)\log \frac{M}{m} & \text{Unbounded} \\  1 &\left(\frac{289 z^4}{720} + \frac{275 z^2}{216}+\frac{179}{1350}\right)\log \frac{M}{m} & \text{Unbounded} \\ \hline 
    \end{array}$$
        \caption{
Reduced coefficients $\Delta\bC$ and  extremal black hole decay bounds on $z$ if $\bC_{\rm UV} = 0$ in $d=8$.        
        }
    \label{tab:WC_D8_BlackHole}
\end{table}

\begin{table}[h]
    $$\begin{array}{|c||c|c|c|} \hline 
    \text{Spin} & \Delta\bC   &  \text{Bound if }\bC_{\rm UV} = 0 \,\, & \text{Bound if }\bC_{\rm UV}\gg 1 \,\, \\ \hline\hline 
    0 & -\frac{7 z^4}{720} - \frac{(61-150 \xi) z^2}{3780} - \frac{13500 \xi ^2-4500 \xi +947}{396900} & \text{Figure }\ref{fig:BH} & z\leq 3.18 \, \bC_{\rm UV}^{1/4}\\ 
    \frac{1}{2} & -\frac{4 z^4}{45}-\frac{34 z^2}{135}-\frac{2591}{99225} & \textit{Excluded} & z\leq 1.83  \, \bC_{\rm UV}^{1/4} \\ 
     1 &-\frac{37 z^4}{90}-\frac{1669 z^2}{1890}-\frac{15023}{198450} & \textit{Excluded} & z\leq 1.25  \, \bC_{\rm UV}^{1/4}\\ \hline
    \end{array}$$
        \caption{Reduced coefficients $\Delta\bC$ and extremal black hole decay bounds on $z$ if $\bC_{\rm UV}=0$ and $\bC_{\rm UV}\gg 1$ in $d=9$.}
    \label{tab:Delta_bar_IR_UV_9D_BlackHole}
\end{table}

\begin{table}[h]
$$
\begin{array}{|c||c|c|c|}
\hline
 \text{Spin } & \Delta\bC &  \text{Bound if }\bC_{\rm UV} = 0  \,\, & \text{Bound if }\bC_{\rm UV}\gg 1 \,\, \\ \hline \hline

0 &-\left(\frac{7 z^4}{720} + \frac{ (2 - 5 \xi) z^2}{160} + \frac{378 \xi ^2-126 \xi +25}{ 16128 }\right)\log \frac{M}{m} & \text{Figure  }\ref{fig:BH}  & z\leq \left(\frac{720}{7\log\frac{M}{m}}\bC_{\rm UV} \right)^{\frac{1}{4}}\\ 

\frac{1}{2} & -\left(\frac{8 z^4}{45}+\frac{23 z^2}{60}+\frac{19}{576}\right)\log \frac{M}{m} & \textit{Excluded}  & z\leq \left(\frac{45}{8\log\frac{M}{m}}\bC_{\rm UV} \right)^{\frac{1}{4}}\\ 

1 & -\left(\frac{101 z^4}{240} + \frac{329 z^2}{480}+\frac{809}{16128}\right)\log \frac{M}{m} & \textit{Excluded}  & z\leq \left(\frac{240}{101\log\frac{M}{m}}\bC_{\rm UV} \right)^{\frac{1}{4}} \vspace{0.05cm} \\ \hline
\end{array}
$$
        \caption{Reduced coefficients $\Delta\bC$ and extremal black hole decay bounds on $z$ if $\bC_{\rm UV}=0$ in $d=10$.}
    \label{tab:Delta_bar_IR_UV_10D_BlackHole}
\end{table}

\begin{table}[t]
    $$\begin{array}{|c||c|c|} \hline
    \text{Spin} & \Delta\bC & \text{Bound if }\bC_{\rm UV} = 0  \,\,\\ \hline\hline 
   0 &  \frac{7 z^4}{1080}+\frac{(83-210 \xi) z^2}{12150} + \frac{5880 \xi ^2-1960 \xi +373}{510300} & \text{Unbounded} \\ 
   \frac{1}{2} & \frac{16 z^4}{135}+\frac{1256 z^2}{6075}+\frac{3881}{255150}  & \text{Unbounded}\\ 
   1 & \frac{31 z^4}{108} +\frac{458 z^2}{1215}+\frac{493}{20412} & \text{Unbounded} \\ \hline
    \end{array}$$
        \caption{Reduced coefficients $\Delta\bC$ and extremal black hole decay bounds on $z$ if $\bC_{\rm UV}=0$ in $d=11$. }
    \label{tab:WC_D11_BlackHole}
\end{table}

\begin{figure}[h]
    \centering

    \includegraphics[scale=0.58]{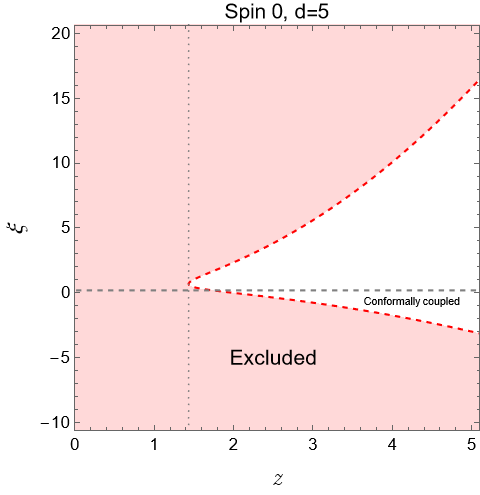}
    
    \includegraphics[scale=0.58]{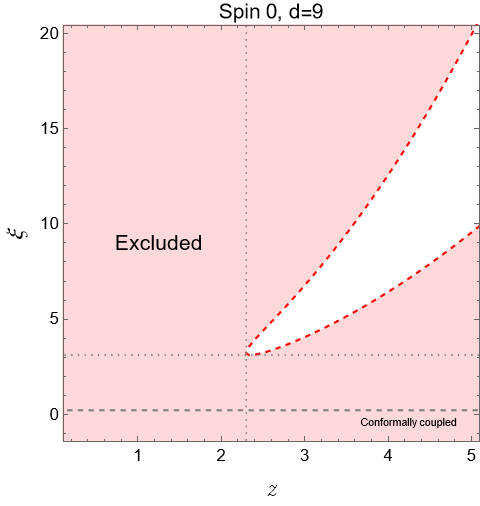}
    \includegraphics[scale=0.58]{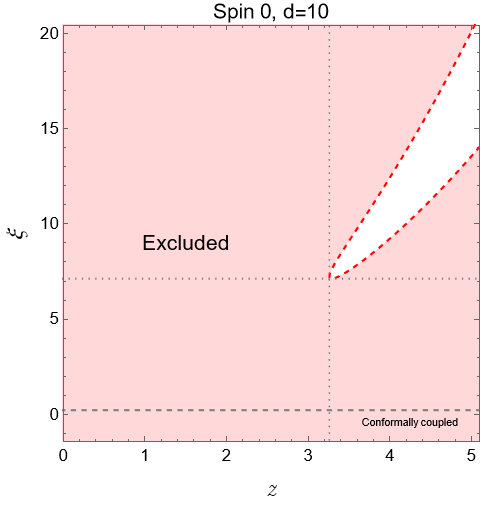}

    \caption{Extremal black hole decay bounds  on the charged spin $0$ particle if $\bC_{\rm UV} = 0$ in $d=5, 9,10$.}
    \label{fig:BH}
\end{figure}

\FloatBarrier

\section{The Heat Kernel Coefficients}
\label{app:HK}

The general expressions for the coefficients appearing in \eqref{eq:Gam1_b} and \eqref{eq:Leff_oneloop}
are \cite{Vassilevich:2003xt,Gilkey_original}
\begin{align}
b_0&=I \nn \\
b_2&=\frac{1 }{6}RI-X \nn \\ \nn 
b_4&=\frac{1}{360}\Big(
12 \square R+5 R^2-2R_{\mu\nu}R^{\mu\nu}+2R_{\mu\nu\rho\sigma}R^{\mu\nu\rho\sigma} \Big)I \label{eq:b4}
\\ 
 & \quad - \frac{1}{6} \square X - \frac{1}{6} R X + \frac{1}{2} X^2 + \frac{1}{12} \Omega_{\mu\nu}\Omega^{\mu\nu}
\end{align}
\begin{align}
b_6 &=  \frac{1}{360}\bigg(
8 D_{\rho}\Omega_{\mu\nu}D^{\rho}\Omega^{\mu\nu}
+2 D^{\mu}\Omega_{\mu\nu} D_{\rho}\Omega^{\rho\nu}
+12 \Omega_{\mu\nu}\square \Omega^{\mu\nu}
-12 \Omega_{\mu\nu}\Omega^{\nu\rho}\Omega^{~~\mu}_{\rho}  \nn  \\ \nn 
& \quad +6 R_{\mu\nu\rho\sigma}\Omega^{\mu\nu}\Omega^{\rho\sigma}
-4 R_{\mu}^{~\nu}\Omega^{\mu\rho}\Omega_{\nu\rho} + 5 R\Omega_{\mu\nu}\Omega^{\mu\nu} \\ &\quad \nn
- 6 \square^2 X +60 X\square X+ 30 D_\mu X D^\mu X - 60 X^3
\\ &\quad \nn
- 30 X \Omega_{\mu\nu}\Omega^{\mu\nu} - 10 R \square X - 4 R_{\mu\nu} D^\nu  D^\mu X - 12 D_\mu R D^\mu X + 30 XX R \\ &\quad \nn 
- 12 X \square R  - 5 X R^2 + 2 X R_{\mu\nu}R^{\mu\nu} - 2 X R_{\mu\nu\rho\sigma}R^{\mu\nu\rho\sigma}
\bigg)   
\\ &\quad \nn
+\frac{1}{7!}\bigg(
18 \square^2  R + 17 D_\mu R D^\mu R  - 2D_\rho R_{\mu\nu} D^\rho R^{\mu\nu}
- 4 D_\rho R_{\mu\nu} D^\mu R^{\rho\nu}  \\ &\quad \nn
+9 D_\rho R_{\mu\nu\sigma\lambda} D^\rho R^{\mu\nu\sigma\lambda}  + 28 R \square R - 8 R_{\mu\nu} \square R^{\mu\nu} 
 \\ &\quad \nn
+ 24 R_{\mu\nu} D_\rho D^\nu  R^{\mu\rho} + 12 R_{\mu\nu\sigma\lambda} \square R^{\mu\nu\sigma\lambda} +35/9 R^3 
\\ &\quad \nn
-14/3 R R_{\mu\nu} R^{\mu\nu} 
+ 14/3 R R_{\mu\nu\rho\sigma}R^{\mu\nu\rho\sigma} -208/9 R_{\mu\nu} R^{\mu\rho} R_{~~\rho}^{\nu} 
\\ &\quad \nn
+ 64/3 R_{\mu\nu}R_{\rho\sigma} R^{\mu\rho\nu\sigma}  
-16/3 R^{\mu}_{~\nu } R_{\mu\rho\sigma\lambda}R^{\nu\rho\sigma\lambda}
\\ &\quad 
+ 44/9 R^{\mu\nu}_{~~\alpha\beta} R_{\mu\nu\rho\sigma} R^{\rho\sigma \alpha\beta }   + 80/9 R_{\mu~~\rho~~}^{~~\nu~~\sigma} R^{\mu\alpha\rho \beta } R_{\nu\alpha \sigma \beta} 
\bigg) I 
\label{eq:b6full}
\end{align}
with $I$ the identity matrix for internal indexes.

\bibliographystyle{JHEP}
\bibliography{biblio}

\providecommand{\href}[2]{#2}\begingroup\raggedright\begin{thebibliography}{10}

\bibitem{Pham:1985cr}
T.~N. Pham and T.~N. Truong, {\it {Evaluation of the Derivative Quartic Terms
  of the Meson Chiral Lagrangian From Forward Dispersion Relation}},  {\em
  Phys. Rev. D} {\bf 31} (1985) 3027.

\bibitem{Ananthanarayan:1994hf}
B.~Ananthanarayan, D.~Toublan, and G.~Wanders, {\it {Consistency of the chiral
  pion pion scattering amplitudes with axiomatic constraints}},  {\em Phys.
  Rev. D} {\bf 51} (1995) 1093--1100,
  [\href{http://arxiv.org/abs/hep-ph/9410302}{{\tt hep-ph/9410302}}].

\bibitem{Adams:2006sv}
A.~Adams, N.~Arkani-Hamed, S.~Dubovsky, A.~Nicolis, and R.~Rattazzi, {\it
  {Causality, analyticity and an IR obstruction to UV completion}},  {\em JHEP}
  {\bf 10} (2006) 014, [\href{http://arxiv.org/abs/hep-th/0602178}{{\tt
  hep-th/0602178}}].

\bibitem{Arkani-Hamed:2020blm}
N.~Arkani-Hamed, T.-C. Huang, and Y.-t. Huang, {\it {The EFT-Hedron}},  {\em
  JHEP} {\bf 05} (2021) 259, [\href{http://arxiv.org/abs/2012.15849}{{\tt
  arXiv:2012.15849}}].

\bibitem{Alberte:2020jsk}
L.~Alberte, C.~de~Rham, S.~Jaitly, and A.~J. Tolley, {\it {Positivity Bounds
  and the Massless Spin-2 Pole}},  {\em Phys. Rev. D} {\bf 102} (2020), no.~12
  125023, [\href{http://arxiv.org/abs/2007.12667}{{\tt arXiv:2007.12667}}].

\bibitem{Alberte:2020bdz}
L.~Alberte, C.~de~Rham, S.~Jaitly, and A.~J. Tolley, {\it {QED positivity
  bounds}},  {\em Phys. Rev. D} {\bf 103} (2021), no.~12 125020,
  [\href{http://arxiv.org/abs/2012.05798}{{\tt arXiv:2012.05798}}].

\bibitem{Henriksson:2021ymi}
J.~Henriksson, B.~McPeak, F.~Russo, and A.~Vichi, {\it {Rigorous bounds on
  light-by-light scattering}},  {\em JHEP} {\bf 06} (2022) 158,
  [\href{http://arxiv.org/abs/2107.13009}{{\tt arXiv:2107.13009}}].

\bibitem{Henriksson:2022oeu}
J.~Henriksson, B.~McPeak, F.~Russo, and A.~Vichi, {\it {Bounding violations of
  the weak gravity conjecture}},  {\em JHEP} {\bf 08} (2022) 184,
  [\href{http://arxiv.org/abs/2203.08164}{{\tt arXiv:2203.08164}}].

\bibitem{Bellazzini:2020cot}
B.~Bellazzini, J.~Elias~Mir\'o, R.~Rattazzi, M.~Riembau, and F.~Riva, {\it
  {Positive moments for scattering amplitudes}},  {\em Phys. Rev. D} {\bf 104}
  (2021), no.~3 036006, [\href{http://arxiv.org/abs/2011.00037}{{\tt
  arXiv:2011.00037}}].

\bibitem{Caron-Huot:2021enk}
S.~Caron-Huot, D.~Mazac, L.~Rastelli, and D.~Simmons-Duffin, {\it {AdS Bulk
  Locality from Sharp CFT Bounds}},
  \href{http://arxiv.org/abs/2106.10274}{{\tt arXiv:2106.10274}}.

\bibitem{Bellazzini:2021oaj}
B.~Bellazzini, M.~Riembau, and F.~Riva, {\it {IR side of positivity bounds}},
  {\em Phys. Rev. D} {\bf 106} (2022), no.~10 105008,
  [\href{http://arxiv.org/abs/2112.12561}{{\tt arXiv:2112.12561}}].

\bibitem{Caron-Huot:2020cmc}
S.~Caron-Huot and V.~Van~Duong, {\it {Extremal Effective Field Theories}},
  {\em JHEP} {\bf 05} (2021) 280, [\href{http://arxiv.org/abs/2011.02957}{{\tt
  arXiv:2011.02957}}].

\bibitem{Caron-Huot:2021rmr}
S.~Caron-Huot, D.~Mazac, L.~Rastelli, and D.~Simmons-Duffin, {\it {Sharp
  boundaries for the swampland}},  {\em JHEP} {\bf 07} (2021) 110,
  [\href{http://arxiv.org/abs/2102.08951}{{\tt arXiv:2102.08951}}].

\bibitem{Davighi:2021osh}
J.~Davighi, S.~Melville, and T.~You, {\it {Natural selection rules: new
  positivity bounds for massive spinning particles}},  {\em JHEP} {\bf 02}
  (2022) 167, [\href{http://arxiv.org/abs/2108.06334}{{\tt arXiv:2108.06334}}].

\bibitem{deRham:2021fpu}
C.~de~Rham, S.~Melville, and J.~Noller, {\it {Positivity bounds on dark energy:
  when matter matters}},  {\em JCAP} {\bf 08} (2021) 018,
  [\href{http://arxiv.org/abs/2103.06855}{{\tt arXiv:2103.06855}}].

\bibitem{Caron-Huot:2022ugt}
S.~Caron-Huot, Y.-Z. Li, J.~Parra-Martinez, and D.~Simmons-Duffin, {\it
  {Causality constraints on corrections to Einstein gravity}},  {\em JHEP} {\bf
  05} (2023) 122, [\href{http://arxiv.org/abs/2201.06602}{{\tt
  arXiv:2201.06602}}].

\bibitem{Tolley:2020gtv}
A.~J. Tolley, Z.-Y. Wang, and S.-Y. Zhou, {\it {New positivity bounds from full
  crossing symmetry}},  {\em JHEP} {\bf 05} (2021) 255,
  [\href{http://arxiv.org/abs/2011.02400}{{\tt arXiv:2011.02400}}].

\bibitem{deRham:2021bll}
C.~de~Rham, A.~J. Tolley, and J.~Zhang, {\it {Causality Constraints on
  Gravitational Effective Field Theories}},  {\em Phys. Rev. Lett.} {\bf 128}
  (2022), no.~13 131102, [\href{http://arxiv.org/abs/2112.05054}{{\tt
  arXiv:2112.05054}}].

\bibitem{deRham:2022hpx}
C.~de~Rham, S.~Kundu, M.~Reece, A.~J. Tolley, and S.-Y. Zhou, {\it {Snowmass
  White Paper: UV Constraints on IR Physics}},  in {\em {Snowmass 2021}}, 3,
  2022.
\newblock \href{http://arxiv.org/abs/2203.06805}{{\tt arXiv:2203.06805}}.

\bibitem{Chiang:2022ltp}
L.-Y. Chiang, Y.-t. Huang, L.~Rodina, and H.-C. Weng, {\it {De-projecting the
  EFThedron}},  \href{http://arxiv.org/abs/2204.07140}{{\tt arXiv:2204.07140}}.

\bibitem{Caron-Huot:2022jli}
S.~Caron-Huot, Y.-Z. Li, J.~Parra-Martinez, and D.~Simmons-Duffin, {\it
  {Graviton partial waves and causality in higher dimensions}},  {\em Phys.
  Rev. D} {\bf 108} (2023), no.~2 026007,
  [\href{http://arxiv.org/abs/2205.01495}{{\tt arXiv:2205.01495}}].

\bibitem{Haring:2022sdp}
K.~H\"aring, A.~Hebbar, D.~Karateev, M.~Meineri, and J.~a. Penedones, {\it
  {Bounds on photon scattering}},  \href{http://arxiv.org/abs/2211.05795}{{\tt
  arXiv:2211.05795}}.

\bibitem{Hamada:2023cyt}
Y.~Hamada, R.~Kuramochi, G.~J. Loges, and S.~Nakajima, {\it {On (Scalar QED)
  Gravitational Positivity Bounds}},
  \href{http://arxiv.org/abs/2301.01999}{{\tt arXiv:2301.01999}}.

\bibitem{Bellazzini:2023nqj}
B.~Bellazzini, G.~Isabella, S.~Ricossa, and F.~Riva, {\it {Massive gravity is
  not positive}},  {\em Phys. Rev. D} {\bf 109} (2024), no.~2 024051,
  [\href{http://arxiv.org/abs/2304.02550}{{\tt arXiv:2304.02550}}].

\bibitem{Eichhorn:2024wba}
A.~Eichhorn, A.~O. Pedersen, and M.~Schiffer, {\it {Application of positivity
  bounds in asymptotically safe gravity}},
  \href{http://arxiv.org/abs/2405.08862}{{\tt arXiv:2405.08862}}.

\bibitem{Knorr:2024yiu}
B.~Knorr and A.~Platania, {\it {Unearthing the intersections: positivity
  bounds, weak gravity conjecture, and asymptotic safety landscapes from
  photon-graviton flows}},  \href{http://arxiv.org/abs/2405.08860}{{\tt
  arXiv:2405.08860}}.

\bibitem{Kats:2006xp}
Y.~Kats, L.~Motl, and M.~Padi, {\it {Higher-order corrections to mass-charge
  relation of extremal black holes}},  {\em JHEP} {\bf 12} (2007) 068,
  [\href{http://arxiv.org/abs/hep-th/0606100}{{\tt hep-th/0606100}}].

\bibitem{Cheung:2018cwt}
C.~Cheung, J.~Liu, and G.~N. Remmen, {\it {Proof of the Weak Gravity Conjecture
  from Black Hole Entropy}},  {\em JHEP} {\bf 10} (2018) 004,
  [\href{http://arxiv.org/abs/1801.08546}{{\tt arXiv:1801.08546}}].

\bibitem{Loges:2019jzs}
G.~J. Loges, T.~Noumi, and G.~Shiu, {\it {Thermodynamics of 4D Dilatonic Black
  Holes and the Weak Gravity Conjecture}},  {\em Phys. Rev. D} {\bf 102}
  (2020), no.~4 046010, [\href{http://arxiv.org/abs/1909.01352}{{\tt
  arXiv:1909.01352}}].

\bibitem{Goon:2019faz}
G.~Goon and R.~Penco, {\it {Universal Relation between Corrections to Entropy
  and Extremality}},  {\em Phys. Rev. Lett.} {\bf 124} (2020), no.~10 101103,
  [\href{http://arxiv.org/abs/1909.05254}{{\tt arXiv:1909.05254}}].

\bibitem{Jones:2019nev}
C.~R.~T. Jones and B.~McPeak, {\it {The Black Hole Weak Gravity Conjecture with
  Multiple Charges}},  {\em JHEP} {\bf 06} (2020) 140,
  [\href{http://arxiv.org/abs/1908.10452}{{\tt arXiv:1908.10452}}].

\bibitem{Loges:2020trf}
G.~J. Loges, T.~Noumi, and G.~Shiu, {\it {Duality and Supersymmetry Constraints
  on the Weak Gravity Conjecture}},  {\em JHEP} {\bf 11} (2020) 008,
  [\href{http://arxiv.org/abs/2006.06696}{{\tt arXiv:2006.06696}}].

\bibitem{Arkani-Hamed:2021ajd}
N.~Arkani-Hamed, Y.-t. Huang, J.-Y. Liu, and G.~N. Remmen, {\it {Causality,
  unitarity, and the weak gravity conjecture}},  {\em JHEP} {\bf 03} (2022)
  083, [\href{http://arxiv.org/abs/2109.13937}{{\tt arXiv:2109.13937}}].

\bibitem{Cao:2022iqh}
Q.-H. Cao and D.~Ueda, {\it {Entropy Constraint on Effective Field Theory}},
  \href{http://arxiv.org/abs/2201.00931}{{\tt arXiv:2201.00931}}.

\bibitem{DeLuca:2022tkm}
V.~De~Luca, J.~Khoury, and S.~S.~C. Wong, {\it {Implications of the Weak
  Gravity Conjecture for Tidal Love Numbers of Black Holes}},
  \href{http://arxiv.org/abs/2211.14325}{{\tt arXiv:2211.14325}}.

\bibitem{Arkani-Hamed:2006emk}
N.~Arkani-Hamed, L.~Motl, A.~Nicolis, and C.~Vafa, {\it {The String landscape,
  black holes and gravity as the weakest force}},  {\em JHEP} {\bf 06} (2007)
  060, [\href{http://arxiv.org/abs/hep-th/0601001}{{\tt hep-th/0601001}}].

\bibitem{vanBeest:2021lhn}
M.~van Beest, J.~Calder\'on-Infante, D.~Mirfendereski, and I.~Valenzuela, {\it
  {Lectures on the Swampland Program in String Compactifications}},  {\em Phys.
  Rept.} {\bf 989} (2022) 1--50, [\href{http://arxiv.org/abs/2102.01111}{{\tt
  arXiv:2102.01111}}].

\bibitem{Grana:2021zvf}
M.~Gra\~na and A.~Herr\'aez, {\it {The Swampland Conjectures: A Bridge from
  Quantum Gravity to Particle Physics}},  {\em Universe} {\bf 7} (2021), no.~8
  273, [\href{http://arxiv.org/abs/2107.00087}{{\tt arXiv:2107.00087}}].

\bibitem{Agmon:2022thq}
N.~B. Agmon, A.~Bedroya, M.~J. Kang, and C.~Vafa, {\it {Lectures on the string
  landscape and the Swampland}},  \href{http://arxiv.org/abs/2212.06187}{{\tt
  arXiv:2212.06187}}.

\bibitem{Cheung:2014ega}
C.~Cheung and G.~N. Remmen, {\it {Infrared Consistency and the Weak Gravity
  Conjecture}},  {\em JHEP} {\bf 12} (2014) 087,
  [\href{http://arxiv.org/abs/1407.7865}{{\tt arXiv:1407.7865}}].

\bibitem{Bellazzini:2015cra}
B.~Bellazzini, C.~Cheung, and G.~N. Remmen, {\it {Quantum Gravity Constraints
  from Unitarity and Analyticity}},  {\em Phys. Rev. D} {\bf 93} (2016), no.~6
  064076, [\href{http://arxiv.org/abs/1509.00851}{{\tt arXiv:1509.00851}}].

\bibitem{Cheung:2016wjt}
C.~Cheung and G.~N. Remmen, {\it {Positivity of Curvature-Squared Corrections
  in Gravity}},  {\em Phys. Rev. Lett.} {\bf 118} (2017), no.~5 051601,
  [\href{http://arxiv.org/abs/1608.02942}{{\tt arXiv:1608.02942}}].

\bibitem{Hamada:2018dde}
Y.~Hamada, T.~Noumi, and G.~Shiu, {\it {Weak Gravity Conjecture from Unitarity
  and Causality}},  {\em Phys. Rev. Lett.} {\bf 123} (2019), no.~5 051601,
  [\href{http://arxiv.org/abs/1810.03637}{{\tt arXiv:1810.03637}}].

\bibitem{Bellazzini:2019xts}
B.~Bellazzini, M.~Lewandowski, and J.~Serra, {\it {Positivity of Amplitudes,
  Weak Gravity Conjecture, and Modified Gravity}},  {\em Phys. Rev. Lett.} {\bf
  123} (2019), no.~25 251103, [\href{http://arxiv.org/abs/1902.03250}{{\tt
  arXiv:1902.03250}}].

\bibitem{Chen:2019qvr}
W.-M. Chen, Y.-T. Huang, T.~Noumi, and C.~Wen, {\it {Unitarity bounds on
  charged/neutral state mass ratios}},  {\em Phys. Rev. D} {\bf 100} (2019),
  no.~2 025016, [\href{http://arxiv.org/abs/1901.11480}{{\tt
  arXiv:1901.11480}}].

\bibitem{Bern:2021ppb}
Z.~Bern, D.~Kosmopoulos, and A.~Zhiboedov, {\it {Gravitational effective field
  theory islands, low-spin dominance, and the four-graviton amplitude}},  {\em
  J. Phys. A} {\bf 54} (2021), no.~34 344002,
  [\href{http://arxiv.org/abs/2103.12728}{{\tt arXiv:2103.12728}}].

\bibitem{Gilkey_original}
P.~B. Gilkey, {\it {The spectral geometry of a Riemannian manifold}},  {\em
  Journal of Differential Geometry} {\bf 10} (1975), no.~4 601 -- 618.

\bibitem{Vassilevich:2003xt}
D.~V. Vassilevich, {\it {Heat kernel expansion: User's manual}},  {\em Phys.
  Rept.} {\bf 388} (2003) 279--360,
  [\href{http://arxiv.org/abs/hep-th/0306138}{{\tt hep-th/0306138}}].

\bibitem{vandeVen:1984zk}
A.~E.~M. van~de Ven, {\it {Explicit Counter Action Algorithms in Higher
  Dimensions}},  {\em Nucl. Phys. B} {\bf 250} (1985) 593--617.

\bibitem{Fradkin:1983jc}
E.~s. Fradkin and A.~a. Tseytlin, {\it {QUANTIZATION AND DIMENSIONAL REDUCTION:
  ONE LOOP RESULTS FOR SUPERYANG-MILLS AND SUPERGRAVITIES IN D \ensuremath{>}=
  4}},  {\em Phys. Lett. B} {\bf 123} (1983) 231--236.

\bibitem{Metsaev:1987ju}
R.~R. Metsaev and A.~A. Tseytlin, {\it {ON LOOP CORRECTIONS TO STRING THEORY
  EFFECTIVE ACTIONS}},  {\em Nucl. Phys. B} {\bf 298} (1988) 109--132.

\bibitem{Bastianelli:2008cu}
F.~Bastianelli, J.~M. Davila, and C.~Schubert, {\it {Gravitational corrections
  to the Euler-Heisenberg Lagrangian}},  {\em JHEP} {\bf 03} (2009) 086,
  [\href{http://arxiv.org/abs/0812.4849}{{\tt arXiv:0812.4849}}].

\bibitem{Ritz:1995nt}
A.~Ritz and R.~Delbourgo, {\it {The Low-energy effective Lagrangian for photon
  interactions in any dimension}},  {\em Int. J. Mod. Phys. A} {\bf 11} (1996)
  253--270, [\href{http://arxiv.org/abs/hep-th/9503160}{{\tt hep-th/9503160}}].

\bibitem{Hoover:2005uf}
D.~Hoover and C.~P. Burgess, {\it {Ultraviolet sensitivity in higher
  dimensions}},  {\em JHEP} {\bf 01} (2006) 058,
  [\href{http://arxiv.org/abs/hep-th/0507293}{{\tt hep-th/0507293}}].

\bibitem{Abe:2023anf}
Y.~Abe, T.~Noumi, and K.~Yoshimura, {\it {Black hole extremality in nonlinear
  electrodynamics: a lesson for weak gravity and Festina Lente bounds}},  {\em
  JHEP} {\bf 09} (2023) 024, [\href{http://arxiv.org/abs/2305.17062}{{\tt
  arXiv:2305.17062}}].

\bibitem{Barbosa:2025smt}
S.~Barbosa, S.~Fichet, and L.~de~Souza, {\it {On The Black Hole Weak Gravity
  Conjecture and Extremality in the Strong-Field Regime}},
  \href{http://arxiv.org/abs/2503.20910}{{\tt arXiv:2503.20910}}.

\bibitem{Palti:2019pca}
E.~Palti, {\it {The Swampland: Introduction and Review}},  {\em Fortsch. Phys.}
  {\bf 67} (2019), no.~6 1900037, [\href{http://arxiv.org/abs/1903.06239}{{\tt
  arXiv:1903.06239}}].

\bibitem{Manohar:1996cq}
A.~V. Manohar, {\it {Effective field theories}},  {\em Lect. Notes Phys.} {\bf
  479} (1997) 311--362, [\href{http://arxiv.org/abs/hep-ph/9606222}{{\tt
  hep-ph/9606222}}].

\bibitem{Manohar:2018aog}
A.~V. Manohar, {\it {Introduction to Effective Field Theories}},
  \href{http://arxiv.org/abs/1804.05863}{{\tt arXiv:1804.05863}}.

\bibitem{Misner:1973prb}
C.~W. Misner, K.~S. Thorne, and J.~A. Wheeler, {\em {Gravitation}}.
\newblock W. H. Freeman, San Francisco, 1973.

\bibitem{ZWIEBACH1985315}
B.~Zwiebach, {\it Curvature squared terms and string theories},  {\em Physics
  Letters B} {\bf 156} (1985), no.~5 315--317.

\bibitem{Hiscock:1990ex}
W.~A. Hiscock and L.~D. Weems, {\it {Evolution of Charged Evaporating Black
  Holes}},  {\em Phys. Rev. D} {\bf 41} (1990) 1142.

\bibitem{Barbosa:2025uau}
S.~Barbosa, P.~Brax, S.~Fichet, and L.~de~Souza, {\it {Running Love Numbers and
  the Effective Field Theory of Gravity}},
  \href{http://arxiv.org/abs/2501.18684}{{\tt arXiv:2501.18684}}.

\bibitem{CarrilloGonzalez:2023cbf}
M.~Carrillo~Gonz\'alez, C.~de~Rham, S.~Jaitly, V.~Pozsgay, and A.~Tokareva,
  {\it {Positivity-causality competition: a road to ultimate EFT consistency
  constraints}},  {\em JHEP} {\bf 06} (2024) 146,
  [\href{http://arxiv.org/abs/2307.04784}{{\tt arXiv:2307.04784}}].

\bibitem{Remmen:2019cyz}
G.~N. Remmen and N.~L. Rodd, {\it {Consistency of the Standard Model Effective
  Field Theory}},  {\em JHEP} {\bf 12} (2019) 032,
  [\href{http://arxiv.org/abs/1908.09845}{{\tt arXiv:1908.09845}}].

\bibitem{Strathdee:1986jr}
J.~A. Strathdee, {\it {EXTENDED POINCARE SUPERSYMMETRY}},  {\em Int. J. Mod.
  Phys. A} {\bf 2} (1987) 273.

\bibitem{Fichet:2013ola}
S.~Fichet and G.~von Gersdorff, {\it {Anomalous gauge couplings from composite
  Higgs and warped extra dimensions}},  {\em JHEP} {\bf 03} (2014) 102,
  [\href{http://arxiv.org/abs/1311.6815}{{\tt arXiv:1311.6815}}].

\bibitem{Drummond80}
I.~T. Drummond and S.~J. Hathrell, {\it Qed vacuum polarization in a background
  gravitational field and its effect on the velocity of photons},  {\em Phys.
  Rev. D} {\bf 22} (Jul, 1980) 343--355.

\bibitem{EH_Lag}
W.~Heisenberg and H.~Euler, {\it Consequences of dirac theory of the positron},
  .

\bibitem{Dunne:2004nc}
G.~V. Dunne, {\em {Heisenberg-Euler effective Lagrangians: Basics and
  extensions}}, pp.~445--522.
\newblock 6, 2004.
\newblock \href{http://arxiv.org/abs/hep-th/0406216}{{\tt hep-th/0406216}}.

\bibitem{Fichet:2014uka}
S.~Fichet, G.~von Gersdorff, B.~Lenzi, C.~Royon, and M.~Saimpert, {\it
  {Light-by-light scattering with intact protons at the LHC: from Standard
  Model to New Physics}},  {\em JHEP} {\bf 02} (2015) 165,
  [\href{http://arxiv.org/abs/1411.6629}{{\tt arXiv:1411.6629}}].

\bibitem{birrell1984quantum}
N.~Birrell and P.~Davies, {\em Quantum Fields in Curved Space}.
\newblock Cambridge Monographs on Mathematical Physics. Cambridge University
  Press, 1984.

\bibitem{Grinstein:2014xba}
B.~Grinstein, D.~Stone, A.~Stergiou, and M.~Zhong, {\it {Challenge to the $a$
  Theorem in Six Dimensions}},  {\em Phys. Rev. Lett.} {\bf 113} (2014), no.~23
  231602, [\href{http://arxiv.org/abs/1406.3626}{{\tt arXiv:1406.3626}}].

\bibitem{Grinstein:2015ina}
B.~Grinstein, A.~Stergiou, D.~Stone, and M.~Zhong, {\it {Two-loop
  renormalization of multiflavor $\phi^3$ theory in six dimensions and the
  trace anomaly}},  {\em Phys. Rev. D} {\bf 92} (2015), no.~4 045013,
  [\href{http://arxiv.org/abs/1504.05959}{{\tt arXiv:1504.05959}}].

\bibitem{SCHWARZ1982223}
J.~H. Schwarz, {\it Superstring theory},  {\em Physics Reports} {\bf 89}
  (1982), no.~3 223--322.

\bibitem{Polchinski:1998rr}
J.~Polchinski, {\em {String theory. Vol. 2: Superstring theory and beyond}}.
\newblock Cambridge Monographs on Mathematical Physics. Cambridge University
  Press, 12, 2007.

\bibitem{deRoo:2003xv}
M.~de~Roo and M.~G.~C. Eenink, {\it {The Effective action for the four point
  functions in Abelian open superstring theory}},  {\em JHEP} {\bf 08} (2003)
  036, [\href{http://arxiv.org/abs/hep-th/0307211}{{\tt hep-th/0307211}}].

\bibitem{Fradkin:1985qd}
E.~S. Fradkin and A.~A. Tseytlin, {\it {Nonlinear Electrodynamics from
  Quantized Strings}},  {\em Phys. Lett. B} {\bf 163} (1985) 123--130.

\bibitem{Davila:2013wba}
J.~M. Davila, C.~Schubert, and M.~A. Trejo, {\it {Photonic processes in
  Born-Infeld theory}},  {\em Int. J. Mod. Phys. A} {\bf 29} (2014) 1450174,
  [\href{http://arxiv.org/abs/1310.8410}{{\tt arXiv:1310.8410}}].

\bibitem{Born:1934gh}
M.~Born and L.~Infeld, {\it {Foundations of the new field theory}},  {\em Proc.
  Roy. Soc. Lond. A} {\bf 144} (1934), no.~852 425--451.

\bibitem{Fichet:2013gsa}
S.~Fichet, G.~von Gersdorff, O.~Kepka, B.~Lenzi, C.~Royon, and M.~Saimpert,
  {\it {Probing new physics in diphoton production with proton tagging at the
  Large Hadron Collider}},  {\em Phys. Rev. D} {\bf 89} (2014) 114004,
  [\href{http://arxiv.org/abs/1312.5153}{{\tt arXiv:1312.5153}}].

\end{thebibliography}\endgroup

\end{document}